%% file: ebikeThesis.tex
\documentclass[thesis,twoside,a4paper,12pt]{title} 
\makeindex
\title{A situation aware electric bike for smart cities}	       
\author{Shaun Sweeney}                
\copyrightYear{2017}            
\graduationMonth{April}           
\supervisor{Prof. Robert Shorten \\ Prof. David Timoney}	    
\nomProf{My Nominating Professor}	    
\headSchool{Prof. David Timoney} 
\major{Energy Systems Engineering}	
\degree{Master of Energy Systems Engineering}	    
\college{Engineering and Architecture}           
\dept{Electrical, Electronic and Communications Engineering}	
\university{University College Dublin}	
\numberOfCommitteeMembers{2} 
\coSupervisorA {My Co-Supervisor}	
\committeeMemberB {Committee Member 2}	
\committeeMemberC {Committee Member 3}	
\committeeMemberD {Committee Member 4}	
\committeeMemberE {Committee Member 5}  
\usepackage{nomencl}                    
\usepackage{float}  
\usepackage{graphicx}					

\graphicspath{ {figures/}{figures/eps/}{figures/pdf/} }
\usepackage{epstopdf}					
\usepackage{fancyhdr}                   
\usepackage{textcomp}
\usepackage{fixltx2e}
\usepackage{url}  
\usepackage[inactive]{srcltx}		 	
\usepackage{relsize}                    
\usepackage{booktabs}                   
\usepackage[config, font={it}, labelfont={bf}]{caption} 
\usepackage{subcaption}
\usepackage{mathrsfs}
\usepackage{hyphenat}                   
\usepackage[Conny]{fncychap}	
\usepackage{rotating}
\usepackage{pdflscape}
\usepackage{rotating}
\usepackage{multirow}
\usepackage{booktabs}
\usepackage[parfill]{parskip}

\let\oldfootnote\footnote
\renewcommand{\footnote}{\unskip\oldfootnote}

\usepackage[style=ieee,citestyle=numeric,backend=biber,backref=true]{biblatex}
\usepackage{listings}
\usepackage{color}

\definecolor{dkgreen}{rgb}{0,0.6,0}
\definecolor{gray}{rgb}{0.5,0.5,0.5}
\definecolor{mauve}{rgb}{0.58,0,0.82}

\usepackage{titlesec}
\usepackage{lipsum}
\usepackage{afterpage}

\titleformat{\chapter}{\Huge\bfseries}{\chaptername\ \thechapter}{0pt}{\vskip 10pt\raggedright}%
\titlespacing{\chapter}{0pt}{0pt}{20pt}%
\titlespacing{\section}{0pt}{20pt}{10pt}%
\titlespacing{\subsection}{0pt}{10pt}{10pt}%

\newcommand\blankpage{%
	\null
	\thispagestyle{empty}%
	\addtocounter{page}{-1}%
	\newpage}

\begin{document}
    \pagenumbering{alph} 
    %
    %
    \makeTitlePage 
    \pagenumbering{roman}
    \setcounter{page}{1}
    %
    %
    %
    \afterpage{\blankpage}
    \include{front-matter/copyright}
    \afterpage{\blankpage}
    \include{front-matter/acknowledgements} 
    %
    %
    \include{front-matter/abstract} 
    \include{front-matter/outputs}
    %
    \tableofcontents 
    %
    %
    \listoffigures 
    %
    %
    \newpage
    \pagenumbering{arabic}
    \setcounter{page}{1}

    \include{chapters/ch_introduction}

    \include{chapters/ch_literature_review}

    \include{chapters/ch_system_design}
     \include{chapters/ch_modelling}
     \include{chapters/ch_open_loop_applications}

     \include{chapters/ch_closed_loop_applications}
	\include{chapters/ch_conclusions}

	\printbibliography

%

\end{document}

%% file: front-matter/copyright.tex
\chapter*{Copyright Declaration}\label{ch:copyright}

This  thesis  has  been  composed  by  the author and has not been previously submitted for examination which has led to the award of a degree. \\

Due  acknowledgement  must  always  be made for the use of any material contained in, or derived from this thesis. \\

%% file: front-matter/acknowledgements.tex
\chapter*{Acknowledgements}

This project was completed with the help of a number of people. I was fortunate to be jointly supervised by Prof. Robert Shorten and Prof. David Timoney. Thank you to Bob for his enthusiasm throughout the project, for always making time and for his ten new ideas per week. Thank you to David for always having an open door, a lot to say about torque and a newfound expertise in spirometry equipment. Thank you to both for having complementary temperaments.

The contributions of Liam, Luke and Declan in the UCD Electronics workshop were essential for the hardware components of this project with Liam commenting that I had ``everyone in the workshop working on the project'' at one point. A special word of thanks to Luke for showing me how to solder properly. Brian Mulkeen was on hand to deal with multiple hardware crises which were all resolved with thorough, exhaustive logic. Thank you to John Gahan in the UCD Mechanical Engineering department for his stories, for providing somewhere to keep the bike for the first semester and for always knowing how to fix everything. Thank you to Giovanni Russo from IBM Research for his advice in the design of the control system.

Thank you to Rodrigo~Ord\'o\~nez-Hurtado who played a big role in trying to figure out how to use maths to describe an electric bike and then use that maths to get the electric bike to do what we wanted. Thank you to Wynita Griggs for all of her help with Android Studio, SUMO and in getting ready for the demonstration in Trinity College in March. Thank you to Francesco Pilla for his input on the air pollution use case.

To my parents, Tommy and Elizabeth, who were my first educators, thank you for 23 years of advice, and 19 years of support during formal education. I promise I will get a job now. A sincere word of thanks to JP McManus, the All Ireland Scholarship programme and the UCD Ad Astra Academy who helped to support me financially during my first four years of study in UCD.

%% file: front-matter/abstract.tex
\chapter*{Abstract}\label{ch:abstract}
	There are a number of problems in the area of transportation in cities around the world today. Problems relating to air pollution, energy sustainability and energy security have initiated a widespread transition to implementing e-mobility solutions. Electric bikes have the potential to contribute to a number of transport solutions. The objective of this project is to research and implement different services that an electric bike can provide to help and protect people in cities. A commercially available electric bike is modified by equipping it with sensors to enable it to gather data from its surroundings. The electric bike is then connected to a smartphone which provides the electric bike with more data. It is demonstrated that these insights can be used to modify the behaviour of the electric bike in a responsive way. The modified electric bike has the potential to provide a number of different services in the areas of energy management, health connectivity, smart navigation and reducing the harmful effects of air pollution. In this project, a new method which reduces a cyclist's inhalation of air pollutants while cycling through a polluted area is proposed and validated. This method relies heavily on an electric bike energy model that is derived here. The implementation is based on closed loop control which regulates the proportion of power that a cyclist is providing to move the wheel of the bike. It is shown that a specific cyclist's breathing rate can be approximately halved while cycling through a polluted area by implementing this method. As such, the cyclist's inhalation of harmful air pollutants will be reduced, which will have a positive impact on their health.\\
	\vspace{30pt}
	
	This work was in part supported by SFI grant 11/PI/1177.

%

%% file: front-matter/outputs.tex
\newpage
\section*{Outputs} \label{ch:outputs}
	\begin{itemize}
		\item Project demonstrated for the Fraunhofer Society as part of Morgenstadt city of the future initiative in Dublin, March 2017.\\ http://www.morgenstadt.de/en/events/morgenstadt--city-insights-meeting-dublin.html
		\item Project presented at the Eighteenth Yale Workshop on Adaptive and Learning Systems at Yale University, USA in June 2017. \\
		https://www.eng.yale.edu/css/ \\
		Conference paper: https://arxiv.org/abs/1706.00646 
		\item Video demonstrating the system \\
		https://youtu.be/265u9KO-9QE
		\item IBM Research Blog \\``Hacking an e-bike to help cyclists avoid breathing in polluted air''\\
		https://www.ibm.com/blogs/research/2017/08/hacking-an-e-bike/
	\end{itemize}

\vspace{50pt}	

\section*{Referencing} \label{ch:referencing}
\begin{itemize}
	\item Web pages are referenced using footnotes as they are mentioned in the text.
	\item Journal and conference papers are referenced using the IEEE citation style when mentioned in the text and have a full citation in the Bibliography at the end of the document.
\end{itemize}

%% file: chapters/ch_introduction.tex
\chapter{Introduction} \label{ch:introduction}
\setstretch{1.3}

\section{Motivation} \label{sec:motivation} 

	There are a number of transportation problems in cities today. Cars that rely on fossil fuels release pollutants into the atmosphere which impacts negatively on people's health. In recent times, major cities such as Paris and Stuttgart have banned cars at certain times because air pollutant levels were unsafe for humans. Other major cities such as Madrid, Athens and Mexico City have plans to phase out diesel engine vehicles by 2025. There are also other problems with cars in cities such as congestion and parking. These, and other reasons are motivating a transition to e-mobility. E-mobility refers to transportation that is powered by electricity. 
	
	Electric bikes could help to solve a number of the above-outlined problems. They are powered by electricity and therefore do not emit incremental air pollutants. They are also physically smaller than cars and do not require dedicated charging infrastructure since the battery is removable and can be charged at a standard wall socket. The other trend that is important to understand is the Internet of Things trend. Devices are becoming connected thanks to developments in sensor technology and the availability of fast internet. This trend adds the potential for electric bikes to provide new context aware e-mobility services.
	
 \section{About the project} \label{sec:about}
 
	 The project aimed to demonstrate different services that electric bikes can provide to help and protect people in cities. Chapter \ref{ch:litreview} reviews theory from previous academic studies that is relevant to the project. 
	 
	 The first part of the project was to define, design and implement the concept of a smart electric bike. The two main tasks involved were to 1) create a situation aware electric bike by equipping it with sensors 2) enable the bike to respond to data from the sensors. These tasks involved both hardware and software components and are discussed in Chapter \ref{ch:sysdesign}. 
	 
	 The next part of the project was to gain an understanding of the behaviour of the system and to demonstrate specific services that the electric bike can provide. A model of the system is developed in Chapter \ref{ch:modelling}. Use cases are introduced and implemented in Chapter \ref{ch:openloop} Open Loop Applications and Chapter \ref{ch:closedloop} Closed Loop Applications. Chapter \ref{ch:conclusions} summarises the achievements of the project and outlines areas for future work.

\section{Objectives} \label{sec:objectives}
The objectives of the project can be split in two parts which are listed below.
\subsection{Part A: System design} \label{sec:sysdesign}
	\begin{itemize}
		\item Research and purchase a suitable electric bike that can accomplish the objectives of the project
		\item Design and implement a system that enables the electric bike to be situation aware
		\item Design and implement a system that enables the electric bike to respond to data gathered from its surroundings to deliver services
	\end{itemize}
\subsection{Part B: Use cases} \label{sec:usecases}
	\begin{itemize}
		\item Research potential services that electric bikes can provide in cities to help and protect people
		\item Produce a model of the system to understand its behaviour 
		\item Demonstrate some of the potential use cases that have been identified
	\end{itemize}

%% file: chapters/ch_literature_review.tex
\chapter{Literature Review} \label{ch:litreview}
\setstretch{1.4}

\section{Structure of the literature review} \label{sec:lit_review_structure}

Chapter \ref{ch:introduction} outlined the motivation behind the work to follow and the objectives of this project. Much of the design, and many of the proposed use cases can benefit from existing domain specific knowledge. This was gathered by carrying out a review of previous studies. This chapter discusses methods and useful results from those studies. Of significant interest, is to model energy in cycling and manage the way that energy is consumed, this topic is discussed in Section \ref{sec:lr_energy_models}. A central use case is to minimise a cyclist's inhalation of air pollutants while cycling, therefore it is a priority to have an understanding of issues such as 1) the link between air pollutants and human health 2) a cyclist's exposure to air pollutants while cycling 3) existing methods to minimise a cyclist's inhalation of pollutants and 4) methods to validate that a cyclist's inhalation of pollutants has been minimised. These and related issues are discussed in Sections \ref{sec:lr_air_pollution_health}, \ref{sec:lr_cycling_pollution}, \ref{sec:lr_cycling_minimise_pollution} and \ref{sec:lr_validation_hr}. A final concern of this review is with regards to smart routing which is useful for a number of different use cases. This idea is that a cyclist may take a number of different routes to arrive at their final destination, but that somehow some of these routes are better than the others. The best route accomplishes some optimisation objective e.g. the route which minimises the cyclist's inhalation of pollutants. Smart routing is discussed in Section \ref{sec:lr_routing}. This section particularly draws on important results from graph theory.

\section{Energy models for cycling} \label{sec:lr_energy_models}
	Many of the proposed use cases involve managing the way in which energy from the electric bike battery is provided to the cyclist. As such, it is useful to have an appreciation of the variables that influence the amount of power that is required for cycling. Cycling requires power to be provided to the wheel of the bike. Power provided to the wheel of the bike is calculated as
	
	\begin{equation*}
		\text{Power (Watts)} = \text{Torque (Nm)} \times \text{Wheel Speed (rad/s)}.
	\end{equation*}
	
	This power can be thought of as the power that is required to overcome a number of resistive forces. Having an understanding of these forces is therefore essential to being able to predict the amount of power that is required to travel on a journey. These resistive forces are discussed in Martin's 1998 paper \cite{martin_validation_1998}. In this paper, aerodynamic drag, rolling resistance, friction in the bicycle drive system and changes in kinetic and potential energy are modelled. These are external factors that change the energy input required from a cyclist. It is found that of these factors, aerodynamic drag is the biggest factor. Aerodynamic drag varies with air velocity (wind), air density and exposed frontal area and shape.
	 
	 The net cycling power input from a cyclist is modelled in \cite{martin_validation_1998} as	 
	 \begin{equation*}
		 P_{\text{Net}} = P_{AT} + P_{RR} + P_{WB} + P_{PE} + P_{KE},
	 \end{equation*}
	 where $P_{AT}$ is aerodynamic power, $P_{RR}$ is rolling resistance power, $P_{WB}$ is power lost to bearing friction torque, $P_{PE}$ is power associated with changes in potential energy and $P_{KE}$ is power associated with changes in kinetic energy. Equations are derived for each of these terms and parameter values are established. The model was verified by comparing predictions made by the model to measurements made by a measurement system (SRM training system) using a linear regression model and it was found that the two were highly correlated with $R^2=0.97$.
	 
\begin{figure}[H]
	 		\centering
	 		\includegraphics[width=3in]{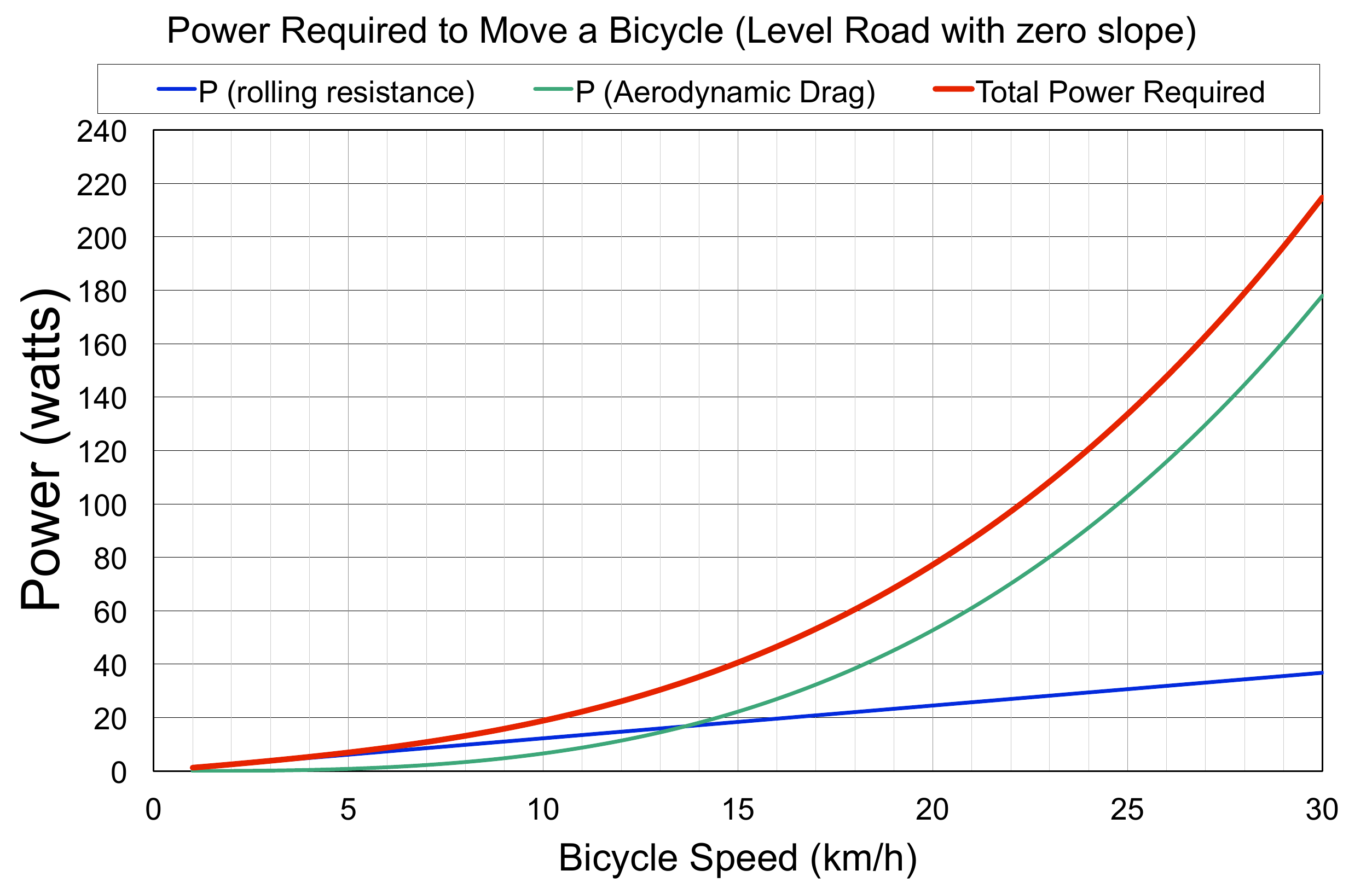}
	 		\caption{Power required to overcome forces of aerodynamic drag and rolling resistance.} 
	 		\label{fig:david_power}
	 	\end{figure}

	 Graphically, power that is required to overcome the forces of aerodynamic drag $P_{AT}$ and rolling resistance $P_{RR}$ are plotted in Figure \ref{fig:david_power} \footnote{Graph reproduced by Prof. David Timoney based on data from http://www.avdweb.nl/solar-bike/hub-motor/hub-motor-simulation.html}.
	 
	 Reference \cite{Le2009} (Le's 2009 paper) proposes another model for power involved in cycling as part of a group. Energy involved in group cycling may be of interest for many use cases. In group training, the group speed must be the same for all cyclists. But, individual cyclists have different goals and abilities in terms of cycling performance. This paper discusses some of the energy considerations involved in group cycling. 
	 
	 Like Martin's paper \cite{martin_validation_1998}, Le's paper \cite{Le2009} models cycling power in terms of the power required to overcome a number of resistive forces presented by the cycling environment. These forces include air resistance, rolling resistance, gravity, intertia and frictional losses from the drivechain and wheel bearings. The air resistance $F_W$ is modelled as
	 \begin{equation*} \label{eqn:air_resistance}
		 F_W = \frac{1}{2}\rho c_d A_p (v_b + v_W)^2,
	 \end{equation*}
	 where $\rho$ is the air density, $c_d$ is the drag coefficient, $A_p$ is the projected frontal area of the cyclist and bicycle, $v_b$ is the bicycle speed, $v_W$ is the wind speed.	 
	 
\begin{figure}[H]
    	\centering
    	\includegraphics[width=0.7\textwidth]{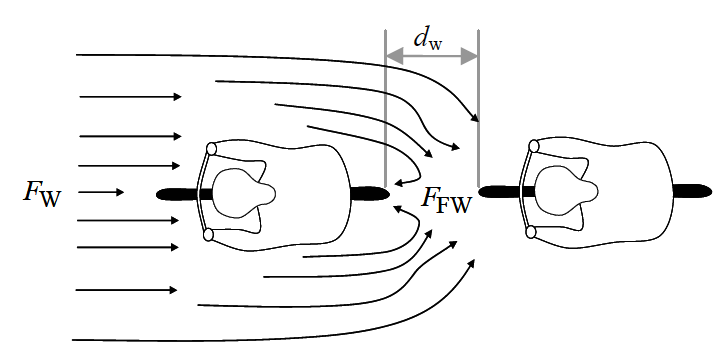}
    	\caption{Slip stream of the lead cyclist from \cite{Le2009}.} 
    	\label{fig:group_cycling_slip_stream}
	\end{figure}

	 In group cycling, the air resistance force to overcome depends on the cyclist's position in the group. The leading cyclist must overcome a larger air resistance force than cyclists cycling in the wake (behind the leading cyclist). Figure \ref{fig:group_cycling_slip_stream} from \cite{Le2009} illustrates the slip stream of the leading cyclist.
	 
	 This effect is called `drafting' and a correction factor for the reduction in air resistance force by drafting is used in \cite{Le2009} (which is taken from \cite{Olds1998}) as
	\begin{equation} \label{eqn:cf_draft}
	 	  CF_{\text{draft}} = 0.62 - 0.0104d_w + 0.0452d_w^2,
	\end{equation}
	 where $d_W$ is the wheel-to-wheel distance between two cyclists. As such, the air resistance with drafting is given in as
	 \begin{equation*} \label{eqn:air_resistance_drafting}
	 	 F_{FW} = F_WCF_{\text{draft}}.
	 \end{equation*}
	 The next force, the force to overcome rolling resistance, can be modelled as
	 \begin{equation*} \label{eqn:rolling_resistance}
		 F_R = c_R(M_c + M_b)\text{cos}(\text{arctan}(G_R)),
	 \end{equation*}
	 where $c_R$ is the rolling resistance coefficient, $M_c$ is the weight of the cyclist, $M_b$ is the weight of the bicycle and $G_R$ is the road gradient.
	 
	 If the cycling route is not flat, additional force is required to overcome the earth's gravity $F_G$,. It depends on the mass of the rider and bike and is given as 
	 \begin{equation*} \label{eqn:gravity_force}
		 F_G = (M_c + M_b)g \text{sin}(\text{arctan}(G_R)),
	 \end{equation*}
	 where $g$ is the acceleration due to gravity.
	 
	 $F_A$ is the force due to speed variation given as
	 \begin{equation*} \label{eqn:force_speed_variation}
		 F_A = (M_c + M_b)a,
	 \end{equation*}
	 where $a$ is the acceleration due to the bike speed variation.

	The mechanical efficiency $c_m$ is introduced to model resistance caused by the spokes of the wheel slicing through the air and frictional losses in the wheel bearings and drive chain. These losses are typically small. Combining the equations that have been shown so far, the total power output of the leading cyclist is therefore modelled as $P_{LT}$ which is defined as
	  \begin{equation*} \label{eqn:power_leading_cyclist}
		  P_{LT} = (F_W + F_R + F_G +F_A)v_b/c_m.
	  \end{equation*}
	  
	 The total power output $P_{DT}$ of the cyclists in drafting is modelled as
	 \begin{equation*} \label{eqn:power_drafting_cyclist}
		 P_{DT} = (F_WCF_{\text{draft}}+F_R+F_G+F_A)v_b/c_m,
	 \end{equation*}
	 which contains the extra $CF_{\text{draft}}$ factor as defined in Equation \eqref{eqn:cf_draft}.
	 
	 In order to enable each cyclist to achieve their personal cycling goal while cycling in a group, three control signals can be defined. These control signals are 1) the cyclist's position in the group 2) the group speed and 3) the number of subgroups in the group. The paper goes on to show how these control signals can be managed to enable an individual cyclist to achieve their personal goal, while keeping the group speed the same. 
	 
\begin{figure}[H]
		\centering
		\includegraphics[width=0.8\textwidth]{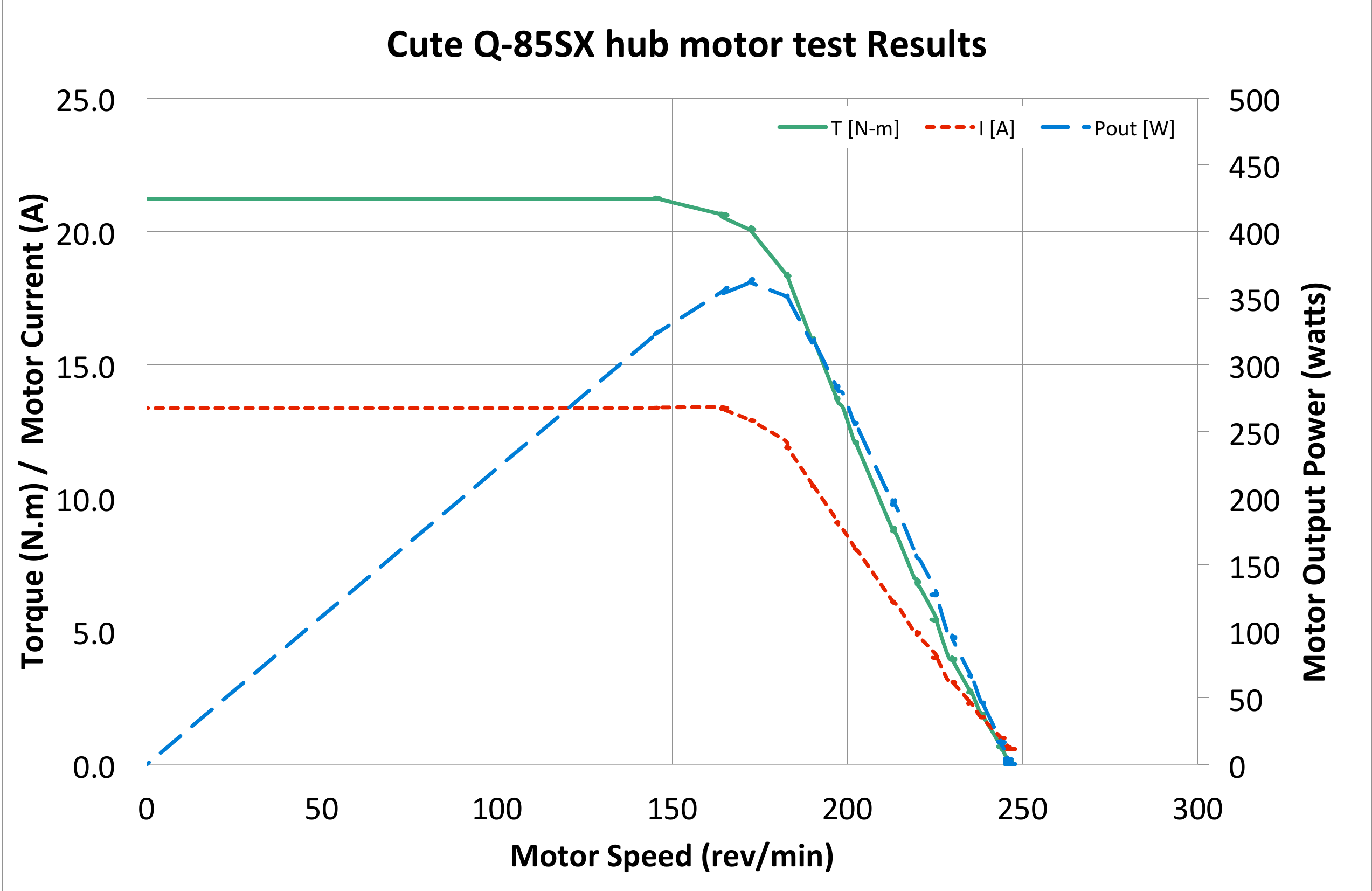}
		\caption{Plot of motor torque (Nm) and output power (W) plotted against motor speed (RPM).} 
		\label{fig:david_motor}
	\end{figure}

	 The knowledge gathered from \cite{martin_validation_1998} and \cite{Le2009} provides a basis for understanding power that is required for cycling. The work to follow is concerned with electric bikes, as such, this power can be provided from one of two energy reservoirs 1) the electric bike battery or 2) the human cyclist. In an electric bike, chemical energy stored in a battery is converted to electrical energy which is converted by the motor to mechanical energy. A number of different types of electric bike motor exist but this review will not focus on these details. The general relationship between speed, torque and power in an electric bike motor is of interest. Figure \ref{fig:david_motor} \footnote{Plot reproduced by Prof. David Timoney based on data from  http://www.avdweb.nl/Article\_files/Solarbike/Q-85SX-test/Q-85SX-torque-speed.jpg} shows this characteristic relationship.
	 
	Also of interest is deciding which energy source in the electric bike should provide power in different situations. An electric bike is a hybrid electric vehicle. Analogous methods have been developed for hybrid electric cars which model energy management systems in hybrid electric vehicles. One such method based on ``Lagrange's calculus'' is proposed in \cite{morchin_energy_1998}. This paper presents a method to optimise the load division between the car's engine and battery by tracking the energy content of the battery over a specified route and travel time. This method could have applications in this project.


\section{Air pollution and human health} \label{sec:lr_air_pollution_health}
	The link between human health and air pollution has been well studied in numerous academic journal publications.
	Kampa's 2008 paper \cite{Kampa2008} recalls the definition of an air pollutant as ``any substance which may harm humans, animals, vegetation or material''. Noting that, harm for a human is determined based on the results of clinical, epidemiological or animal studies which have demonstrated that exposure to a substance is associated with health effects. The paper states that changes in atmospheric composition (addition of pollutants) are primarily caused by anthropogenic activities, namely the combustion of fossil fuels. The paper goes on to categorise these pollutants into: 
	\begin{enumerate}
		\item Gaseous pollutants e.g. SO$_2$, NOx, CO, ozone, VOCs;
		\item Persistent organic pollutants e.g. dioxins;
		\item Heavy metals e.g. lead, mercury;
		\item Particulate Matter (PM).
	\end{enumerate}
	
	The paper states that diverse health effects have been observed in humans due to the variation in compositions of air pollutants, the dose and time of exposure and the fact that humans are usually exposed to pollutant mixtures rather than to a single pollutant. Anatomical systems that are known to have been affected by atmospheric pollutants include the respiratory system, the cardiovascular system, the nervous system, the urinary system and the digestive system, of which the cardiovascular and respiratory systems have been observed to be worst affected. Specific symptoms and diseases have included minor upper respiratory irritation, chronic lung disease, chronic heart disease, lung cancer, acute respiratory infections in children and chronic bronchitis in adults, aggravating pre-existing heart and lung disease or asthma attacks. In addition, short and long term exposures have also been linked with premature mortality and reduced life expectancy. 
	
	The link between PM and human health has been the subject of a number of recent studies. PM is a generic term used for a type of pollutant that consist of a complex and varied mix of particles suspended in breathing air. The paper notes ``major sources of particulate pollution are factories, power plants, refuse incinerators, motor vehicles, construction activity, fires and natural windblown dust''. The size of the particles varies, PM$_{10}$ defines a category of PM with aerodynamic diameter smaller than 10$\mu$m, PM$_{2.5}$ refers to PM with an aerodynamic diameter smaller than 2.5$\mu$m. Other definitions for these pollutants include referring to them as ultrafine, fine or coarse particles, which is again based on particle size. The major components of PM are metals, organic compounds, material of biological origin, ions, reactive gases and the particle carbon core. There is strong evidence to suggest that ultrafine and fine particles are more harmful to human health than larger ones as these particles can travel further into the respiratory tract. Coarse particles deposit mainly in the upper respiratory tract but fine and ultrafine particles can travel further and reach the lung alveoli. \cite{Kampa2008} 
	
	Seaton's 1995 paper \cite{Seaton1995} published in the Lancet is in agreement with the harmful effects of PM noting that ``epidemiological studies have consistently shown an association between particulate air pollution and not only exacerbations of illness in people with respiratory disease but also rises in the number of deaths from cardiovascular and respiratory disease among older people''. A hypothesis is proposed  for the mechanism through which particulate air pollution can cause illness in humans. It is hypothesised that ultrafine particle characteristics of air pollution provoke alveolar inflammation which causes changes in blood coagulability (ability of the blood to clot) and releases mediators which are able to provoke attacks of acute respiratory illness. These blood changes result in an increase, in the exposed population's susceptibility to acute episodes of cardiovascular disease.
	
	Of particular note in recent times is a 2017 study by Chen \cite{Chen2017} which was published in the Lancet. This study found that living near major roads was associated with a higher incidence of dementia. The study aimed to determine the association between traffic proximity and incident dementia, Parkinson's disease and multiple sclerosis. The paper introduces by stating that living near traffic is a multi-faceted exposure representing heightened exposure to nitrogen oxides, ultrafine particles, fine particulate matter, heavy metals, polycyclic aromatic hydrocarbons, volatile organic components, noise and other factors. 
	
	The study was a large population based study of all adults aged 20-85 years living in Ontario, Canada. The study was split into two cohorts based on the ages of the participants. The first cohort was made up of adults aged 55 years or older as this corresponds to the ages that dementia and Parkinson's disease are most likely to occur. The second cohort was made up of adults younger than 50 years of age because multiple sclerosis is most common in this age group. The database was populated using Ontario's Registered Participant's database which is a database of all residents who have ever had health insurance. Individuals with any of the three disorders at baseline were excluded. This yielded 4,372,720 in cohort 1 and 2,165,268 in cohort 2.
	
	Proximity to major roadways or highways was calculated based on 6 character postal codes. PM$_{2.5}$ ground level concentrations were derived from satellite observations of aerosol optical depth. Diagnoses of diseases was obtained using validated databases that were created from the Canadian Institute for Health Information, physician service claims from the Ontario Health Insurance Plan database and prescription medication claims from the Ontario Drug benefits programme database. The cohorts were linked to the database using encrypted unique identifiers.
	
	Statistical analyses and sensitivity analyses were conducted and it was found that those living in major urban centres were observed to have an increased incidence of dementia. No association was found between roadway proximity and incidence of Parkinson's disease or multiple sclerosis. In the discussion, the paper states that the mechanism through which traffic exposure might affect brain health are unknown, but comments that systemic inflammation arising from traffic-related air pollution is probably important.
		
	The literature on the links between air pollution and human health is extensive and consistently shows the strong negative impacts that air pollution has on human health that have been outlined above. For further reading, two review papers present comprehensive reviews of the important literature 1) Englert's 2004 review paper \cite{Englert2004} published in Toxicology letters and 2) Brunekreef's 2002 review paper \cite{Brunekreef2002} published in the Lancet. 
	

%
%
%
%
	 


\section{Cycling and air pollution} \label{sec:lr_cycling_pollution}
	Having established from Section \ref{sec:lr_air_pollution_health} that air pollutants have extremely negative impacts on human health, it is then of interest to study the relationship between air pollutants and cycling. This section aims to establish if a cyclist is particularly at risk of inhaling increased amounts of air pollutants by cycling than other road users. 

	Cycling is a form of exercise and as such instinctively should be associated with health benefits from increased cardiovascular activity. However, cycling in urban environments also poses a number of risks and potential negative consequences for human health. Two of these risks are:
	\begin{enumerate}
		\item Increased probability of inhaling harmful air pollutants due to cyclists having elevated breathing rates.
		\item Safety concerns from increased vulnerability and likelihood of being involved in a road traffic accident. 
	\end{enumerate}	
	De Hartog's 2010 paper \cite{de_hartog_health_2010} tries to address the net benefit of cycling by aiming to answer the question ``Do the health benefits of cycling outweigh the risks?'', where societal and individual benefits and disadvantages are considered. The paper questioned the assumption that there will be a net positive societal benefit due to a transport mode shift from car to bike through reduced emissions from motor vehicles, emphasising that the benefit to individual cyclists is more subtle. The benefits such as increased physical activity had to be quantified and compared with the potential negative impacts of cycling such as the increased risk of being involved in an accident and the risk of inhaling increased amounts of harmful pollutants. Overall the study found a net positive impact to individuals by shifting to cycling which was principally driven by the benefits of physical activity to an individual such as decreased risk of cardiovascular disease.
	
	De Hartog's paper establishes there is a net positive benefit to cycling. However, it is logical that the overall net benefit of cycling can be increased by minimising a cyclist's inhalation of pollutants. Int Panis's 2010 paper \cite{IntPanis2010} investigated the relationship between the amount of PM that cyclists inhaled compared to car passengers. The study was conducted in three different Belgian regions - Brussels, Louvain-la-Neuve and Mol. Subjects in the study were first driven by car and then cycled along identical routes in a pairwise design with the aim of comparing lung deposition of particle number concentrations (PNC) and PM between car trips and biking trips. The concentrations of atmospheric PNC, PM$_{10}$ and PM$_{2.5}$ and respiratory behaviour of the participant were simultaneously measured. PNC concentrations were made using P-Trak UFP Counters which measures particles in the size range 0.02 - 1$\mu$m. PM$_{10}$ and PM$_{2.5}$ were measured using the TSI DustTrak DRX model 8534. Respiratory measurements of the subjects were recorded	using a portable cardiopulmonary indirect breath-by-breath calorimetry system. The respiratory variable that was used for comparison was minute ventilation (VE) which was calculated as breathing frequency $\times$ tidal volume.
	
	The subjects were healthy, commuter cyclists aged between 18-65 years old. 55 test subjects took part in the study in total. After the test was conducted, some results were excluded due to unreliable measurements or equipment failure. The final dataset included 24 cyclists in Brussels (Male = 15, Female = 9), 6 cyclists in Louvain-la-Neuve (Male = 5, Female = 1) and 13 cyclists in Mol (Male = 8, Female = 5).
	
	Atmospheric concentrations of PNC and PM measurements were similar for both cycling and car journeys across all three locations. However, breathing frequency, breathing volume and journey time were all greater for cyclists than for cars. The study also made use of the result from \cite{Daigle2003} which showed that lung deposition fractions increases strongly with exercise. This meant that even though the atmospheric concentrations were similar, quantities of particles inhaled by cyclists were between 400\% and 900\% higher compared to car passengers on the same route. The longer duration of the cycling trip also increased the inhaled doses.
	
\section{Minimising a cyclist's inhalation of air pollutants} \label{sec:lr_cycling_minimise_pollution}

	Having established that atmospheric pollutants from cars are harmful and that cyclists inhale larger amounts of these pollutants while cycling, it is then of interest to understand how a cyclist can minimise their inhalation of pollutants while cycling. There are a number of ways that this objective can be achieved. These ways can be either passive or active. A sensible approach would be to reduce the concentration of pollutants in the environment. This can be achieved in a number of ways e.g. by replacing motor vehicles that run on fossil fuels with electric and hybrid-electric vehicles. Hybrid-electric vehicles can minimise their release of pollutants by ensuring that they are operating in electric engine mode when they enter an area where cyclists are present, this can be done by making the cars context aware \cite{Herrmann2017}.
	
	Another approach would be to enable the cyclist to avoid areas with high concentrations of pollutants. This could also be done in a number of different ways. Some options would be 1) making real-time pollution concentration data publicly available and letting the cyclist check the data and choose their route based on this data or 2) through smart routing where a cyclist's bicycle would automatically use this data to plan the route for the cyclist to take to minimise their inhalation of pollutants. Literature relating to smart routing is presented in Section \ref{sec:lr_routing}. A third approach would be to optimise the cyclist's cycling performance through a polluted area in order to minimise their inhalation of pollutants. Again, there are a number of ways to do this 1) a new approach based on situation aware bikes is developed in this project and is discussed in Chapter  \ref{ch:openloop} and Chapter \ref{ch:closedloop} or 2) changing the cyclist's speed of travel through a polluted area to minimise their inhalation of pollutants, this approach will now be discussed.
	
	Bigazzi published two papers in 2016 (\cite{bigazzi_breath_2016} and \cite{AlexanderYBigazzi2017}) which aimed to determine 1) biomarkers that could be used to assess a cyclist's uptake of pollutants in urban environment and 2) optimal cycling speeds to minimise a cyclist's inhalation of those biomarkers. Bigazzi's paper \cite{bigazzi_breath_2016} aimed to determine a way to assess the inhalation of harmful pollutants by cyclists for trips through different kinds of areas. The study identified 26 volatile organic compounds (VOCs) such as CFCs, benzene, styrene and carbon disulfide and compared the amounts of these compounds that were present in ambient air with the amounts that were present in a cyclist's breath after cycling through an area. The study identified 8 of these VOCs as being potentially useful breath biomarkers. 
	
	The study found statistically significant increases (compared to background levels) in the concentration of these biomarkers in the breath of cyclists after cycling in high-traffic streets and industrial areas. No statistically significant increases in concentrations of these breath biomarkers were detected after cycling through low traffic streets or off-street paths. Interestingly all 8 biomarkers that were proposed (benzene, toluene, ethylbenzene, xylenes etc.) are known to be present in emissions from motor vehicle exhausts and are known to present some level of toxicity in humans. 
	
	In \cite{AlexanderYBigazzi2017} the ideas from \cite{bigazzi_breath_2016} are built upon in order to propose optimal cycling speeds in order to minimise exposure to harmful pollutants. Higher cycling speeds leads to higher breathing rates but shorter exposure duration for a given journey which affects levels of pollutants that are inhaled. The paper proposed the concept of minimum-dose speeds (MDS) which is the optimal speed for a cyclist to cycle at to minimise their inhalation of pollutants. 
	
	Different MDS are proposed for different demographics and different types of cycling terrain. For example the MDS for a female under 20 years of age on flat terrain was found to be 12.5km/hr and the MDS for a male aged 20-60 years was found to be 15km/hr on the same flat terrain. Bigazzi found that cycling at speeds 10km/hr greater than the MDS can cause double the amount of pollutants to be inhaled by cyclists over a fixed distance.
		
	\section{Validation of minimised pollution} \label{sec:lr_validation_hr}
	
	Section \ref{sec:lr_cycling_minimise_pollution} discussed methods that can be used to minimise a cyclist's inhalation of these pollutants. This is a motif that will be central throughout this project. Validating that a cyclist's inhalation of pollutants has been minimised using spirometry equipment would involve measuring the cyclist's breathing which would not be practical for real life cycling situations. As such, it is of interest to have an alternative validation approach that can be used for real life cycling situations. It is well known in the medical community that the cardiac system and the respiratory system behave as coupled biological oscillators \cite{Bahmed2016}. Simply put, during exercise the muscles of the body are working harder and need to produce more energy (ATP) for muscle contraction. To do this they need more oxygen and so more air must be breathed in. This can be roughly interpreted that a cyclist's pulse rate should increase as their breathing rate increases while cycling and vice versa.
	
	A cyclist's heart rate (pulse rate) is much easier and less intrusive to record during cycling than measuring the volume of air that a cyclist has inhaled. This is particularly so due to the growth of wearable activity trackers  (See \cite{NikolicPopovic2011}). These activity trackers typically record pulse rates measured at the wrist. For these reasons, heart rate will be used as an approximation to validate that a cyclist's inhalation of pollutants has been minimised. The author does wish to note that the precise relationship between heart rate and breathing rate is complicated and involves frequency domain analysis. More can be read about heart rate and breathing rate in \cite{Bahmed2016}, \cite{SchipkeJD}, \cite{EvgenyVaschillo2004} and \cite{Wallaart1997} which focus on the behaviour of heart rate and breathing rate during exercise. 

	It is therefore of interest to understand the behaviour of heart rate and particularly how it can be modelled during exercise. In \cite{Le2009} and \cite{Le2007} a model for heart rate under exercise is proposed. The proposed model aims to predict the heart rate of a cyclist according to the individual cyclist's blood lactate threshold. Eight parameters are used in the model to calculate future heart rates from current values, training duration and training loading. The future heart rate is modelled as	
	\begin{equation} \label{eqn:hr}
		HR(k) = HR_S + \Delta HR(k),	
	\end{equation} 	
	where $HR(k)$ is the cyclist's heart rate at sampling step k and $HR_S$ is the heart rate of the cyclist before the training started. $\Delta HR(k)$ is defined as	
	\begin{align*}\label{eqn:delta_hr}
		\Delta HR(k) = & K_1 P(k) + K_2 \Delta HR(k-1)\\
		+ & K_3(1-exp(-T_Ak/\tau))P(k) \\
		+ & K_4 \sum_{i=1}^{k-1}T_A[HR(i) - HR_{iAT}]\sigma[HR(i)-HR_{iAT}]\\
		+ & K_5 \sum_{j=K_{\text{on}}}^{k-1}T_A[HR_{iAT}-HR(j)]\sigma[HR_{iAT}-HR(j)],\\
	\end{align*} 
	where $\sigma(x)$ is the unit step function with
	\begin{equation*}
		\sigma(x) =
		\begin{cases}
			&1, \forall x \geq 0 \\
			&0, \text{otherwise}.
		\end{cases}
	\end{equation*}
	
	The meaning of each variable and parameter are as shown in Table \ref{table:hr_variables}. The individual's anaerobic threshold ($HR_{iAT}$) is the physiological point in the exercise at which anaerobic processes become more dominant and lactic acid starts to accumulate in the muscles. Importantly (and intuitively), it is seen that heart rate is a function of the power output of the cyclist and a number of other personalised parameters that are experimentally derived. This level of understanding is sufficient for the work involved in this project.
	\begin{table}[H]
		\centering
    	\begin{tabular}{||c|c||}
    		\hline
    		\textbf{Variable} & \textbf{Explanation} \\ [0.5ex]
    		\hline\hline
    		$HR_{iAT}$ &  Heart rate at individual anaerobic threshold \\
    		\hline\hline
    		$K_1 ... K_5$ &  Heart rate model parameters \\
    		\hline\hline
    		$K_{\text{on}}$ &  Time point at which $HR(k)$ is greater than  $HR_{iAT}$ for the first time\\
    		\hline\hline
    		$T_A$ & Sampling time \\
    		\hline\hline
    		$P(k)$ & Power output of cyclist at k \\
    		\hline\hline
    		$\tau$ & Time constant related to training duration  \\ [1ex]
    		\hline
    	\end{tabular}
    	\caption{Variables used in \cite{Le2007} and \cite{Le2009} to model heart rate during cycling.}
    	\label{table:hr_variables}
	\end{table} 
	 
	 Another point to note that is not accounted for in the model above is raised in Gold's 2000 paper \cite{Gold2000} which discusses the link between air pollution and heart rate variability (HRV). Heart rate variability is the variation in time interval (RR interval) between heart beats (the heart does not beat at a constant frequency). Gold's paper found that exposure to particulate matter may reduce HRV. A reduction in HRV is linked to stress and fatigue in humans. This effect will not be considered in the use cases and validation to follow.

\section{Routing algorithms for bikes} \label{sec:lr_routing}

	As introduced in Section \ref{sec:lit_review_structure} and Section \ref{sec:lr_cycling_minimise_pollution}, routing is concerned with somehow identifying the best route in a network to accomplish some objective. An example would be that a cyclist wishes to travel on a route over which their inhalation of atmospheric pollutants is minimised. This section discusses the formal problem of smart routing. Routing algorithms largely involve determining the route that a user should take to travel from point A to point B in a journey in order to optimise some objective e.g. quickest journey time. 
	 
	 Perhaps the most well established routing algorithm is Dijkstra's algorithm \cite{e.w._note_1959}. Dijkstra presents an algorithm to calculate the shortest path from a source node to a sink node in a graph. The algorithm considers all options that can be taken at a node and assigns weighted distances between nodes. The algorithm stops when the destination node is reached by the shortest total distance from source node to sink node. In \cite{crisostomi_google-like_2011}, a model of road network dynamics based on Markov chain theory is presented. Markov chain theory is used by Google to rank page results for its search algorithm. The model constructs a transition matrix that contains all information about the network e.g. travel times along a road and probability functions for turning at a junction. The key benefit of modelling road network dynamics in this way is that it means that many complex road dynamical concepts e.g. traffic lights and interactions between cars do not need to be modelled directly.	
	 
\begin{figure}[H]
	    	\centering
	    	\includegraphics[width=0.8\textwidth]{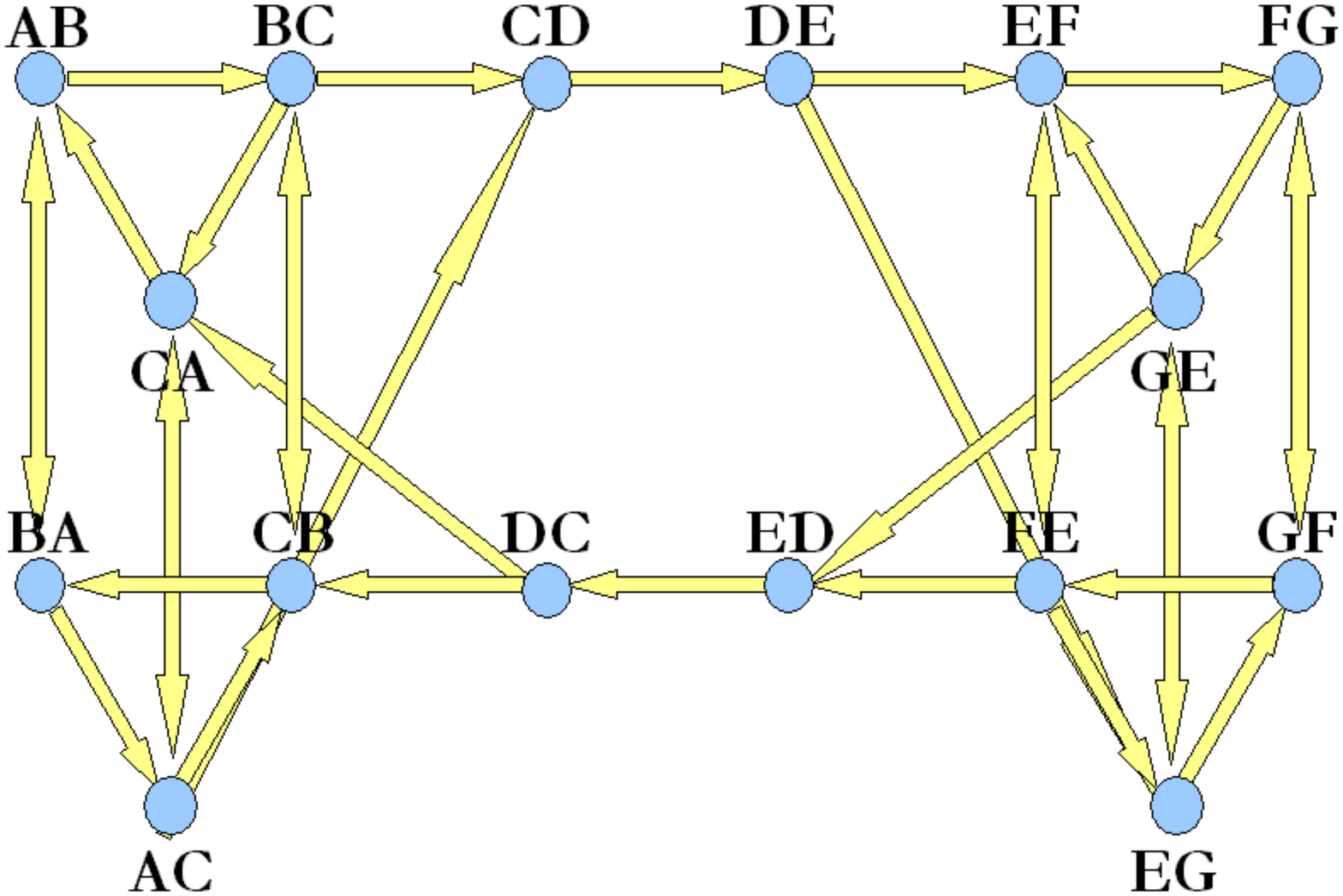}
	    	\caption{Example of modelling a road network from \cite{crisostomi_robust_2011}.} 
	    	\label{fig:robust_road_network}
	 \end{figure}

	 Using Markov chain theory for routing is further explored in \cite{crisostomi_robust_2011}. This paper proposes robust and risk-averse strategies that ``discriminate'' among a set of routes that have similar journey times in order to mathematically model ``common-sense'' aspects of driver behaviour. Figure \ref{fig:robust_road_network} shows an example of a road network where a decision must be taken at each node regarding which route to follow.
	 
	 The road network in Figure \ref{fig:robust_road_network} can be mathematically represented as a transition matrix $\mathbb{P}$ as \footnote{equation from \cite{crisostomi_robust_2011}} 
	 \begin{equation*}
		 \mathbb{P} =
		 \begin{bmatrix}
		 \boldsymbol{P_{AB \rightarrow AB}} & P_{AB \rightarrow AC} & P_{AB \rightarrow BA} & \cdots & P_{AB \rightarrow GF} \\
		 P_{AC \rightarrow AB} & \boldsymbol{P_{AC \rightarrow AC}} & P_{AC \rightarrow BA} & \cdots & P_{AC \rightarrow GF} \\
		 P_{BA \rightarrow AB} & P_{BA \rightarrow AC} & \boldsymbol{P_{BA \rightarrow BA}} & \cdots & P_{BA \rightarrow GF} \\
		 \vdots & \vdots & \vdots & \ddots & \vdots \\
		 P_{GF \rightarrow AB} & P_{GF \rightarrow AC} & P_{GF \rightarrow BA} & \cdots & \boldsymbol{P_{GF \rightarrow GF}} \\
		 \end{bmatrix},
		 \label{eq:transition_matrix}
	 \end{equation*} 
	 where an element $P_{ij}$ is the probability that at the end of road i a car will turn into road j. The diagonal elements of the matrix $P_{ii}$ are related to the average travel time $t_i$ for a particular road i.	
	 
	 The algorithms proposed in this paper can be described as robust and risk-averse because they attempt to model human behaviour by explicitly adjusting probabilities in the transition matrix that a driver might do things such as 1) take the wrong turn at a crucial junction 2) prioritise a route that contains detours from the optimal so that there will be options if an unexpected event takes place e.g. car accident, road works etc. 
	 
\begin{figure}[H]
	    	\centering
	    	\includegraphics[width=0.8\textwidth]{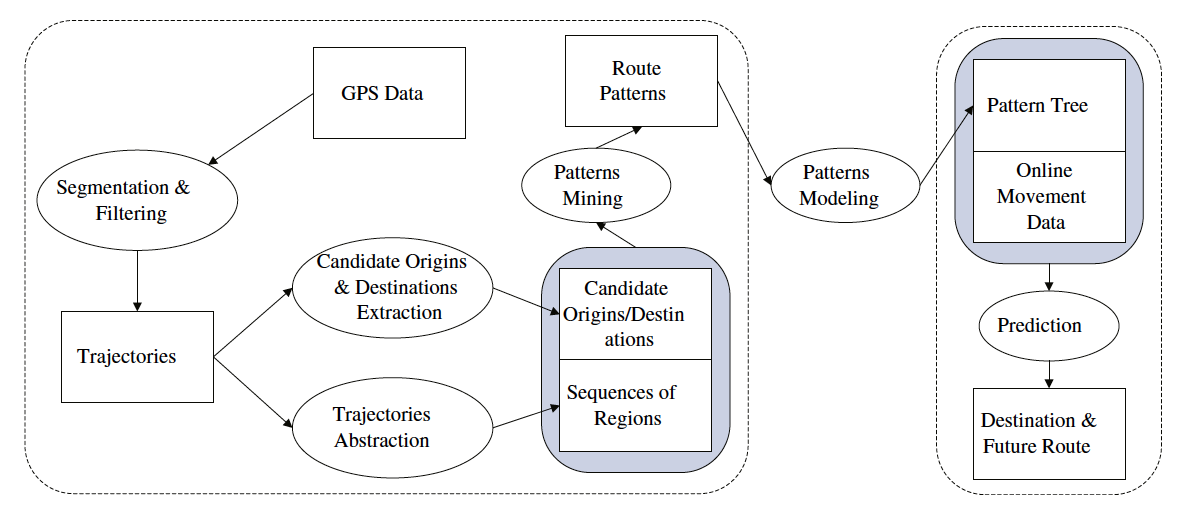}
	    	\caption{Architecture of trajectory data mining prediction system from \cite{chen_system_2010}.} 
	    	\label{fig:data_mining_architecture}
	 \end{figure}

	 Ideas presented in \cite{chen_system_2010} and \cite{chen_personal_2011} propose a system based on trajectory data mining in order to predict both the intended destination and the future route of a person in an integrated way. In the papers discussed up to now the final destination of a user has been assumed to be known. The method does this by probabilistically comparing historical routes with a user's current trajectory (based on GPS data) to establish their most likely destination and future route. The architecture of this approach is shown in Figure \ref{fig:data_mining_architecture}. 
	 
%

%

%

%% file: chapters/ch_system_design.tex
\chapter{System Design} \label{ch:sysdesign}
\setstretch{1.4}

In order to achieve the objectives outlined in Section \ref{sec:objectives}, a system involving both hardware and software components had to be designed and implemented. Section \ref{sec:design_considerations} first introduces some of the factors that had to be considered in designing the system and the final system is discussed in Section \ref{sec:hardware} and Section \ref{sec:software}.

	    \section{Design considerations} \label{sec:design_considerations} 
	    This section serves to introduce some of the key issues that were considered in designing the system. \\
	    \textbf{Situation aware:} The system should be able to gather data from its environment and use that data to provide useful services for the cyclist.\\
	    \textbf{Waterproofing:} The system should be waterproof to the same extent as a standard electric bike.\\
	    \textbf{Safety:} The system should not put the cyclist at increased risk of injury or harm.\\
	    \textbf{Legal:} Final implementation of all use cases should abide by all relevant jurisdictional legislation.\\
	    \textbf{Reliable:} The system should be consistent and fail-safe where possible.\\
	    \textbf{Appearance:} The system should attempt to minimise the number of additional hardware components that are required.\\

	    \section{Hardware} \label{sec:hardware}
	    There were three main functional objectives that had to be achieved through hardware, these are listed below. 
	    \begin{enumerate}
	    	\item Design and implement a situation aware electric bike that can gather data from its surroundings.
	    	\item Design a system that can analyse data gathered by the smart electric bike and use these insights to provide services to help and protect humans.
	    	\item Design a system that accepts a user defined control signal that can be used to modify the output of the electric motor.
	    \end{enumerate}
	    
\begin{figure}[H]
   	    	\centering
   	    	\includegraphics[width=0.8\textwidth]{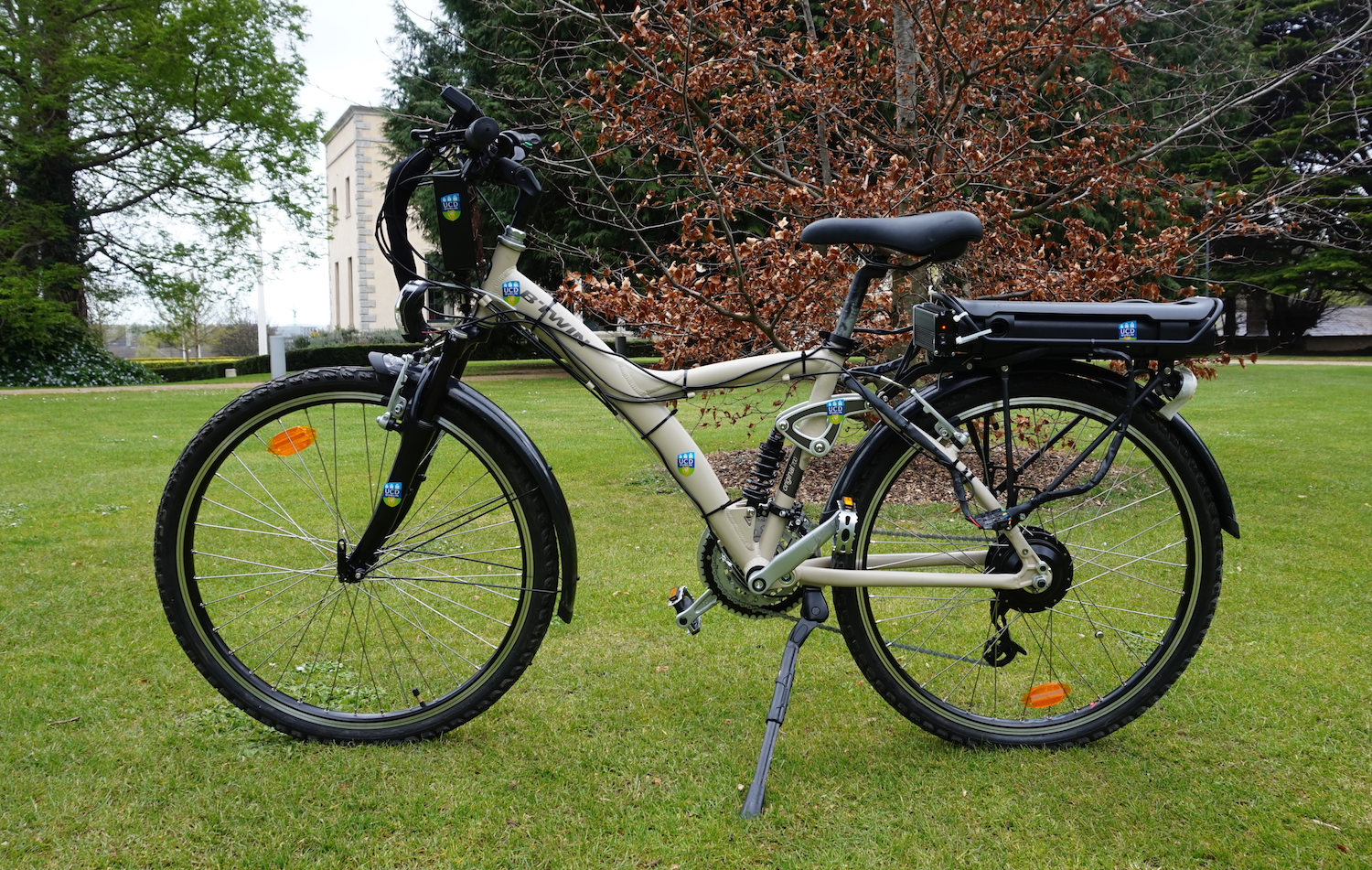}
   	    	\caption{Photograph of fully instrumented electric bike.} 
   	    	\label{fig:photo_bike}
	    \end{figure}

	    A photograph of the final system that was designed and instrumented is shown in Figure \ref{fig:photo_bike}.    
	    
\begin{landscape}
	    	\begin{figure}
	    		\centering
	    		\includegraphics[width=\linewidth]{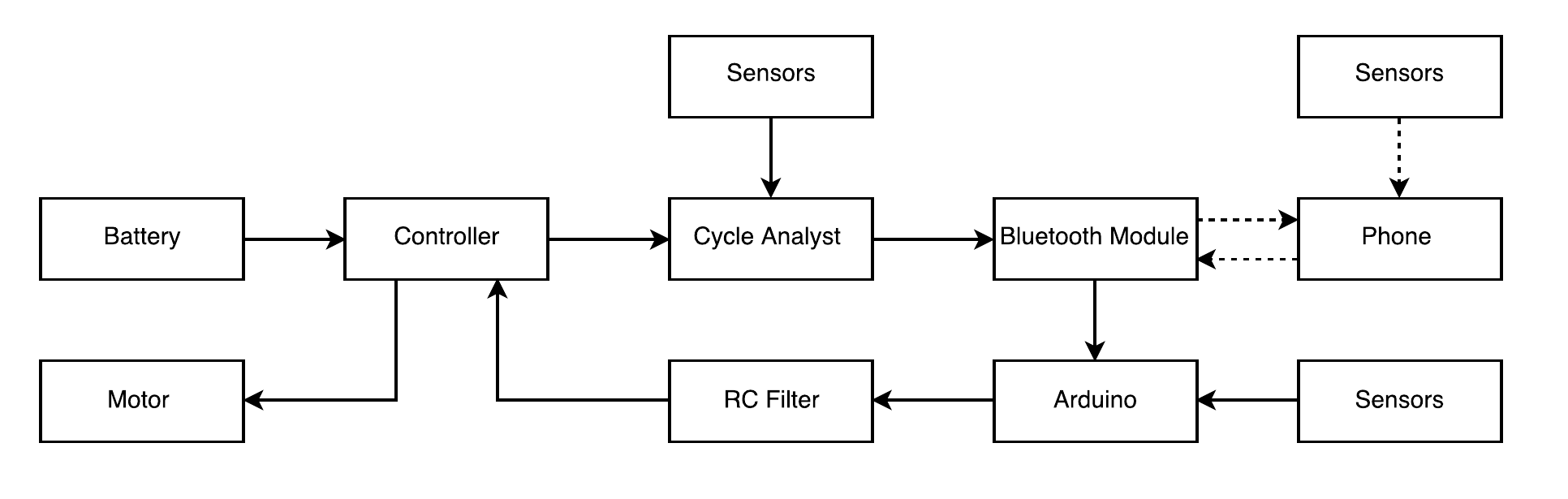}
	    		\caption{Schematic showing final system design.} 
	    		\label{fig:sys_design}
	    	\end{figure}	
	    \end{landscape}

 	    The system is represented as a schematic diagram in Figure \ref{fig:sys_design}. In Figure \ref{fig:sys_design}, solid arrows indicate that the system components are connected by physical cables and dashed arrows indicate that the components are connected wirelessly. Each component of the hardware system will now be discussed in Sections \ref{subsec:elecsystem}, \ref{subsec:dataacquisition}, \ref{subsec:datatransfer} and \ref{subsec:control}.
	    
 	    \subsection{Bike, battery \& motor} \label{subsec:elecsystem}
\begin{figure}[H]
		   	\centering
		   	\includegraphics[width=0.8\textwidth]{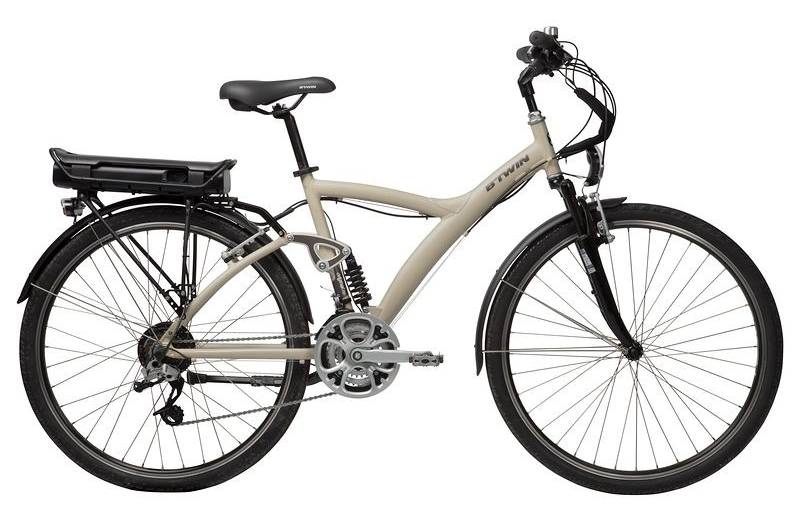}
		   	\caption{Photograph of unmodified electric bike.} 
		   	\label{fig:unmodified_ebike}
		\end{figure}

 	    In order to accomplish the objectives of the project, it was necessary to purchase an electric bike. The Original 700 36V Electric Bike was purchased from Decathlon \footnote{https://www.decathlon.ie/original-700-36v-electric-bike-en-s152571.html}. A photograph of the unmodified electric bike is depicted in Figure \ref{fig:unmodified_ebike}.
 	    
\begin{figure}[H]
			\centering
			\includegraphics[width=0.8\textwidth]{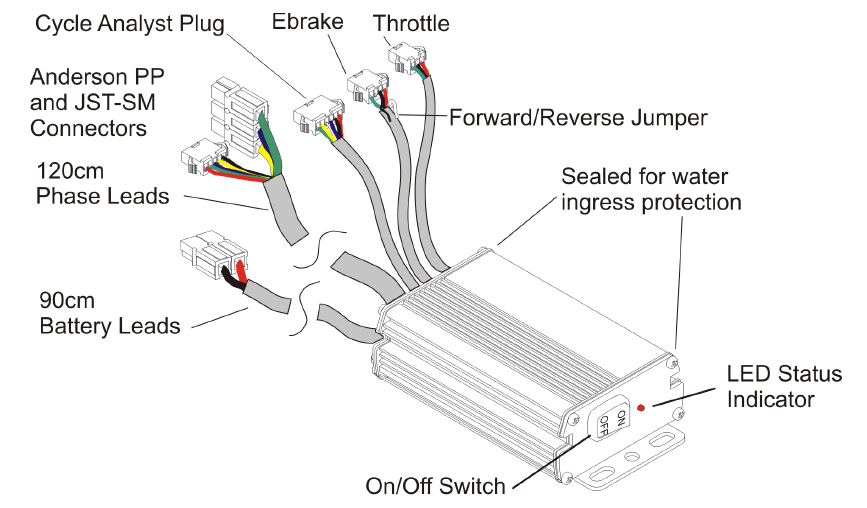}
			\caption{Image of Grinfineon controller which replaced original controller.} 
			\label{fig:grinfineon_controller}
		\end{figure}

 	    The electric bike specifications when purchased were as shown in Table \ref{table:bike_specs}.
 	    \begin{table}[H]
 	    	\centering 
 	    	\begin{tabular}{||c|c||}
 	    		\hline
 	    		\textbf{Variable} & \textbf{Value} \\ [0.5ex]
    			\hline\hline
    			Make \& Model & BTwin Original 700\\
    			\hline\hline
    			Weight & 25.4kg  \\
    			\hline\hline
    			Motor & Brushless 250W electric drive geared motor\\
    			\hline\hline
    			Controller &  Similar to Bafang CR S105.250.SN  \\
    			\hline\hline
    			Battery Voltage & 36V\\
    			\hline\hline
    			Battery Capacity & 7.8Ah / 281Wh\\
    			\hline\hline
    			Battery Chemistry & Li-ion\\
    			\hline\hline
    			Max Torque & 30Nm\\
    			\hline\hline
    			Range & ~30km\\ [1ex]
    			\hline
    		\end{tabular}
    		\caption{Specifications of original 700 36V electric bike purchased from Decathlon.}
    		\label{table:bike_specs}
    	\end{table}
    	The electric bike was then significantly modified so that it could achieve the objectives. An important modification was to replace the controller. The controller is a fundamental component of any electric bike and its basic function is to interface between the battery and the motor. The original controller  \footnote{http://www.szbaf.com/en/components/component/controller/cr-s105250sq.html} could not be easily modified to accept a user defined control input. This meant that the electric bike would be unable to respond to data that it had gathered from its surroundings. As such, it was decided to replace the controller with a Grinfineon C4820-GR \footnote{http://www.ebikes.ca/shop/electric-bicycle-parts/controllers/c4820-gr.html} controller which made it possible to design a control input. An image of the new controller \footnote{From http://www.ebikes.ca/documents/Grinfineon\_V2.0\_Web.pdf} is shown in Figure \ref{fig:grinfineon_controller}.
    	    	
		More information will be given on the control input that was designed in Section \ref{subsec:control}. The new controller was connected between the existing battery and motor. As seen from Figure \ref{fig:grinfineon_controller} the controller has Anderson PP connectors with Blue, Green and Yellow cables and a 5 cable (red, black, blue, green and yellow) JST-SM connector which both had to be connected to the existing motor. The blue, green and yellow cables in the Anderson PP connectors supply a 3 phase current to the motor. The blue, green and yellow wires in the JST-SM connector carry signals from the hall sensors in the motor back to the controller. The black and red wires in the JST-SM connector supply power to the hall sensors. Since cable colours are not standardised for electric bike controllers and motors it was required to test the order that both phase current carrying and hall-sensor cables should be connected in order to achieve smooth operation of the motor. Of the possible 36 combinations, it was found that only 2 combinations corresponded to smooth operation of the motor. The combination that was implemented is shown in Table \ref{table:motor_colour_mappings}.
	
		\begin{table}[H]
			\centering 
			\begin{tabular}{||c|c||c|c||} 
				\hline
				\multicolumn{2}{||c||}{\bfseries Current Carrying Cables} & \multicolumn{2}{|c||}{\bfseries Hall Sensor Cables} \\ 
				\hline
				\bfseries Motor side  & \bfseries Controller side & \bfseries Motor side & \bfseries Controller side \\        \hline
				Yellow & Green  & Yellow &  Blue \\
				Green &  Blue     & Green &  Green \\               
				Blue & Yellow & Blue &  Yellow \\        \hline  
			\end{tabular}
			\caption{Phase and hall sensor mappings for cables between controller and motor.}
			\label{table:motor_colour_mappings}
		\end{table}
		Note that it was also required to replace the connectors at the motor with the appropriate JST-SM and Anderson PP connectors. The operational characteristic of the motor is essential for many use cases and it will be modelled in Chapter \ref{ch:modelling}.
		
		The battery of the electric bike is a Lithium-ion battery with a nominal voltage of 36V. This battery is made up of 10 different cells, each with a nominal voltage of 3.6V. The voltage level of each cell depends on its state of charge. When the battery is fully charged its voltage is approximately 41V and when almost depleted its voltage falls to less than 36V.	
    		    	
    	\subsection{Sensors \& data acquisition} \label{subsec:dataacquisition}
    	In order to enable the bike to become situation aware, it was required to modify the bike so that it could gather data from its surroundings. This was achieved by equipping the bike with a number of sensors. Sensors were added to the bike to measure the  variables shown in Table \ref{table:bike_sensor_variables}.
	   \begin{table}[H]
		   	\centering
 	    	\begin{tabular}{||c | c||}
 	    		\hline
 	    		\textbf{Variable} & \textbf{Units}\\
 	    		\hline\hline
 	    		Battery voltage & Volts\\
 	    		\hline\hline
 	    		Motor current & Amps\\
 	    		\hline\hline
 	    		Wheel speed & $km/hr$\\
 	    		\hline\hline
 	    		Motor temperature & $^\circ C$\\
 	    		\hline\hline
 	    		Pedal speed & RPM\\
 	    		\hline\hline
 	    		Pedal Torque & Nm\\
 	    		\hline
 	    	\end{tabular}
 	    	\caption{Table listing variables that were measured by sensors on electric bike.}
 	    	\label{table:bike_sensor_variables}
	   \end{table}
	   These variables could then be used to calculate a number of other variables of interest such as those shown in Table \ref{table:bike_calculated_variables}.
	   \begin{table}[H]
	   	\centering
	   	   	\begin{tabular}{||c | c||}
	   	   		\hline
	   	   		\textbf{Variable} & \textbf{Units}\\
	   	   		\hline\hline
	   	   		Energy remaining in the battery &Amp-hours\\
	   	   		\hline\hline
	   	   		Electrical power input to motor &Watts\\
	   	   		\hline\hline
	   	   		Distance travelled &meters\\
	   	   		\hline\hline
	   	   		Human power input &Watts\\
	   	   		\hline\hline
	   	   		Acceleration &$m/s^{2} $\\
	   	   		\hline
	   	   	\end{tabular}
   	   		\caption{Table listing variables that could be calculated from data gathered by sensors.}
   	   		\label{table:bike_calculated_variables}
	   \end{table}
	   
\begin{figure}[H]
			\centering
			\includegraphics[width=0.8\textwidth]{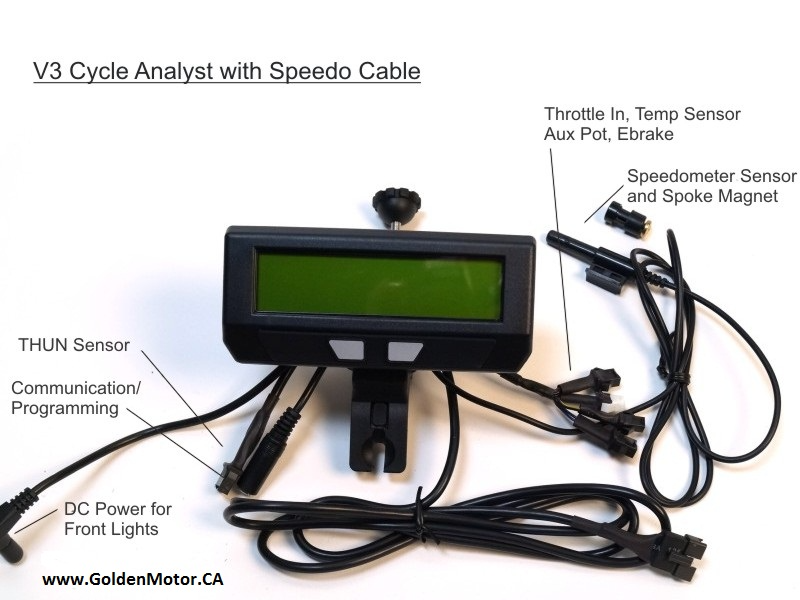}
			\caption{Cycle analyst (data acquisition system).} 
			\label{fig:cycle_analyst}
		\end{figure}

		The data from each of the sensors was read by a data acquisition system called the cycle analyst \footnote{http://www.ebikes.ca/shop/electric-bicycle-parts/cycle-analysts/ca3-dps.html}. The cycle analyst is shown in Figure \ref{fig:cycle_analyst} \footnote{From www.GoldenMotor.CA}.
		
\begin{figure}[H]
			\centering
			\includegraphics[width=0.8\textwidth]{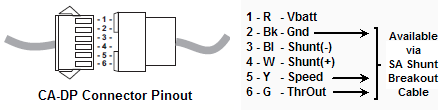}
			\caption{Connector between controller and cycle analyst.} 
			\label{fig:connector_controller_CA}
		\end{figure}

		The cycle analyst was connected directly to the controller of the electric bike. The battery of the electric bike provided power for the cycle analyst. The cycle analyst then provided power for the sensors. The specific sensors that were installed on the electric bike will now be briefly discussed. Battery voltage was measured directly by the cycle analyst. Motor current was also measured directly by the cycle analyst through a shunt connection which corresponds to the S+ and S- wires in Figure \ref{fig:connector_controller_CA}.
		
\begin{figure}[H]
			\centering
			\includegraphics[width=0.6\textwidth]{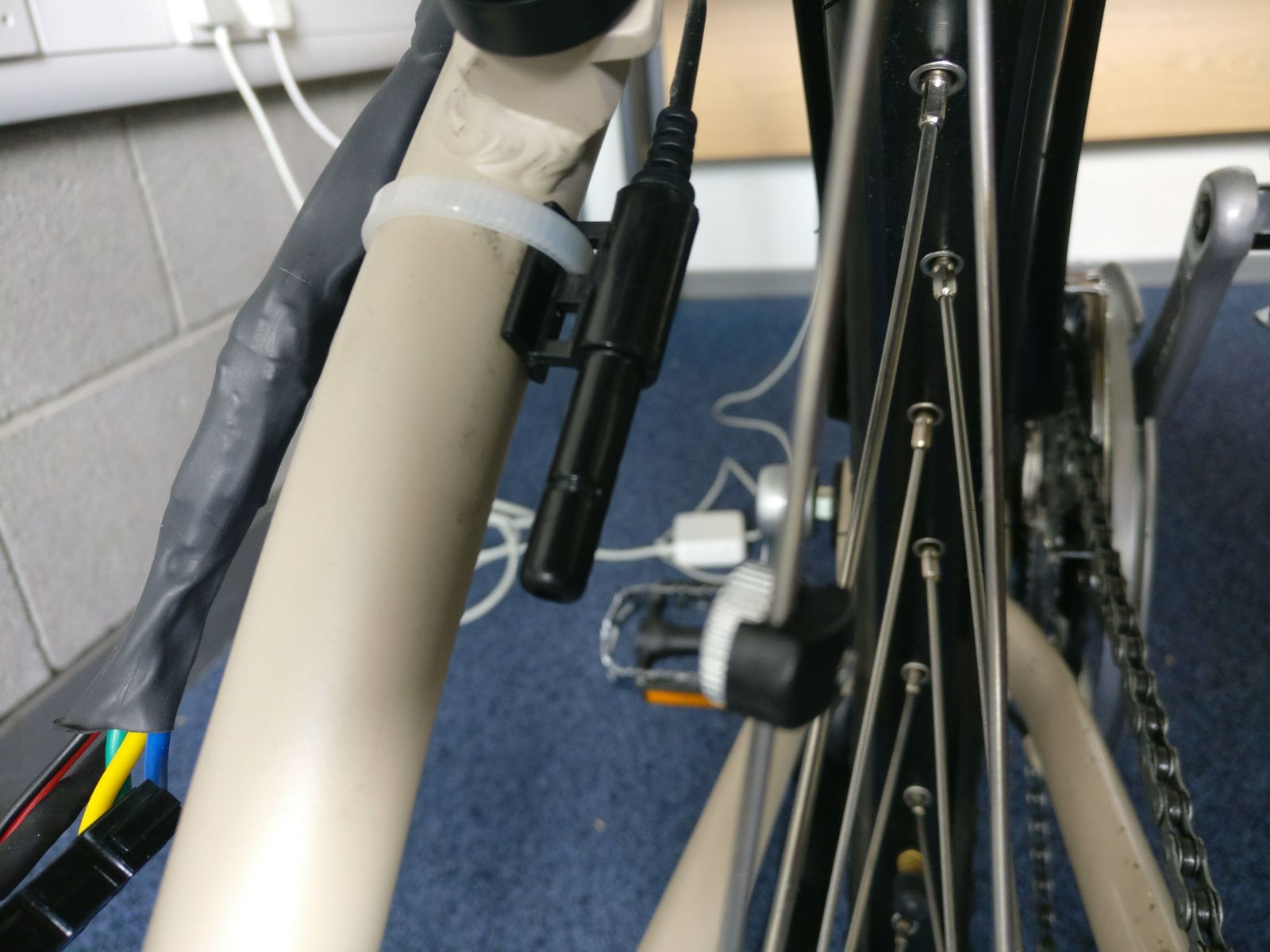}
			\caption{Image showing speed sensor and magnet on wheel.} 
			\label{fig:speed_sensor}
		\end{figure}

		The wheel speed was measured using a speed sensor which was fixed to the frame of the bike, and a magnet which was fixed to the rear wheel of the bike. The speed sensor detected a pulse every time the wheel rotated. These pulses were sent to the cycle analyst which then calculated the speed of the wheel. Both the speed sensor and the magnet are shown in Figure \ref{fig:speed_sensor}.
		
\begin{figure}[H]
	   	\centering
	   	\includegraphics[width=0.6\textwidth]{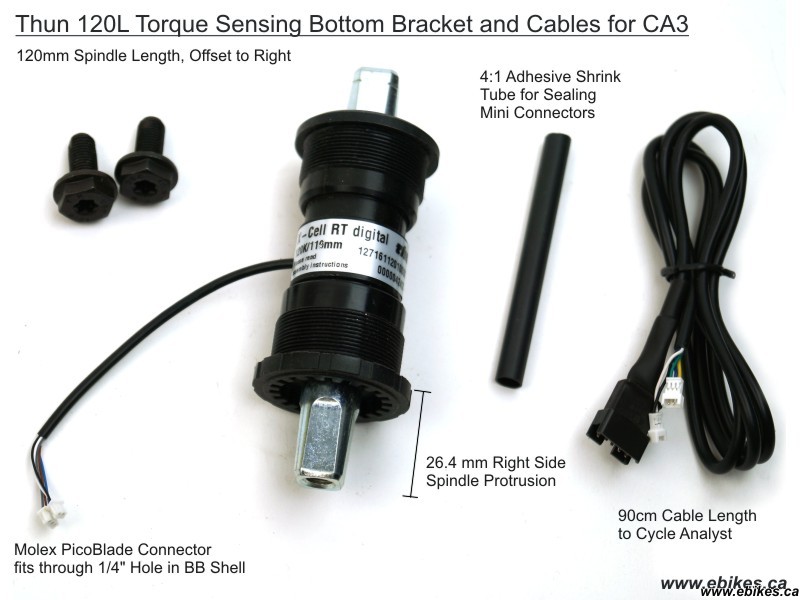}
	   	\caption{THUN X-CELL RT sensor which measured pedal torque and speed.} 
	   	\label{fig:torque_sensor}
	   \end{figure}

	   Pedal speed and pedal torque were measured using a THUN X-CELL RT sensor which is shown in Figure \ref{fig:torque_sensor} \footnote{From http://www.ebikes.ca/catalog/product/gallery/image/undefined/id/24}.
	   
\begin{figure}[H]
		   	\centering
		   	\includegraphics[width=0.7\textwidth]{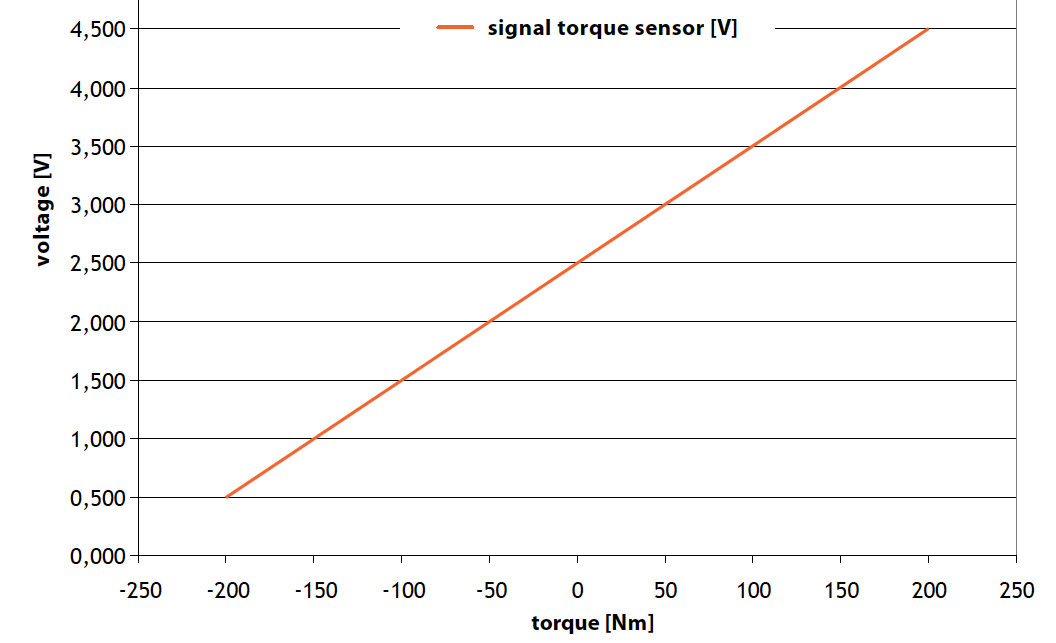}
		   	\caption{Translation of voltage detected to torque in Nm.} 
		   	\label{fig:torqueSensorTorque}
	   \end{figure}

	   The sensor was fitted across the spindle of the bike (with thanks to Mr. John Gahan, UCD Mechanical Engineering Department for his help) and measures the torque being applied to the left pedal by detecting a voltage. The voltage is translated to a torque value using the relationship shown in Figure \ref{fig:torqueSensorTorque} \footnote{From manufacturer http://www.thun.de/project/x-cell-rt/}. The sensor also measured pedal speed (cadence) using two hall sensors. 
	   	   
\begin{figure}[H]
		   	\centering
		   	\includegraphics[width=0.6\textwidth]{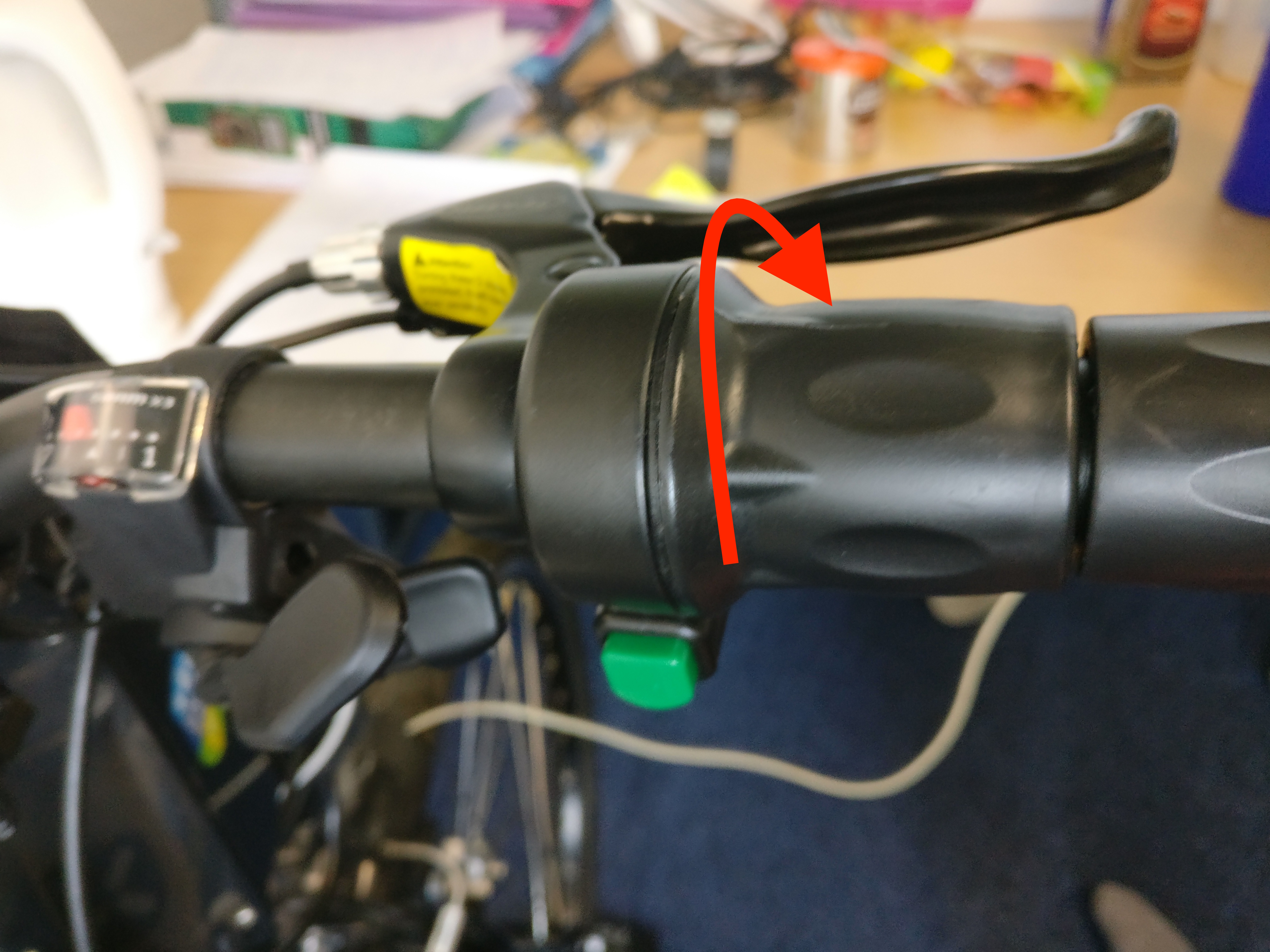}
		   	\caption{Hand throttle fixed on handlebars of bike.} 
		   	\label{fig:hand_throttle}
	   \end{figure}

	   The bike was also equipped with brake sensors which could detect if the brakes were engaged or relaxed. The handlebars of the bike were also equipped with a hand throttle which was a linear hall effect device. A photograph of the hand throttle is shown in Figure \ref{fig:hand_throttle}. The data from the brake sensors and the hand throttle sensor was not read by the cycle analyst, the reason for this design choice will be discussed in Section \ref{subsec:control}.
	   
	   Equipping the bike with the sensors that have been discussed in this section provided excellent insights into what was happening on the electric bike. However, it was clear that other external data streams would be required for a number of the proposed use cases. Getting access to many of these data streams e.g. real time Google maps traffic data or real time pollution sensor data would require an active internet connection. As such, it was decided that the data which was gathered from the sensors on the bike should be transferred to a smartphone where these other data streams were available and all of the data could be analysed centrally. 
	   	    
 	    \subsection{Data transfer} \label{subsec:datatransfer}
 	    
		As stated above, external data that could only be accessed using an internet connection was required for many of the proposed use cases and it was therefore decided that data analysis should be carried out on a smartphone. As such, the system design had to enable two-way data transfer between the electric bike and a smartphone. Two-way data transfer was necessary in order to 1) transfer data that was gathered from the sensors on the bike to the smartphone and 2) send a control signal from the smartphone back to the electric bike that could control the motor.
		
		As such a number of design decisions needed to be made 1) wired vs wireless data transfer 2) choosing a specific wired or wireless technology standard 3) designing a control input that could be sent by the smartphone and accepted by the motor. For ease of usability and safety while cycling it was quickly decided to transfer data wirelessly. Bluetooth was chosen as the technology standard due to its wide prevalence.
		
		As stated in the previous section, most of the data that was measured by the sensors on the bike was read directly by the cycle analyst. The cycle analyst processes this data, and outputs one serial data stream containing data from all of the sensors. \footnote{The cycle analyst outputs a serial data stream operating at 9,600 baud with 8 bits, 1 start bit, 1 stop bit, no parity and no handshake at 0V/+5V TTL levels.} This serial data stream rate could be set to 1Hz or 5Hz through the cycle analyst.
				
\begin{figure}[H]
   	    	\centering
   	    	\includegraphics[width=\textwidth]{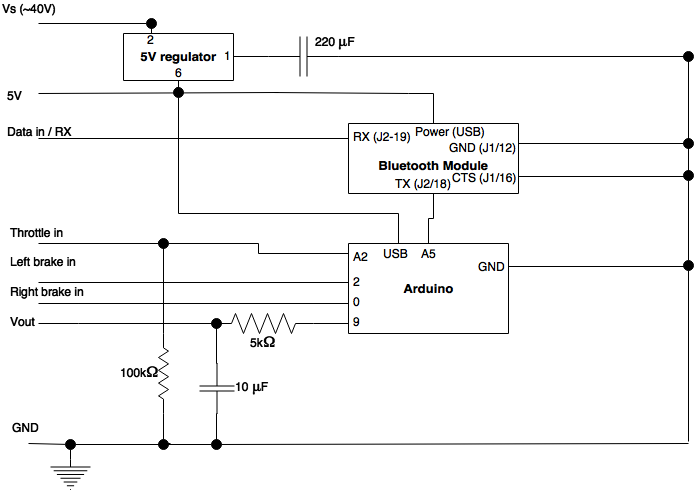}
   	    	\caption{Data transfer circuit schematic that was designed and built.} 
   	    	\label{fig:circuit_schematic}
		\end{figure}

 	    A custom built data transfer system was designed (with thanks to Mr. Brian Mulkeen for his help) and built (with thanks to Liam, Declan and Luke in the UCD electronics workshop for their help) to transfer the serial data stream from the cycle analyst by bluetooth to the user's smartphone. A circuit schematic for the final version of the data transfer system is shown in Figure \ref{fig:circuit_schematic}.
		
\begin{figure}[H]
			\centering
			\includegraphics[width=0.8\textwidth]{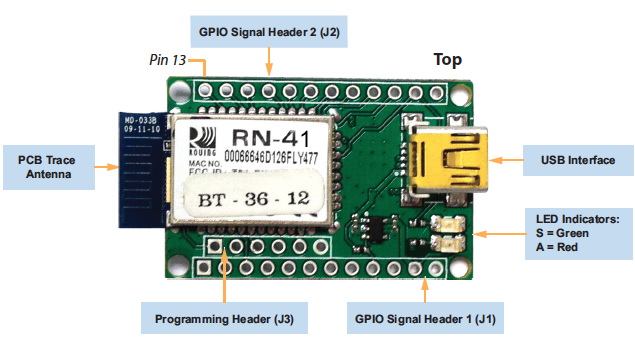}
			\caption{RN-41 bluetooth module from Microchip.} 
			\label{fig:bluetooth_module}
		\end{figure}

		The bluetooth module in use was Microchip's RN-41 bluetooth module which is shown in Figure \ref{fig:bluetooth_module} \footnote{From Microchip's RN41 Evaluation Kit User Guide}. 
		
		The serial data stream from the cycle analyst was wired directly to the RX pin of the RN-41 bluetooth module as shown in Figure \ref{fig:circuit_schematic}. The bluetooth module was powered by using a Traco Power DC/DC converter to step down the battery voltage (nominally 36V) to 5V. It is also very important to note that the CTS (clear to send) pin of the bluetooth module was connected to ground in order for the data to be transferred to the smartphone. It was necessary to code software on both the bluetooth module and the smartphone to enable the transfer of data, this software will be discussed separately in Section \ref{sec:software}. This effectively was all that was required to enable transfer of data from the electric bike to the smartphone.
		
		Transferring data from the smartphone back to the electric bike was more difficult. This was required in order to control the motor of the bike based on the output of the control circuit on the smartphone (See Section \ref{subsec:control}). As introduced in Section \ref{subsec:elecsystem} it was decided to replace the original controller with a Grinfineon controller because it was known that this controller could accept a 0-5V analog voltage input which could be used to control the electrical power input to the motor.
		
		The task then became to convert a signal on the smartphone into an analog voltage in the range 0-5V. The same bluetooth module was used to transfer data back from the smartphone. This signal was then sent to an Arduino. An Arduino has the functionality to convert serial data into a Pulse Width Modulation (PWM) signal. A PWM signal is a type of digital signal that can be easily converted to an analog voltage. Note that the Arduino also received power from the DC/DC converter.
		
\begin{figure}[H]
		   	\centering
		   	\includegraphics[width=0.75\textwidth]{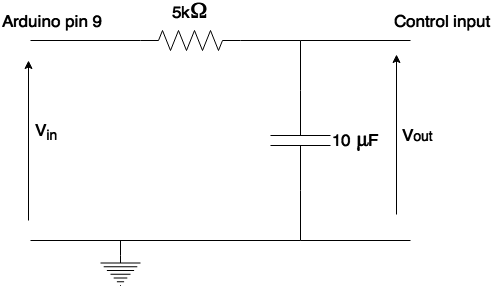}
		   	\caption{Low-pass RC Filter to convert PWM signal to analog voltage.} 
		   	\label{fig:rc_filter}
		\end{figure}

\begin{figure}[H]
			\centering
			\includegraphics[width=0.75\textwidth]{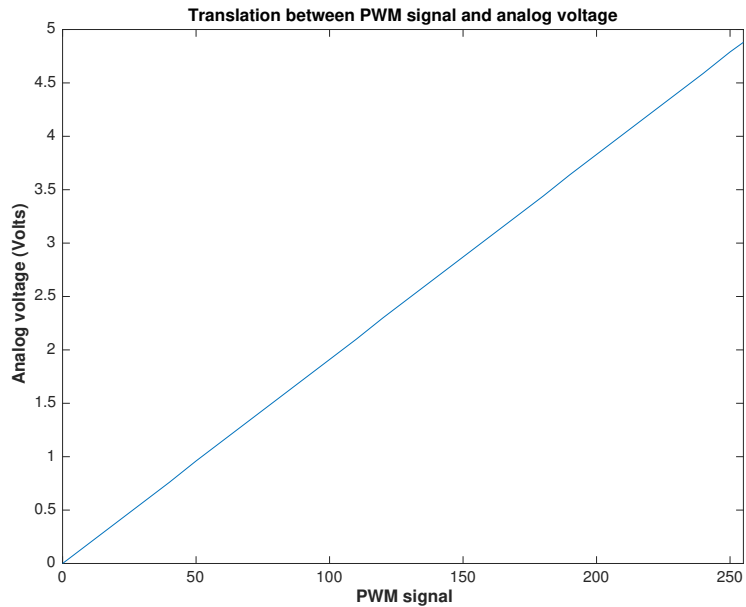}
			\caption{Graph showing relationship between PWM signal and 0-5V analog voltage.} 
			\label{fig:pwm_analog}
		\end{figure}

\begin{figure}[H]
			\centering
			\includegraphics[width=0.8\textwidth]{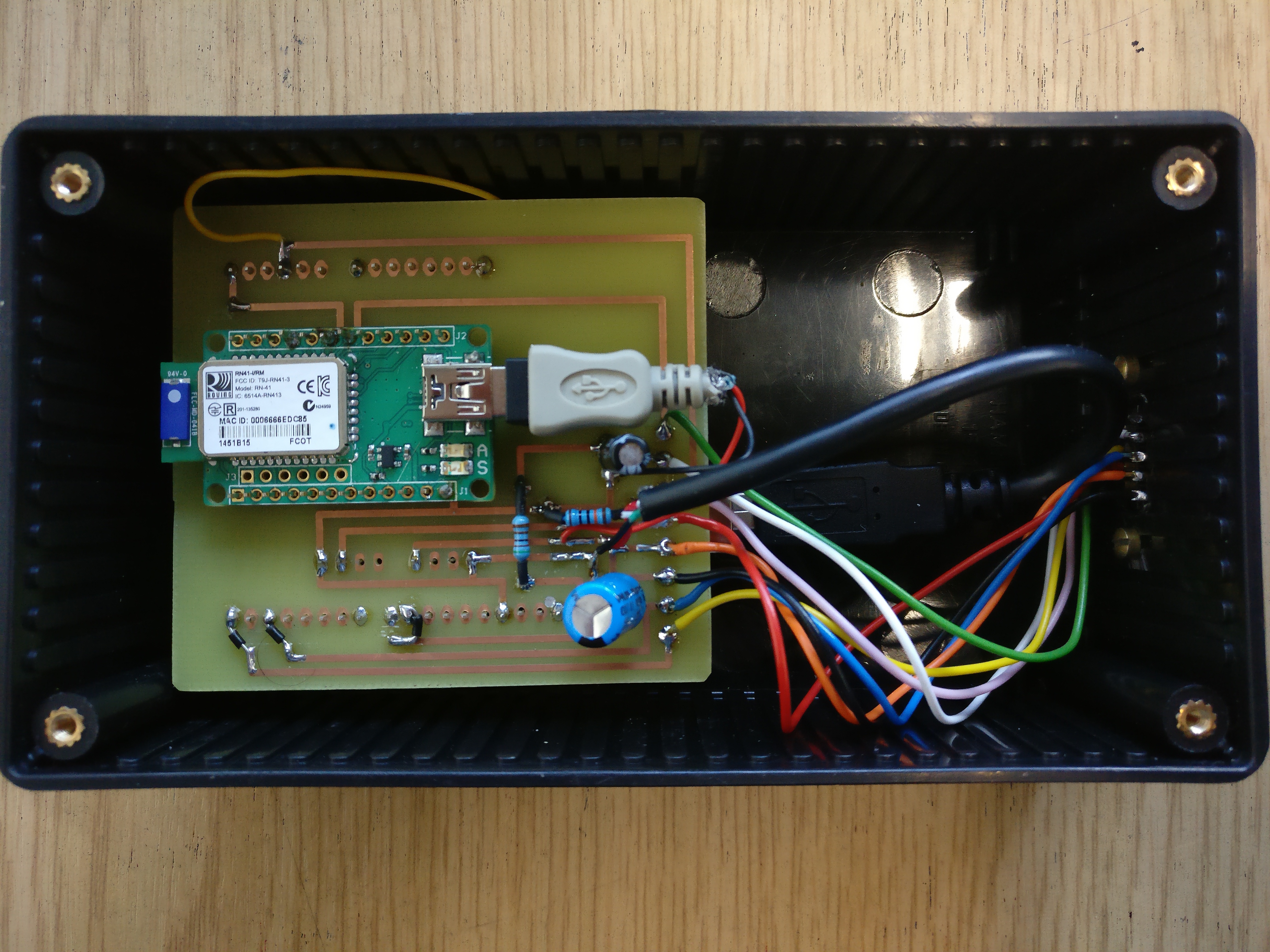}
			\caption{Photograph of data transfer system.} 
			\label{fig:data_transfer_system}
		\end{figure}

		As shown in Figure \ref{fig:circuit_schematic}, data was sent from the TX pin of the bluetooth module to pin A5 of the Arduino. The analog input A5 was configured in software to read a serial stream of data (See Software Section \ref{subsec:arduino}). The Arduino then converted this serial data received at its A5 pin to a PWM signal which it outputted at digital pin 9. This PWM signal was then converted to an analog voltage in the range 0-5V using a low pass RC filter. This specific part of the signal conversion is shown in Figure \ref{fig:rc_filter}.
		The values of the resistor and capacitor in the filter had to be specifically chosen based on the PWM frequency of the Arduino (488Hz) and the desired output range of the analog voltage (0-5V). A 5k$\Omega$ resistor and a 10$\mu$F capacitor were used. The range of the PWM signal from the Arduino was 0-255. A plot showing the translation between a PWM signal and an analog voltage in the range 0-5V is shown in Figure \ref{fig:pwm_analog}.
		The output of the low pass RC filter was then connected to the Grinfineon controller. This achieved the second data transfer objective of transferring data from the smartphone to the motor of the electric bike. A photograph of the complete data transfer system is shown in Figure \ref{fig:data_transfer_system}.	
		
\begin{figure}[H]
			\centering
			\includegraphics[width=0.8\textwidth]{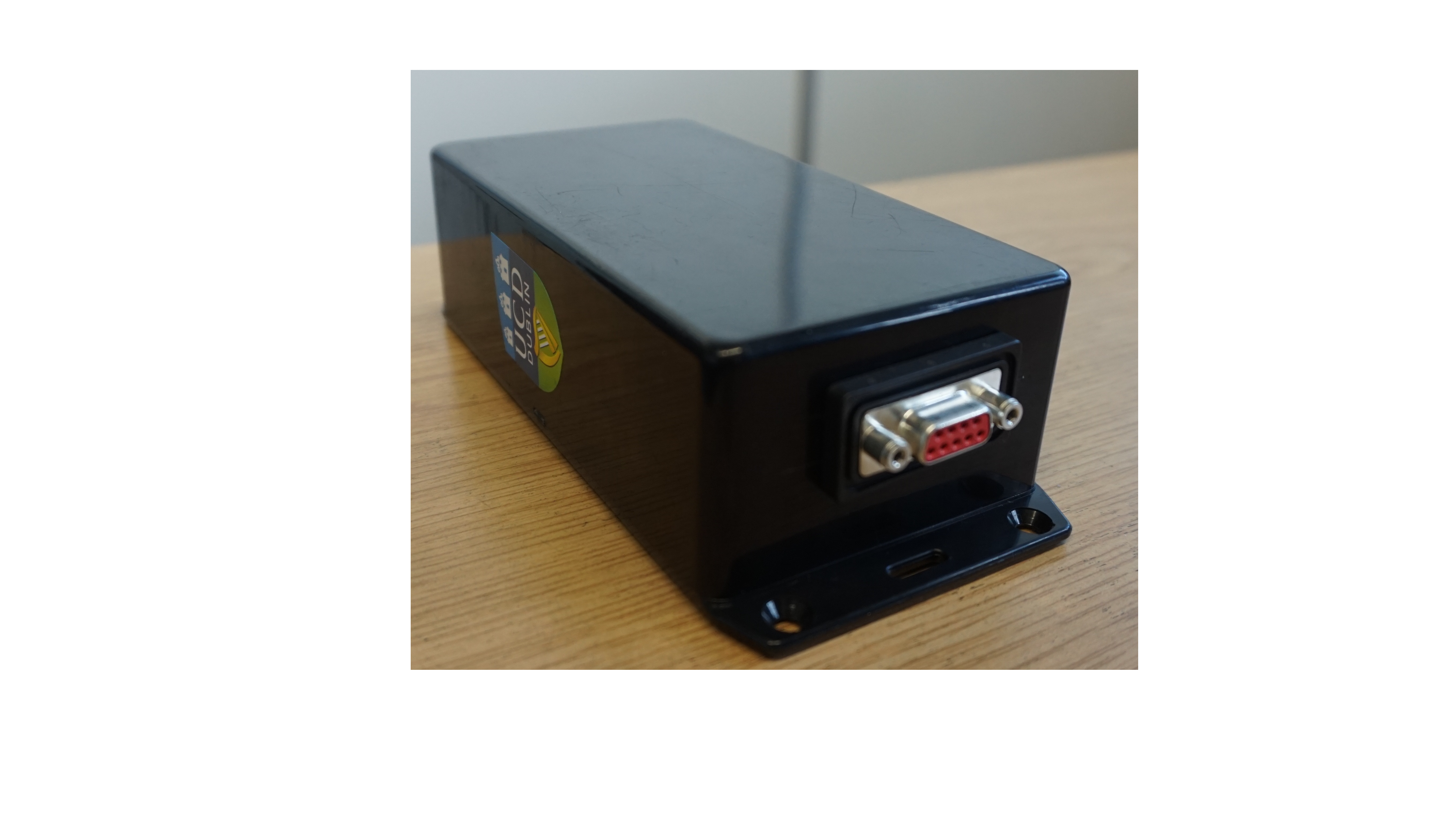}
			\caption{Photograph of waterproof box and connector.} 
			\label{fig:box}
		\end{figure}

		As can be seen from Figure \ref{fig:data_transfer_system}, the data transfer system was implemented on a PCB (printed circuit board). Since it was required to use the data transfer system while cycling, it was a requirement that it had to be waterproofed. This was achieved by protecting the data transfer system by placing it in a waterproof box (waterproof standard IP54). The complete sealed box is shown in Figure \ref{fig:box}.
		
		The box was also made from ABS plastic. A plastic box (as opposed to metal) was required to ensure that data transfer between the smartphone and bluetooth module was not interrupted. The connector for the box was also waterproof (waterproof standard IP67) which can also be seen in Figure \ref{fig:box}. This connector carried 8 different wires which are the 8 wires shown to the left in Figure \ref{fig:circuit_schematic}. The box was mounted on the handlebars of the bike as can be seen in Figure \ref{fig:photo_bike}.

	     	    
 	    \subsection{Analytics \& control} \label{subsec:control}
 	    
\begin{figure}[H]
			\centering
			\includegraphics[width=0.6\textwidth]{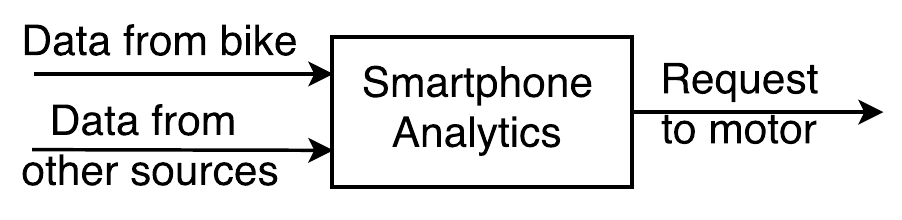}
			\caption{General architecture of data analysis on smartphone.} 
			\label{fig:phone_analytics}
		\end{figure}

 	    As introduced in the previous section, the analytics and control was carried out on a smartphone. The result of this analysis was to calculate the request to send to the bike for the next time step. The specific data analysis and control that was carried out varied by use case. It will be discussed in Chapters \ref{ch:openloop} and \ref{ch:closedloop}. In general, the data analysis architecture can be summarised as shown in Figure \ref{fig:phone_analytics}.
		
\begin{figure}[H]
			\centering
			\includegraphics[width=\textwidth]{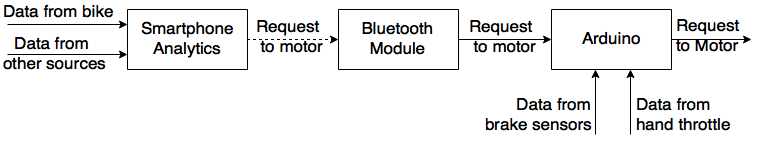}
			\caption{Full architecture of final request to send to motor.} 
			\label{fig:phone_arduino_analytics}
		\end{figure}

		An important design decision was then made. As introduced in Section \ref{subsec:dataacquisition}, the electric bike was also equipped with sensors on the brakes and a hand throttle. It was decided that the output of these sensors should be read directly by the Arduino (as opposed to sent to the smartphone). Figure \ref{fig:phone_arduino_analytics} illustrates this logic.
		
		This decision was made because it was deemed that the output of both of these sensors had a higher priority than the output of the smartphone analytics. The output of the brake sensors was more important for safety reasons i.e. if the cyclist engaged one or both of the brakes, the request to the motor should immediately become zero. It was decided that the output of the hand throttle should overrule the output of the smartphone analytics in order to give the user more control. The hand throttle provided the user with a way to directly increase or decrease the request that was being sent to the motor regardless of the output of the smartphone analytics. This decision may need to be revisited depending on the particular use cases. This logic is implemented in software on the Arduino and is discussed in Section \ref{subsec:arduino}. The important distinction that is made here is that it is the Arduino that sends the final request to the motor, which is not necessarily the same as the output of the smartphone.


	    \section{Software} \label{sec:software}
	    A number of components of the hardware system required software to be written. The two main software systems are 1) the Android app which runs on the smartphone and 2) the Arduino which decides the final request to send to the electric bike motor. A small amount of software also had to be written for the RN-41 bluetooth module. The software for the app will be discussed in Section \ref{subsec:app}, the software for the arduino will be discussed in Section \ref{subsec:arduino} and the software for the bluetooth module will be briefly discussed in Section \ref{subsec:bluetooth}.
	    
	     \subsection{Smartphone software} \label{subsec:app}
	     Software was required to carry out analytics on the user's smartphone. This objective can be subdivided into a number of tasks.
	     \begin{itemize}
	     	\item Task 1: Receive sensor data from the bike by bluetooth
	     	\item Task 2: Receive sensor data from other sensors e.g. activity tracker, Google maps location data 
	     	\item Task 3: Analyse the data that is received
	     	\item Task 4: Send a value by bluetooth back to the bike which corresponds to a request to the motor
	     	\item Task 5: Save all data received by sensors, used in analysis and sent back to the electric bike by bluetooth so that it is recoverable later
	     \end{itemize}
	     
	     There are two main platforms for mobile app development 1) iOS development (iPhones) and 2) Android app development (android phones). It was decided that the software should be coded as an Android app because android phones have a significantly larger market share than iPhones. Android apps are mostly coded in Java with user interface elements coded in XML. A number of different IDEs (Integrated Development Environments) can be used for producing Android apps. The two most prevalent IDEs are 1) Eclipse and 2) Android Studio. It was decided to code the app in Android Studio which is Android's native development environment. The implementation of each of the tasks listed above will now be discussed.
	     
	     \subsubsection{Task 1: Receive sensor data from the bike which is sent by bluetooth}
		     As introduced in Section \ref{subsec:datatransfer}, data that was gathered from the sensors on the electric bike was transferred by bluetooth to the smartphone. Software had to be written on the Android app to receive this data. All of the tasks involved in this process will not be discussed here but the main tasks are listed below:
		     \begin{enumerate}
		     	\item Scan for bluetooth devices
		     	\item Query the local bluetooth adapter for paired bluetooth devices
		     	\item Establish RFCOMM channel
		     	\item Transfer data to and from the device
		     \end{enumerate}
		     
\begin{figure}[H]
		     	\centering
		     	\begin{subfigure}{0.5\textwidth}
		     		\centering
		     		\includegraphics[width=0.8\linewidth]{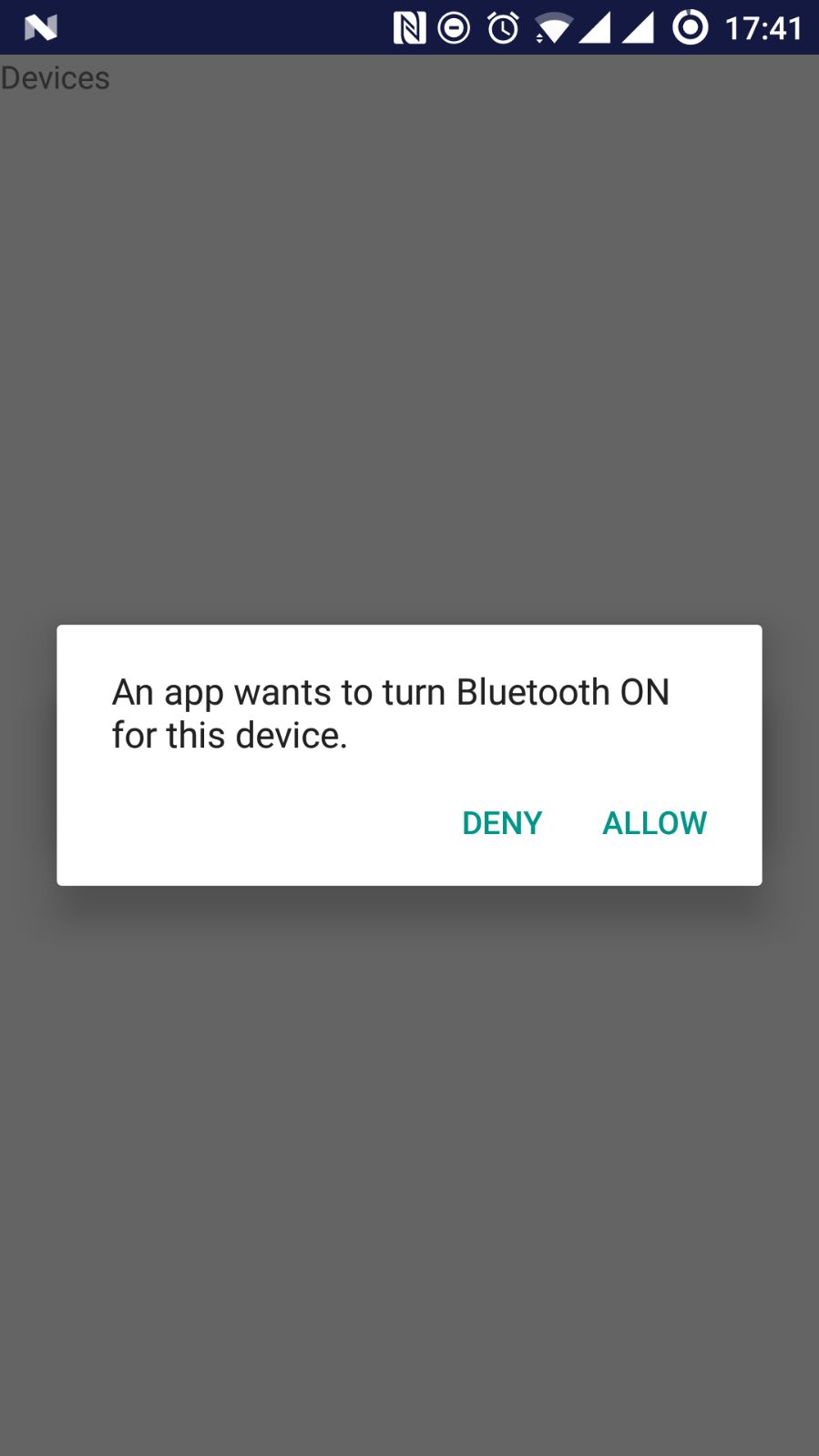}
		     		\label{fig:bluetoothOn}
		     	\end{subfigure}%
		     	\begin{subfigure}{0.5\textwidth}
		     		\centering
		     		\includegraphics[width=0.8\linewidth]{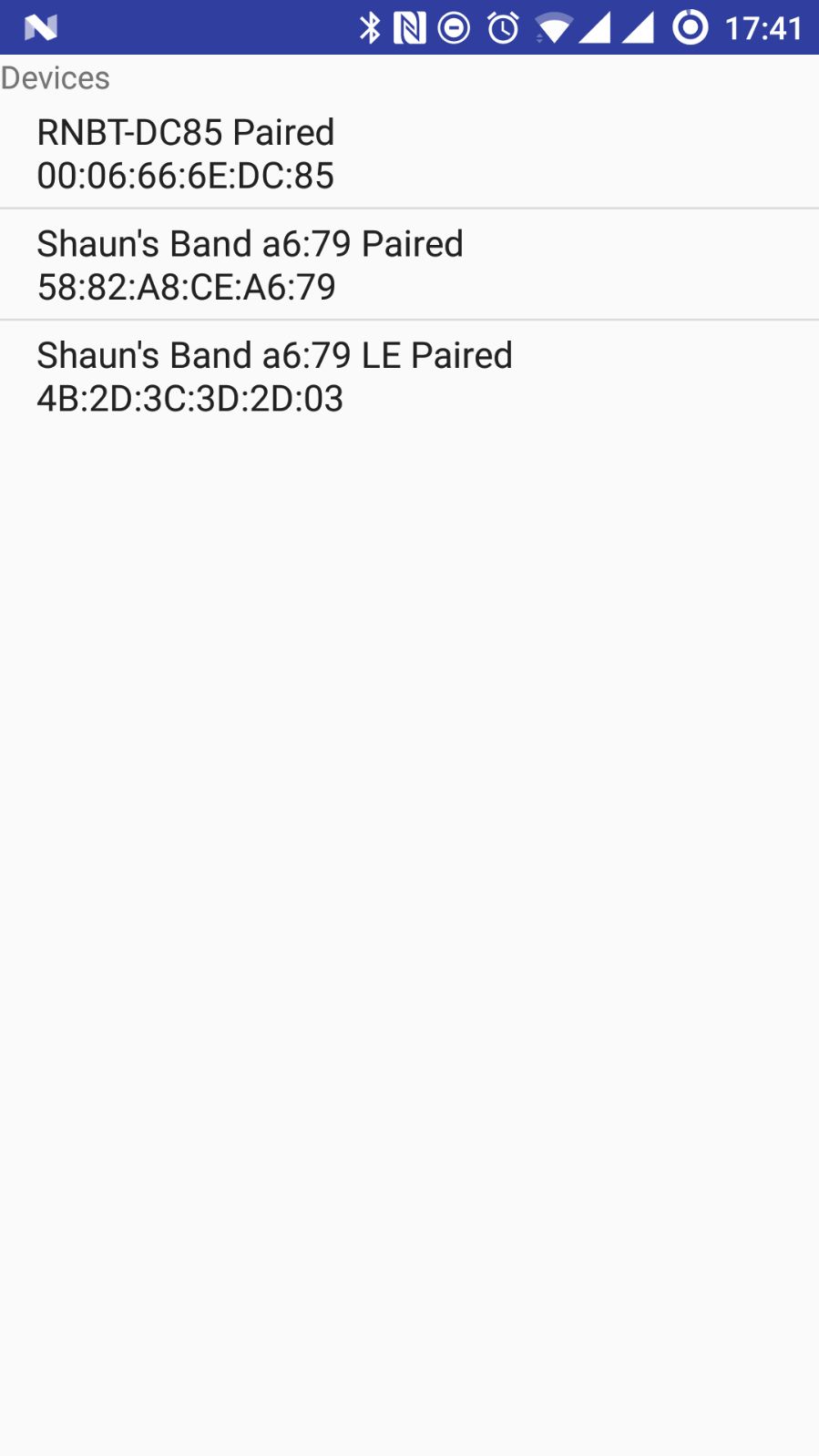}
		     		\label{fig:devices}
		     	\end{subfigure}
		     	\caption{Screenshots from app showing bluetooth connection stage.}
		     	\label{fig:bluetooth_screenshots}
		     \end{figure}

		     In order for the smartphone to receive data from the electric bike (RN-41 bluetooth module), it was required for the smartphone and the bluetooth module to be connected on the same RFCOMM (radiofrequency communication) channel. In order to do this, it was required to be able to control some of the smartphone's fundamental bluetooth tasks. This was done by accessing the smartphone's local bluetooth adapter using Android's BluetoothAdapter class. The following line of code was used to assign the smartphone's bluetooth adapter to an object called btAdapter: 	     
		     
		     \begin{lstlisting}
			     btAdapter = BluetoothAdapter.getDefaultAdapter();
		     \end{lstlisting}
		     
		     This then meant that methods in the BluetoothAdapter class could be used. Methods were used to ask the user to turn their bluetooth on (if it was not already on) and then to query whether there were any bluetooth enabled devices within range. If any bluetooth enabled devices were detected, the names of these devices were printed on the screen as part of an ArrayAdapter in a ListView. Screenshots of some of these steps are shown in Figure \ref{fig:bluetooth_screenshots}.	
			
		     The user was then instructed to select the device that they wished to connect to. To connect to the RN-41 bluetooth module, the user would select ``RNBT-DC85'' in Figure \ref{fig:bluetooth_screenshots}. Once the user selected the device, this initiated the connection process. The main task here was to establish a BluetoothSocket with the RN-41 bluetooth module on the same RFCOMM channel. This was achieved using the following line of code:
		     
		     \begin{lstlisting}
			     mmSocket = device.createRfcommSocketToServiceRecord(MY_UUID);
		     \end{lstlisting}
		     Where mmSocket is a BluetoothSocket object and the MY\_UUID variable is the universally unique identifier of the RN-41 bluetooth module. After this socket object was initialised, it was then possible to receive data that was being sent by the RN-41 bluetooth module. An extract from this code is shown below:
			\begin{lstlisting}
			    String bikeDataString = bufferedReader.readLine();		     
			\end{lstlisting}
		     
		     Note that all of this code was run on different threads, so that the main thread was free for the user to continue to use the app i.e. so that the app did not freeze while the bluetooth connection was being established and bluetooth data transfer was taking place. Note also that bluetooth was implemented as a service (as opposed to an activity). This meant that the bluetooth data transfer could take place across multiple activities in the app without being interrupted i.e. the user could go to a new screen in the app and the data transfer would continue.
		     
		     Data was outputted by the cycle analyst at a data frequency of either 1Hz or 5Hz (configurable). The data was outputted with a tab character (``\textbackslash t'') between each sensor reading, and a new line character (``\textbackslash n'') at the end of each row of data. This data had to be processed before it could be analysed. The BufferedReader class in the code above reads a line of data as a string up to a new line character. This effectively grabbed a row of the data that was being received. The data then needed to be made sense of so that it could be analysed. The data was transferred as an intent to a second service called DatabaseService. An extract from the code to process the data that was received is shown below:
		     \begin{lstlisting}	     	
			     	String[] strings = bikeDataString.split("\\t");
			     	for(int i=0;i<numVariables;i++){
				     	floats[i] = Float.valueOf(strings[i]);
			     	}
			     	float batteryVoltage = floats[0];
			     	...
			     	float pedalTorque = floats[numVariables];    
		   \end{lstlisting}
	   This data was then saved in a database on the user's smartphone where it could be analysed. Please see Task 5 below for more information.
     \subsubsection{Task 2: Receive sensor data from other sensors e.g. activity tracker, Google maps location data}
     
     For many proposed use cases, data streams from other sources were also required. These data streams were generally from either other bluetooth devices or from other data streams available to the smartphone e.g. GPS location data, Google maps real time traffic data. In general the requirement for this data was specific to the use case. The general rule was that data needed to be received and processed. Details on how this was implemented in software is discussed when it is relevant for specific use cases in Chapters \ref{ch:openloop} and \ref{ch:closedloop}.
     
     \subsubsection{Task 3:Analyse the data that is received}
     
     When all data streams had been received, it was then necessary to analyse this data. The purpose of the data analysis step was to decide what request should be sent to the motor of the electric bike for the next time step. The specific analysis of the data that was carried out is use case specific and is discussed in Chapters \ref{ch:openloop} and \ref{ch:closedloop}. 
     \subsubsection{Task 4: Send a value by bluetooth back to the bike which corresponds to a request to the motor}
	     
	     After data analysis was completed, it was necessary to send the output of the analysis back to the electric bike. This was the output of the smartphone analytics and corresponded to a request to the motor as previously shown in Figure \ref{fig:phone_arduino_analytics}. The data was sent as the string representation of an integer from 0-255 which corresponded to the range of the PWM output of the Arduino (see Figure \ref{fig:pwm_analog}). The string also included an exclamation mark character at the end of a command (``!''). The reason why an exclamation mark was included is discussed in the Section \ref{subsec:arduino}. An example of a valid command to send would have been ``110!''. 
%
%
%
	     
		\subsubsection{Task 5: Save all data received by sensors, used in analysis and sent by bluetooth so that it is recoverable later}
		
		Much of the data that was being measured by sensors on the bike were key variables for modelling the system behaviour. Having a reliable model of the system was fundamental to the implementation of the proposed use cases. Therefore, it was required to implement a system in software which saved all relevant data so that it was accessible later for model building. It was necessary to implement a scalable data storage method that could save data in a structured way. It was decided to implement a SQLite database to store the data. SQLite is a relational database management system that is popular in mobile app development and is a standard way of storing structured data in an app that needs to be accessed later.
		
		A SQLite database was created in the app. Within the database a number of tables were created. SQLite uses standard SQL (Structured Query Language). A special custom class called MyDBHandler was created for interacting with the database. This class inherited from Android's SQLiteOpenHelper class. Methods were created in this class for creating tables in the database, adding rows to tables, getting data from a table and deleting data from a table. The following code was used to create a table to store data from the bike.
		
		\begin{lstlisting}
			String bikeDataQuery = "CREATE TABLE " + TABLE_BIKEDATA + "("
			+ COLUMN_ID + " INTEGER PRIMARY KEY AUTOINCREMENT, " +
			COLUMN_TIME + " TIMESTAMP DEFAULT CURRENT_TIMESTAMP, " +
			COLUMN_BATTERYVOLTAGE + " FLOAT, " +
			...
			COLUMN_PEDALTORQUE + " FLOAT " +
			");";
			db.execSQL(bikeDataQuery);
		\end{lstlisting}
		
		Android's ContentValues class was used to add new rows of data to the database. The implementation used getters and setters to access the values of the variables that were received by bluetooth. Using getters and setters encapsulates the variables in accordance with standard object-oriented programming practise. A number of other tables were also created e.g. a table to store all variables involved in data analysis and the final request to be sent to the motor. Since the database was being saved directly onto the smartphone, the amount of memory taken up by the database should be considered. It was not necessary to implement database management techniques for the purposes of this project because more data was seen as being better for getting a better understanding of the system. If memory limitations were ever a concern, it would be easy to routinely batch delete older or less useful data saved in the database.

	     
	     \subsection{Arduino} \label{subsec:arduino}
	     
\begin{figure}[H]
         	\centering
         	\includegraphics[width=0.6\textwidth]{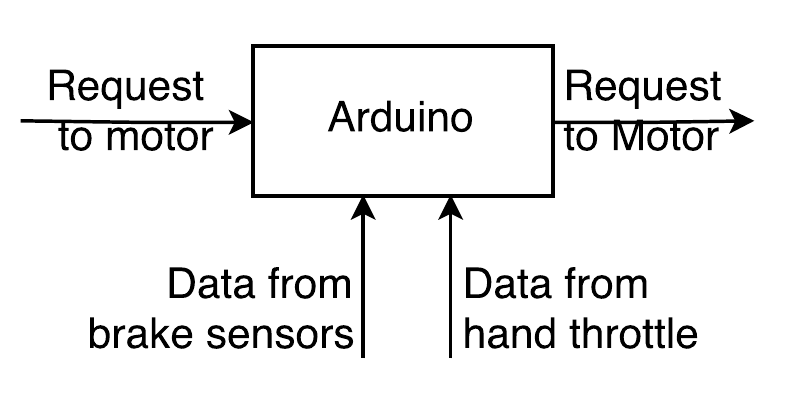}
         	\caption{Architecture showing inputs to Arduino to determine final request to send to motor.} 
         	\label{fig:arduino_analytics}
         \end{figure}

         As discussed in Section \ref{subsec:control}, the Arduino had control over the final request sent to the motor. This was implemented in order to give the user more control. The system architecture is shown in Figure \ref{fig:arduino_analytics}.
         
	     Data from these inputs had to be processed by writing software for the Arduino. Arduino code is written in C/C++ and is uploaded to the Arduino board using Arduino's native IDE. Data from the smartphone (sent to RN-41 bluetooth module and then to the Arduino) was most complicated to read because it was received as a serial stream of data. Arduino provides a SoftwareSerial library which enables serial data to be read. The serial communication needed to be established at a rate of 9,600 baud which was the baud rate that the RN-41 bluetooth module was operating at. This required the following set up code: 
   		 \begin{lstlisting}
			int bluetoothInPin = A5;
			SoftwareSerial mySerial(bluetoothInPin, 11);
			mySerial.begin(9600);
	     \end{lstlisting}
	     Arduino pin A5 was wired to the RN-41 bluetooth module as shown in Figure \ref{fig:circuit_schematic}. As discussed previously, commands sent from the smartphone were sent as the string representation of a number with an exclamation mark character at the end ``110!''. Since the data was being received as serial communication, each bit was read one bit at a time. An example of reading the serial data is shown below.
  	     \begin{lstlisting}
  		     char reading = mySerial.read();
		\end{lstlisting}	     
	     As such, a method was needed to indicate when the last bit in the message had been received. This was why the exclamation mark character was included. Code was implemented to detect when a ``!'' character was received, which indicated that a complete message has been received. The message was then converted to a PWM signal using the analogWrite() function. This outputted a PWM signal at Arduino pin 9. From there the PWM signal was converted to an analog voltage in the range 0-5V using a low pass RC filter which has already been discussed in detail in Section \ref{subsec:datatransfer}.
	     
%

	     It was much easier to read data from the brake inputs than it was to read data being received from the smartphone. The brake sensors were digital sensors that were wired directly to digital input pins 0 and 2 on the Arduino board, the pins were set up to read data using the following code:
	     
  	     \begin{lstlisting}
	     	     int leftBrakeIn = 2;
	     	     int rightBrakeIn = 0;
	     	     pinMode(leftBrakeIn, INPUT_PULLUP);
	     	     pinMode(rightBrakeIn, INPUT_PULLUP);
  	     \end{lstlisting}
  	     
  	     Using INPUT\_PULLUP as the second argument of the pinMode() method made use of a pull-up resistor to implement inverse digital logic which was required for the brake sensors. By default the brake sensors output 1 when they were relaxed (LOW) and 0 when they were engaged (HIGH). Data from the brakes was read using the following method:
		     
	     \begin{lstlisting}
		     leftBrakeValue = digitalRead(leftBrakeIn);
		     rightBrakeValue = digitalRead(rightBrakeIn);
	     \end{lstlisting}
	     
	     Data from the hand throttle was also very straightforward to read. Data read from the hand throttle was an analog voltage in the range 1-4V and was read similarly to data from the brake sensors.
	     
%
%
%
	     
	     Code above has established how data was read and processed from each of the three different types of inputs. Data from all of the sensors was continuously read in a loop. Logic then had to be implemented which related the data from each of the inputs to the final output of the Arduino which corresponded to the final request to send to the motor. This was implemented based on a hierarchy, whereby the following priorities were given:
	     \begin{enumerate}
	     	\item Brake sensors
	     	\item Hand throttle
	     	\item Smartphone analytics
	     \end{enumerate}
	     This meant that the output of the brake sensors could override the output from both the hand throttle and the recommendation from the smartphone analytics. Safety was the main reason for doing this, it was assumed that if the user engaged one or both of the brakes then they immediately wanted the assistance from the motor to stop. The data from the hand throttle had the next highest priority, this meant if the user twisted the hand throttle, this would effectively override the value recommended by the smartphone. This was to give the user more control. This hierarchy possibly may need to be revised for different use cases.

		\subsection{RN-41 bluetooth module} \label{subsec:bluetooth}
		One task had to be achieved in software for the RN-41 bluetooth module. This was to set up the bluetooth module to process a serial data stream at 9,600 baud. This was achieved using open source software called CoolTerm to interface with the bluetooth module and commands specified in Microchip's RN-41 Evaluation Kit User Guide.

\section{Design conclusions} \label{sec:final_design_considerations} 

		A smart electric bike was designed and fully instrumented. A smart electric bike is an electric bike that is situation aware, which means it can gather data from its surroundings and use this data in order to provide useful services to help and protect people. The smart electric bike consists of both hardware and software components. The hardware components of the smart electric bike have been discussed in Section \ref{sec:hardware} and the software components have been discussed in Section \ref{sec:software}.
		
		The system that was designed is flexible, insofar as it can be easily modified to roll out a wide range of electric bike services. A key enabler of this flexibility was achieved by connecting the electric bike and the smartphone. This was a significant achievement because it meant that the electric bike could offer services based on any data stream that was available to the smartphone. This included a huge range of new data streams including weather data, traffic data and location data. This meant that the bike could be truly situation aware.
		
		
		If developing this project in a commercial setting, it would be possible to design a custom controller which directly interfaces between the battery of the electric bike and the motor. This new controller could have its own internet connection which would enable the smart electric bike to get access to data streams from the internet directly without the need for the smartphone. This custom controller could have a bluetooth module incorporated directly which would replace the need for several other system components like the Arduino.
		
		Another improvement that could have been made was with regards to the communication between the electric bike and the smartphone. This communication was implemented using standard bluetooth. This could be improved by implementing bluetooth low energy (BLE). The main benefit here would be reduced power consumption by the bluetooth module. However, that said the consumption of power by the bluetooth module is very small compared to the energy available in the electric bike battery so this was not a significant concern. 
		
		As shown in Figure \ref{fig:circuit_schematic} communication between the RN-41 bluetooth module and the Arduino was achieved by directly wiring the two components together. Each device also had its own power supply. This arrangement could be improved by using an Arduino which has its own inbuilt bluetooth module such as the Arduino Genuino 101 which comes with a BLE module as standard. This would also mean that data read by the brake sensors and hand throttle could also be transferred to the smartphone to include in the smartphone analytics, this would help to give a better picture of how the user is interacting with the system.
		
		All of the objectives were achieved and the design criteria set out in Section \ref{sec:design_considerations} were satisfied. There were a large number of pivot points in the design and instrumentation of the system, the most significant of which was the decision to replace the original controller with the Grinfineon controller.
		
		There were also a number of challenges in designing and implementing the data transfer system (Figure \ref{fig:data_transfer_system}). Several prototypes of this system were built and modified before producing the final design. There were also challenges relating to the cycle analyst which was in use as a data acquisition system to process the data being received from the sensors on the bike. After a few months of work it was discovered that the cycle analyst was improperly processing measurements of pedal torque and pedal speed being received from the THUN X-CELL RT sensor. The original cycle analyst was then replaced.

%% file: chapters/ch_modelling.tex
\chapter{System Modelling} \label{ch:modelling}
\setstretch{1.4}

\section{Modelling introduction} \label{sec:modelling_intro} 
	Each of the use cases that will be discussed in Chapter \ref{ch:openloop} and Chapter \ref{ch:closedloop} involve trying to control the behaviour of the system to accomplish some objective. As such, it was essential to have a model which described the behaviour of the system. The system that has been designed is an example of a cyber-physical system. A cyber-physical system is a system where cyber and physical system components interact with each other. As such, it was necessary to attain a deep understanding of the physical components of the system (the human, the bike) and the cyber components (the algorithmic components). For many of the use cases, modelling the energy interactions between the electric bike and the cyclist was of particular interest. Section \ref{sec:modelling_model} derives a static model of these energy interactions. Note that a model of the system transients has not yet been developed. Section \ref{sec:lr_energy_models} reviewed literature relating to modelling  energy in cycling. However, little research exists, or at least is known of, that models human-bicycle energy interactions in an electric bike. As such, the model in Section \ref{sec:modelling_model} is thought to be one of the first models to exist in this area. A particular word of thanks to Rodrigo~Ord\'o\~nez-Hurtado for his contributions to this section. Note that the model to follow was produced using MATLAB.
		
\section{Energy model of a smart electric bike} \label{sec:modelling_model}

	
	The previous chapter discussed the design and implementation of a situation aware electric bike which could be controlled. Control of the smart electric bike was achieved by designing a control input which will be denoted as $Y$. As discussed in Section \ref{subsec:control}, the logic coded on the Arduino decided the final value of the control input that was sent to the motor. This final value depended on 1) the signal received from the smartphone 2) the brake sensors and 3) the hand throttle. In the model to follow, it was assumed that the human did not engage the brakes or the hand throttle. As such, the final value of the control signal outputted by the Arduino was always equal to the signal received from the smartphone. This meant the only human interaction with the system which is modelled is the energy input of the human at the pedals. Future work may focus on expanding the model to include the brakes and the hand throttle.
	
	As discussed, the control input can take on any integer value in the range 0-255. However, some limitations need to be placed on this range because of the motor. These limitations are listed in Table \ref{table:control_limitations}.
	\begin{table}[H]
		\centering 
		\begin{tabular}{||c|c|c||} 
			\hline
			\bfseries Value of control input $Y$  & \bfseries Validity  & \bfseries Comment\\        \hline
			0-89 & Invalid & Current provided to motor is too small\\
			90-165 &  Valid & Valid operating region\\               
			166-255 &  Invalid & Exceeds motor rated power\\        
			\hline  
		\end{tabular}
		\caption{Limitations on valid range of control input $Y$ due to motor.}
		\label{table:control_limitations}
	\end{table}
	As such the valid range of the control input $Y$ are integer values in the range 90-165. This means that $Y$ can take on 76 distinct values.	It was assumed to be not necessary to test the system for each of the 76 unique values of $Y$. A convenient translation of the control input $Y$ was introduced. This translation of the control input was called $\tilde{Y}$ and the relationship existed between $Y$ and $\tilde{Y}$ could be described as
	\begin{align*}
		&\tilde{Y} = f(Y)\\
		&\tilde{Y} = \frac{Y}{5}-17.
	\end{align*}
	As such, for $Y$=90: $\tilde{Y}$=1 ... $Y$=165: $\tilde{Y}$=16 etc. It was decided that it would be sufficient to test the system for each integer value of $\tilde{Y}$ in the range 1-16 which corresponded to 16 equally spaced tests over the valid range of $Y$ from 90-165.
	
	\subsection{Experiment 1: How the control input $\tilde{Y}$ impacts the motor power with zero human power input} \label{subsec:mod_exp1}
	
	It was desired to understand how the control input $\tilde{Y}$ impacts the motor. As discussed, $\tilde{Y}$ is translated by the low pass RC filter to an analog voltage in the range 0-5V. This analog voltage is an input to the Grinfineon controller. The Grinfineon controller translates the analog voltage to an amount of current that flows to the motor. The electrical power that the motor is provided with is the product of this current and the motor voltage. This electrical power will be denoted as $P_{M_e}$. The motor voltage is equal to the battery voltage (nominally 36V). 
	
	An experiment was designed to test this relationship. In the same experiment, the relationship between $\tilde{Y}$ and the wheel speed was also investigated. The experiment was conducted indoors in a lab environment at University College Dublin. The electric bike was maintained in a stationary position using a Turbo Trainer \footnote{http://www.halfords.ie/cycling/turbo-trainers/trainers/tacx-flow-t2240-smart-turbo-trainer}. The Turbo Trainer simulated the resistance profile a flat road (0 slope).
	
\begin{figure}[H]
		\centering
		\includegraphics[width=0.85\textwidth]{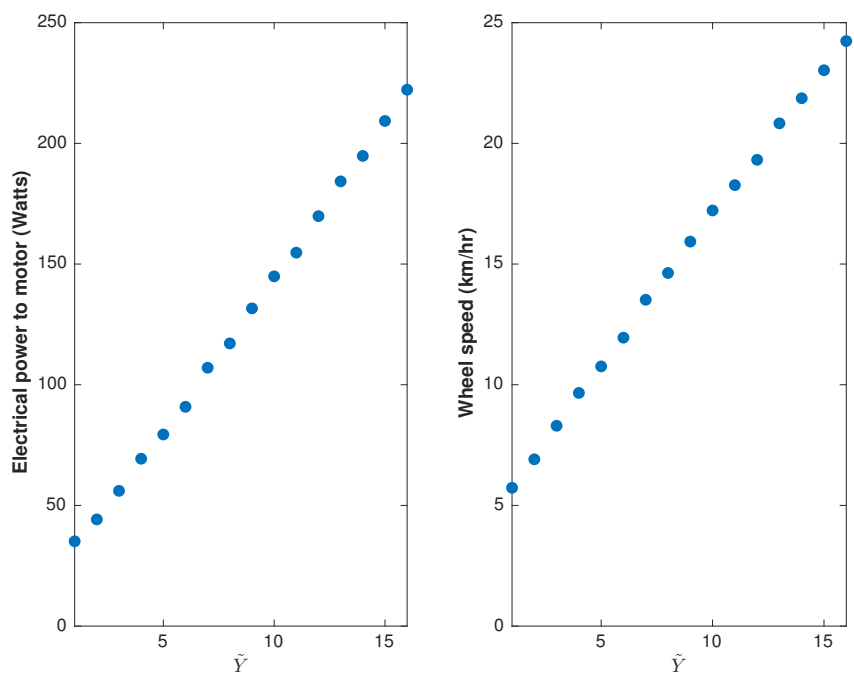}
		\caption{Relationship between control input $\tilde{Y}$ and electrical power input to motor $P_{M_e}$ in Watts (left) and wheel speed in km/hr (right).}
		\label{fig:y_tilde_p_m}
	\end{figure}

	The results of this experiment are shown in Figure \ref{fig:y_tilde_p_m}.	
	
	As can be seen from left graph in Figure \ref{fig:y_tilde_p_m}, as $\tilde{Y}$ increases, the electrical power input provided to the motor $P_{M_e}$ also increases. From the graph on the right, it can be seen that as $\tilde{Y}$ increases, the speed of the wheel increases. The speed of the wheel will be denoted as $S_W$. Note that during this experiment the human was providing zero power input at the pedals.
	
	 \subsection{Experiment 2: How the control input $\tilde{Y}$ impacts the motor power with non-zero human power input} \label{subsec:mod_exp2}
	
	The next experiment \footnote{Note that the experiment conditions were the same as they were for experiment 1 \ref{subsec:mod_exp1}} was to investigate how the results from Experiment \ref{subsec:mod_exp1} would change with non-zero power input from the human. A constant value of $\tilde{Y}$ was selected ($\tilde{Y}$ = 10). The human was instructed to gradually increase the speed of the pedals starting from rest up to the maximum pedal speed possible for them. 
	
\begin{figure}[H]
		\centering
		\includegraphics[width=0.8\textwidth]{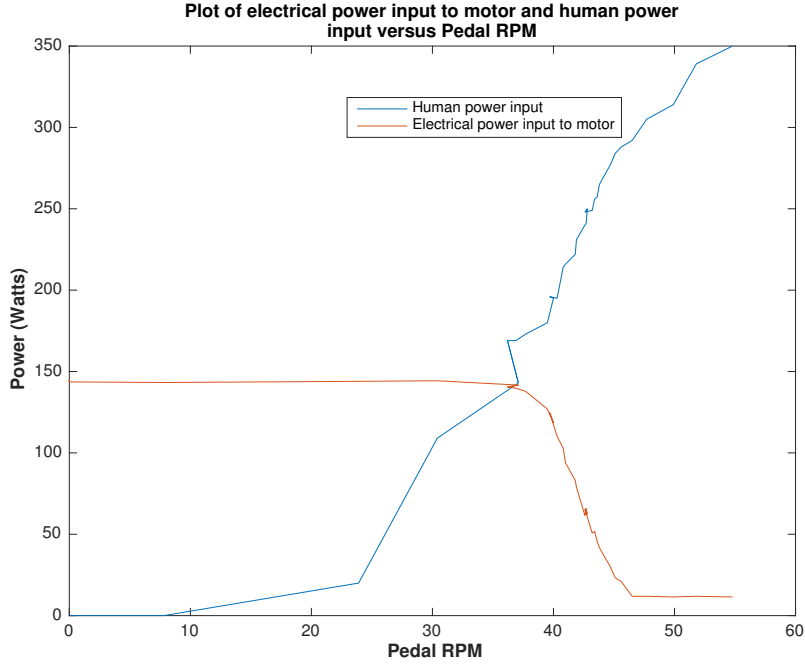}
		\caption{The electrical power input to the motor $P_{M_e}$ in Watts (red) and the human power input at the pedals $P_{H_p}$ in Watts (blue) versus pedal speed in RPM for a fixed value of $\tilde{Y}$=10.}
		\label{fig:motor_human_power}
	\end{figure}

	As discussed in Section \ref{subsec:dataacquisition} the THUN X-CELL RT sensor measures both torque at the pedals and the the speed of the pedals. From those two measurements, the human power input is derived from basic mechanical engineering equations as
	\begin{equation} \label{eqn:p_human}
		P_{H_p} = \tau_p \times \omega_p,
	\end{equation}
	where
	\begin{align*}
		P_{H_p} = &\text{ Human power input at pedals (Watts)} \\
		\tau_p = &\text{ Torque at pedals (Nm)}\\
		\omega_p = &\text{ Angular velocity of pedals (rads$^{-1}$).} 
	\end{align*}
	The results of this experiment are shown in Figure \ref{fig:motor_human_power}.
	
	From Figure \ref{fig:motor_human_power} it is observed that $P_{H_p}$ increases as the human increases the speed of the pedals. It is seen, that when the speed of the pedals is low, the electrical power input to the motor $P_{M_e}$ is constant. As the speed of the pedals increases further, it is seen that $P_{M_e}$ starts to decrease. $P_{M_e}$ continues to decrease and becomes a small constant value as the pedal speed increases further. This reduction in $P_{M_e}$ is explained by the increased human power input $P_{H_p}$. At high values of pedal speed, the human is providing more of the power to move the wheel of the bike, which means that the motor has less of a resistive load which reduces $P_{M_e}$.	
	
	It was then of interest to understand if this relationship holds for different values of $\tilde{Y}$. The experiment was repeated for all values of $\tilde{Y}$ between 1-16. Figure \ref{fig:motor_human_power_all_y_tilde} shows these results plotted versus the pedal speed in RPM. Figure \ref{fig:motor_human_power_all_y_tilde_wheel_speed} shows the same data plotted versus wheel speed $S_W$ in km/hr. As can be seen from Figure \ref{fig:motor_human_power_all_y_tilde} and Figure \ref{fig:motor_human_power_all_y_tilde_wheel_speed} the relationship found from Figure \ref{fig:motor_human_power} holds for all values of $\tilde{Y}$. 
	
	\afterpage{%
		\begin{landscape}
				\begin{figure}
					\centering
					\includegraphics[width=0.8\linewidth]{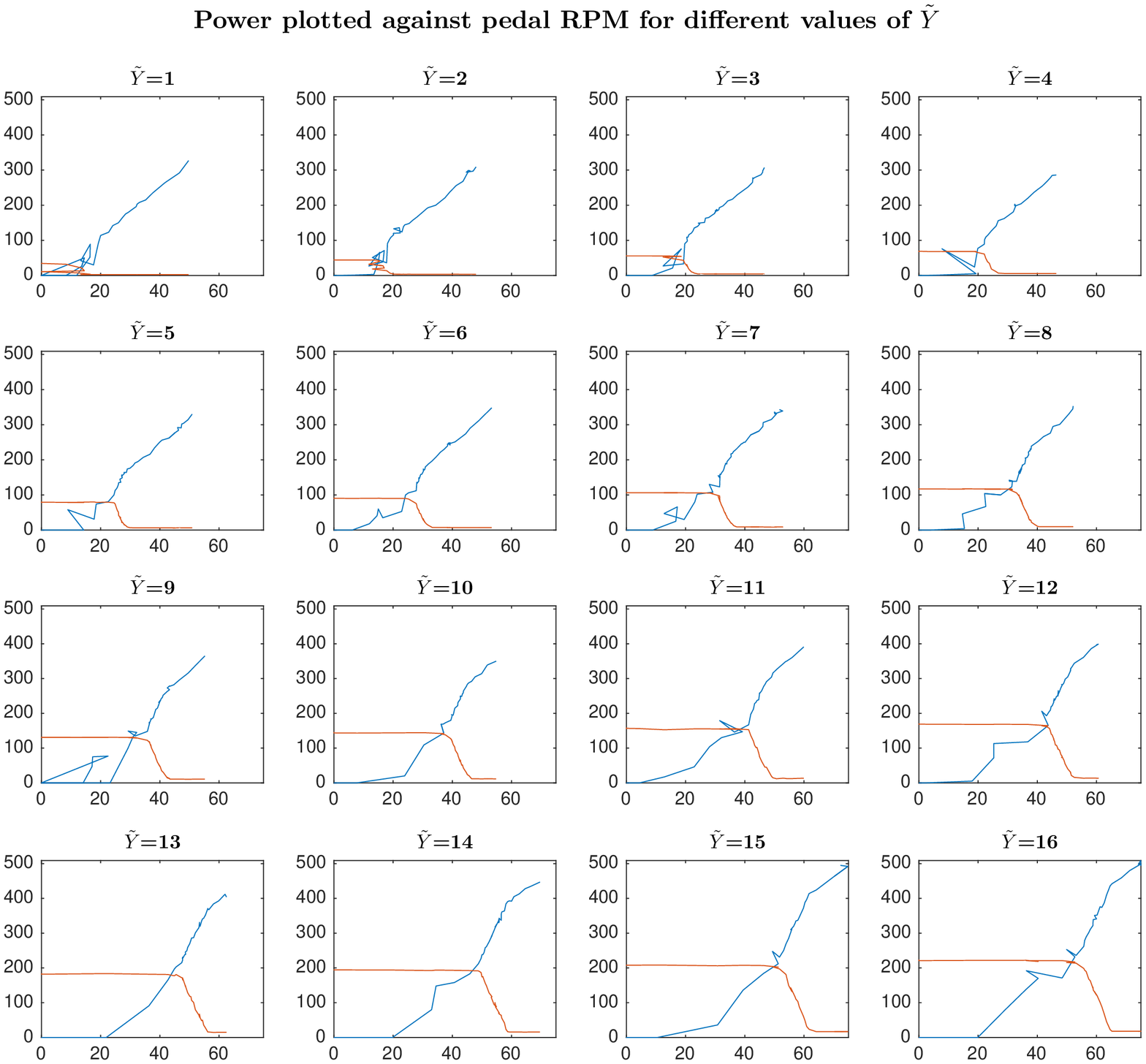}
					\caption{The electrical power input to the motor $P_{M_e}$ in Watts (red) and the human power input at the pedals in Watts $P_{H_p}$ (blue) versus pedal speed in RPM for all values of $\tilde{Y}$ between 1-16.}
					\label{fig:motor_human_power_all_y_tilde}
				\end{figure}
			
			\begin{figure}
				\centering
				\includegraphics[width=0.8\linewidth]{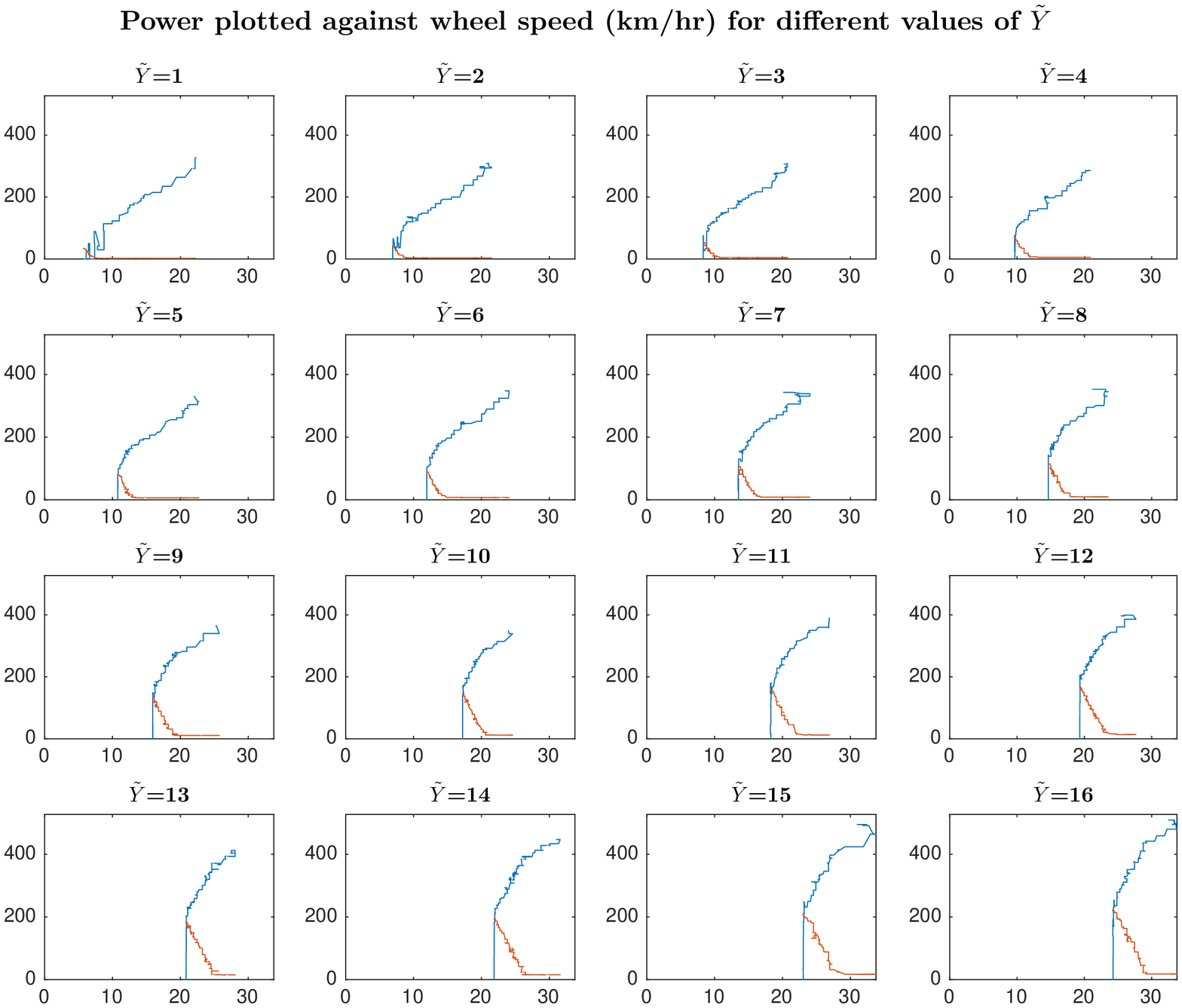}
				\caption{Plot showing electrical power input to the motor $P_{M_e}$ in Watts (red) and the human power input at the pedals $P_{H_p}$ in Watts (blue) versus wheel speed in km/hr for all values of $\tilde{Y}$.}
				\label{fig:motor_human_power_all_y_tilde_wheel_speed}
			\end{figure}
		\end{landscape}
		\clearpage
	}	
	
	\subsection{Modelling variables 1: Approximating $P_W$, $P_{M_w}$ and $P_{H_w}$}	\label{subsec:mod_mod1}
	
	In an electric bike, power to move the wheel of the bike can be provided by 1) the human only 2) the motor only or 3) some combination of the two. The power provided to move the wheel of the bike will be defined as
	\begin{equation}
		P_W = P_{M_w} + P_{H_w},
		\label{eqn:pow_to_wheel}
	\end{equation}
	where
	\begin{align*}
		P_W = &\text{ Total power being provided to the wheel of the bike (Watts)} \\
		P_{M_w} = &\text{ Power from the motor being used to move the wheel (Watts)}\\
		P_{H_w} = &\text{ Power from the human being used to move the wheel (Watts).} 
	\end{align*}
	
	However, neither $P_{M_w}$ or $P_{H_w}$ are known quantities. $P_{M_e}$ and $P_{H_p}$ are measured by sensors on the bike, however, the electrical power input to the motor $P_{M_e}$ does not directly correspond to the amount of motor power used to move the wheel of the bike $P_{M_w}$. Similarly, the human power input at the pedals does not directly correspond to the amount of human power used to move the wheel of the bike $P_{H_w}$. 
	
	The precise relationship between the input powers $P_{M_e}$ and $P_{H_p}$ and the power actually provided to the wheel $P_{M_w}$ and $P_{H_w}$ depends on a number of fundamental characteristics of the system to include the type of motor, the positioning of the motor, the mechanical gear system, the efficiency with which the motor converts electrical energy into mechanical energy, the efficiency with which human power provided at the pedals is converted to mechanical energy at the wheel, etc.
	
	As such, both $P_{M_w}$ and $P_{H_w}$ must be modelled. First, $P_{H_w}$ the human power being used to move the wheel of the bike will be estimated. From basic mechanical engineering, it is known that in order to increase the speed of the bike (while cycling on a road that has the same gradient and resistance profile), more power must be provided to the wheel of the bike. As can be seen from Figure \ref{fig:motor_human_power_all_y_tilde_wheel_speed}, when the human power input at the pedals $P_{H_p}$ is in the low pedal speed range, the wheel speed $S_W$ does not increase. From this, it can be concluded that in the low pedal speed range, all of the power to move the wheel of the bike is being provided by the motor. That is, that $P_{H_w} = 0$ in the low pedal speed range.
	
\begin{figure}[H]
		\centering
		\includegraphics[width=0.8\linewidth]{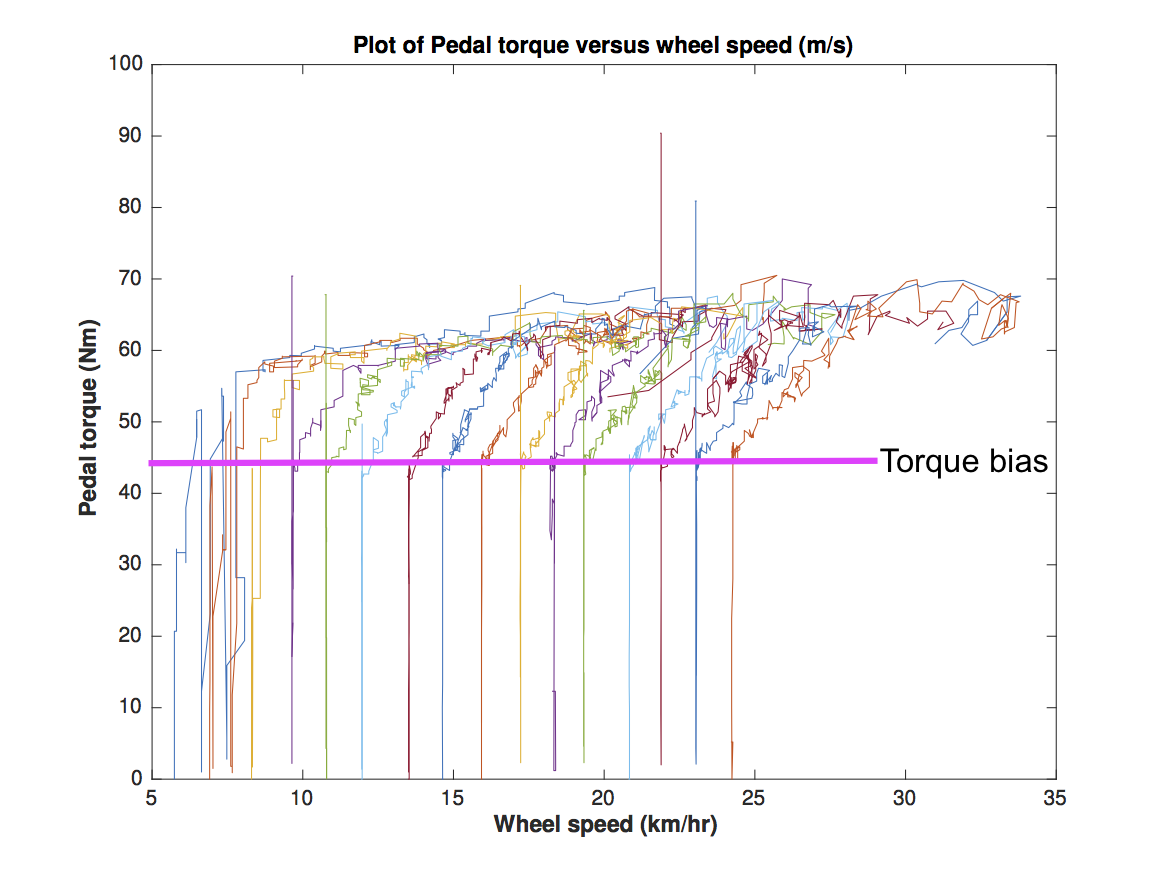}
		\caption{Plot showing pedal torque in Nm plotted versus wheel speed in km/hr for all values of $\tilde{Y}$.}
		\label{fig:torque_wheel_speed_bias}
	\end{figure}

	As introduced in Equation \eqref{eqn:p_human}, the human power input at the pedals $P_{H_p}$ is calculated as the product of the pedal torque and the pedal RPM. Figure \ref{fig:torque_wheel_speed_bias} plots the pedal torque as a function of pedal RPM for all values of $\tilde{Y}$.

	It is found that when $P_{M_e} \ge 0$, there is a minimum amount of pedal torque $\tau_p$ required to increase the wheel speed which is independent of the value of the control input $\tilde{Y}$. This is defined as
	\begin{equation*}
		\tau_{\text{bias}} = 45Nm.
	\end{equation*}	
	In this area, the human cyclist is effectively free-wheeling i.e. they are providing no useful power to the wheel of the bike. As such, the human only begins to provide useful power to the wheel of the bike after this torque threshold is surpassed. Therefore, this torque bias should be subtracted before calculating $P_{H_w}$. An effort should also be made to account for the efficiency of the crankset, this will be denoted as $\eta_C$ with the value
	\begin{equation*}
		\eta_C = 90\%.
	\end{equation*}
	A first approximation for $P_{H_w}$ is given as
	\begin{equation} \label{eqn:p_human_bias}
		P_{H_w} = \text{Max}(\eta_C(\tau_p - \tau_{\text{bias}}) \omega_p, 0) \hspace{20pt} \text{if} P_{M_e} \ge 0,
	\end{equation}
	where	
	\begin{align*}
		P_{H_w} = &\text{ Power from the human being used to move the wheel (Watts)} \\
		\tau_p = &\text{ Torque at pedals (Nm)} \\
		\tau_{\text{bias}} = &\text{ Torque bias introduced by motor (Nm)}\\
		\omega_p =&\text{ Angular velocity of pedals (rads$^{-1}$)}\\
		\eta_C = &\text{ Efficiency of the crankset}\\
	\end{align*}
	The Max(A,0) function is necessary to include because $(\tau_p - \tau_{\text{bias}})$ will be negative in the low pedal speed range which would lead to negative $P_{H_w}$ which is not possible. Now, a model will be derived for $P_{M_w}$. It will be assumed that the motor converts electrical power that it is supplied with $P_{M_e}$, to mechanical power with a constant efficiency $\eta_M$ regardless of the motor load. 	Estimating $\eta_M$ is then required. The efficiency of the 250W motor is quoted as being $\geq 80\%$ \footnote{https://www.aliexpress.com/store/product/Electric-8FUN-SWXH-rear-disc-brake-motor-for-bike-36V-255RPM-250W-ebike-hub-motor/904105\_32699332646.html}. For the purposes of this model, it will be assumed that
	\begin{equation*}
		\eta_M = 80\%.
	\end{equation*}
	In the high pedal speed range, it was observed from Figure \ref{fig:motor_human_power_all_y_tilde} that $P_{M_e}$ decreased to a small value as $P_{H_p}$ increased beyond a certain threshold. It was found that this value was constant for a given value of $\tilde{Y}$. This small value of $P_{M_e}$ was no load electrical power where the motor was effectively free-wheeling. In this high pedal speed region, the motor was providing no useful power to the motor i.e. $P_{M_w} = 0$. This no load power must be subtracted from the estimation of $P_{M_w}$. It was found that this no load power consumption was depdendent on the value of the control input $\tilde{Y}$ and could be estimated using two experimentally derived parameters $\beta_1$ and $\beta_2$. As such, $P_{M_w}$ was estimated as
	\begin{equation*}
		P_{M_w} = P_{M_e}\eta_m - (\tilde{Y}\beta_1 + \beta_2),
	\end{equation*}
	where	
	\begin{align*}
		P_{M_w} = &\text{ Power from the motor being used to move the wheel (Watts)} \\
		P_{M_e} = &\text{ Electrical power input to motor (Watts)}\\
		\eta_m = &\text{ Motor efficiency} \\
		\tilde{Y} = &\text{ Control input}\\
		\beta_1, \beta_2 = &\text{ Modelling parameters for no load power}.\\
	\end{align*}
	$\beta_1$ and $\beta_2$ were estimated using the method of least squares from experimental data. These approximations for $P_{H_w}$ and $P_{M_w}$ were then used to calculate the total power being provided to the wheel $P_W$ according to Equation \eqref{eqn:pow_to_wheel}. The results are shown in Figure \ref{fig:modifiedPowerWheelSpeed}.
	
	\afterpage{%
		\begin{landscape}
			\begin{figure}
				\centering
				\includegraphics[width=0.8\linewidth]{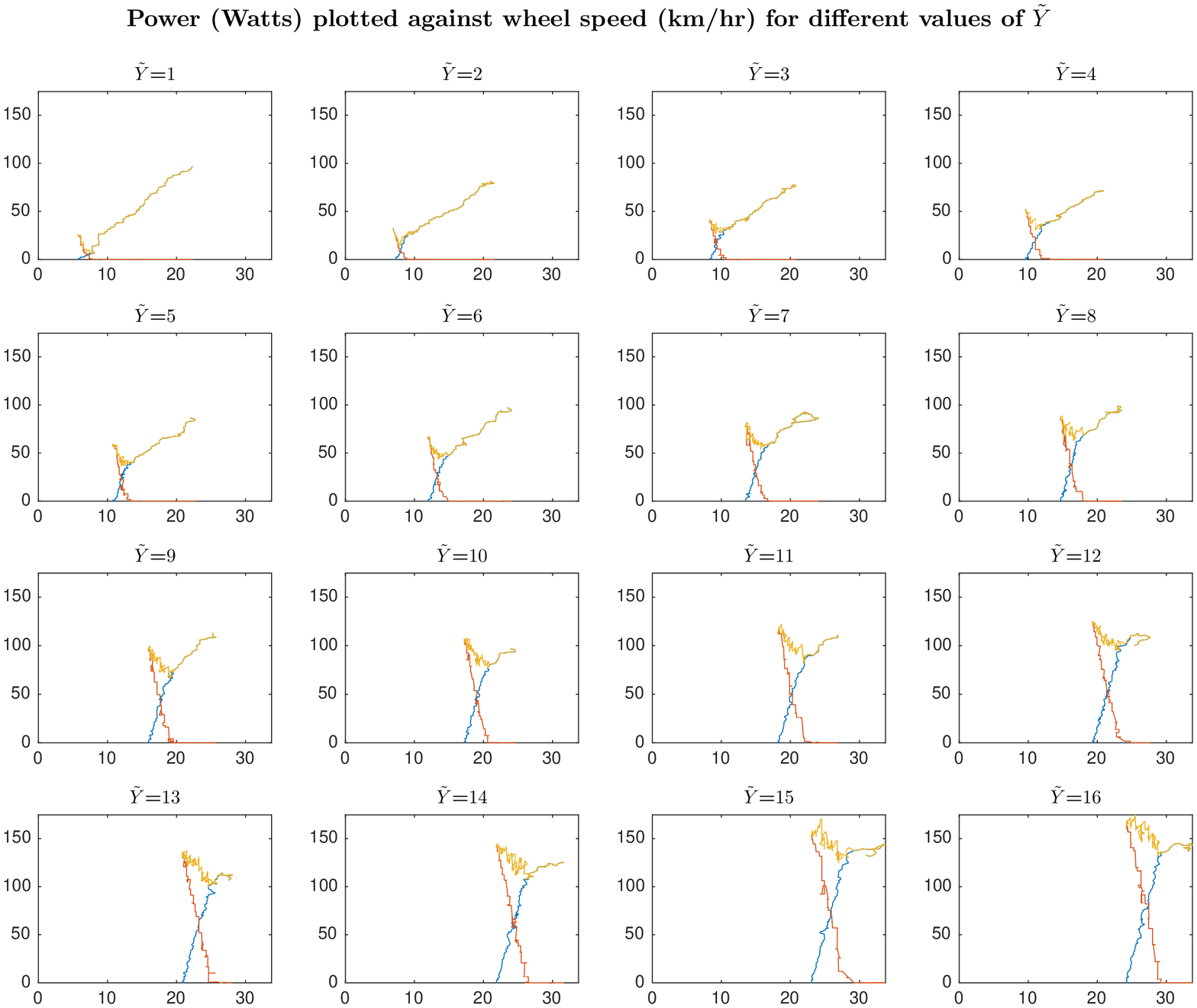}
				\caption{Plot showing motor power provided to the wheel $P_{M_w}$ in Watts (red), human power provided to the wheel $P_{H_w}$ in Watts (blue) and total power provided to the wheel $P_W$ in Watts (yellow) plotted versus wheel speed in km/hr.}
				\label{fig:modifiedPowerWheelSpeed}
			\end{figure}
			
			\begin{figure}
				\centering
				\includegraphics[width=0.7\linewidth]{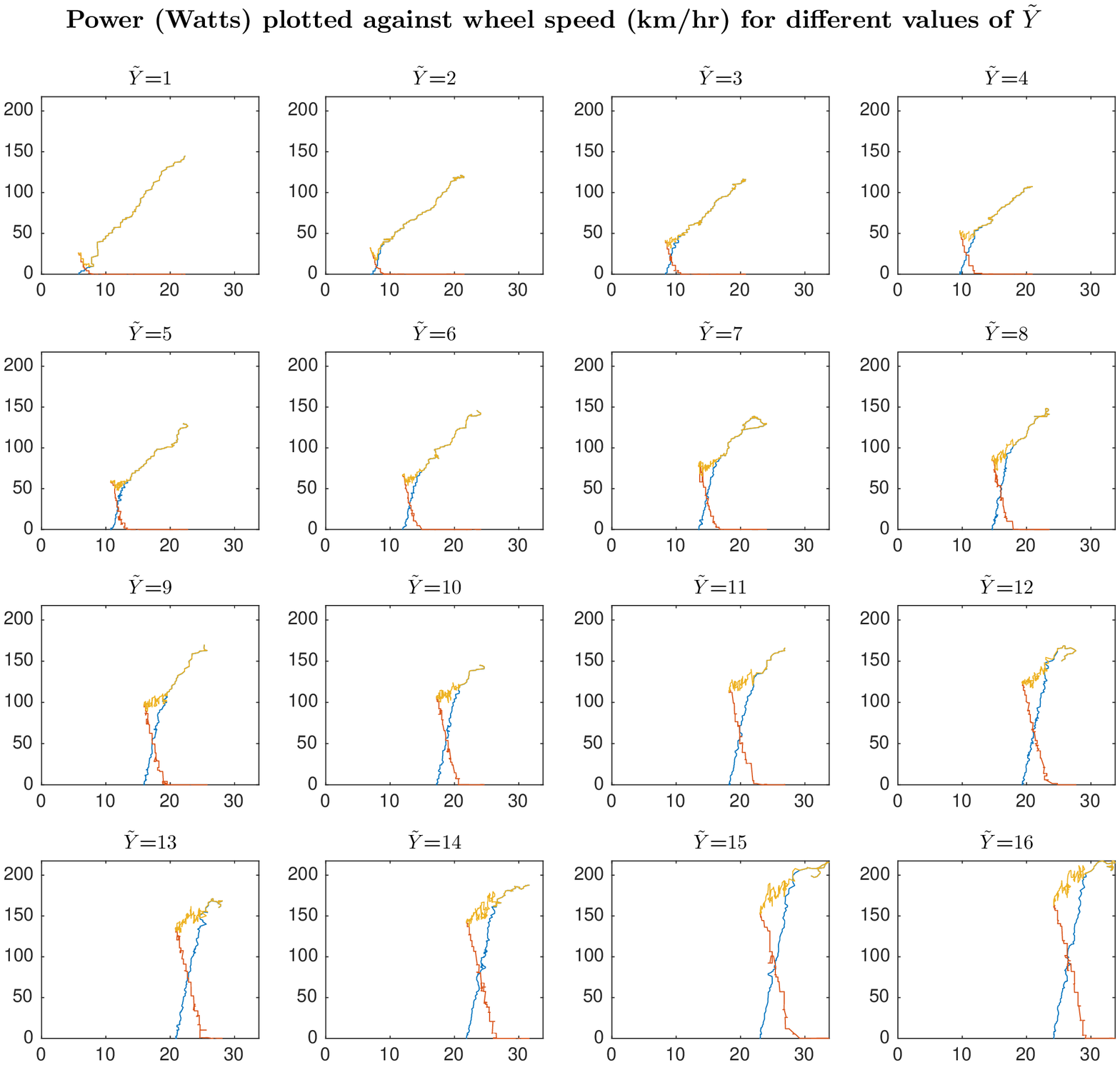}
				\caption{Plot showing motor power provided to the wheel $P_{M_w}$ in Watts (red), human power provided to the wheel $P_{H_w}$ (including scaling factor) in Watts (blue) and total power provided to the wheel $P_W$ in Watts (yellow) plotted versus wheel speed in km/hr.}
				\label{fig:finalModifiedPowerWheelSpeed}
			\end{figure}
		\end{landscape}
		\clearpage
	}
	
	The yellow curve in Figure \ref{fig:modifiedPowerWheelSpeed} is the total power being provided to the wheel $P_W$. It is seen that there are a number of occasions where the wheel speed $S_W$ is increasing, but $P_W$ is decreasing. According to basic physics, this cannot be the case. The model must be modified to take account of this fact. By adding a scaling factor $\alpha$ to the approximation of $P_{H_w}$ this yields
	\begin{equation} \label{eqn:p_human_bias_scaling}
		P_{H_w} = \text{Max}(\alpha\eta_C(\tau_p - \tau_{\text{bias}}) \omega_p, 0) \hspace{20pt} \text{if } P_{M_e} \ge 0,
	\end{equation}
	where
	\begin{align*}
		\alpha = &\text{ Scaling factor}. \\
	\end{align*}
	It is found that $\alpha$ = 1.5 works well. It will be assumed that
	\begin{equation*} 
		\alpha = 1.5.
	\end{equation*}
	
	Recalculating $P_{H_w}$ and $P_{M_w}$ leads to Figure \ref{fig:finalModifiedPowerWheelSpeed}. From Figure \ref{fig:finalModifiedPowerWheelSpeed}, it is seen that $P_W$ increases as $S_W$ increases and vice versa as required by basic physics. These are the approximations of $P_{H_w}$ and $P_{M_w}$ which will be considered for the rest of the modelling section. 
	
	Two variables will now be defined which are useful for characterising the system. The first variable $S_1$ is defined as the lowest wheel speed (km/hr) at which $P_{H_w} > 0$. When $S_W < S_1$, $P_{H_w} = 0$, so $P_W = P_{M_w}$. The second variable to be introduced is $S_2$. $S_2$ is defined as the lowest wheel speed (km/hr) at which $P_{M_w} = 0$. When $S_W > S_2$, $P_{M_w} = 0$, so $P_W = P_{H_w}$. Note that both $S_1$ and $S_2$ depend on the specific value of $\tilde{Y}$ being considered.
	
	Defining these variables helps to characterise the system in a compact way. Equation \eqref{eqn:pow_to_wheel} can be simplified for specific cases as	\\
	\textbf{Sector 1:} \\
	Human free-wheeling: \hspace{20pt} $S_W < S_1,  \hspace{20pt} P_{H_w} = 0$\\
	\textbf{Sector 2:} \\
	Human and motor contributing to $P_W$: \hspace{20pt} $S_1 < S_W < S_2, \hspace{20pt} P_{H_w} > 0, P_{M_w} >0$ \\
	\textbf{Sector 3:} \\
	Motor free-wheeling: \hspace{20pt} $S_W > S_2,  \hspace{20pt} P_{M_w} = 0$\\
%
%
	
	\subsection{Modelling variables 2: Approximating $m$ from $P_W$, $P_{M_w}$ and $P_{H_w}$} \label{subsec:mod_mod2}
	
	For many proposed use cases, knowing the proportion of power provided by both the human $P_{H_w}$ and the motor $P_{M_w}$ to move the wheel of the bike $P_W$ is of interest. The variable $m$ is thus introduced which is defined
	\begin{equation*}
		m = \frac{P_{H_w}}{P_W} = \frac{P_{H_w}}{P_{H_w} + P_{M_w}} \hspace{20pt} 0 < m < 1.
	\end{equation*}	
	By this definition
	\begin{equation*} 
		m =  
		\begin{cases}
			0 & \text{If $P_{H_w}$ = 0 i.e. the motor is providing all of the power to move the wheel } \\	
			1 & \text{if $P_{M_w}$ = 0 i.e. the human is providing all of the power to move the wheel. } \\
		\end{cases}
	\end{equation*}
	
\begin{figure}[H]
		\centering
		\includegraphics[width=0.8\linewidth]{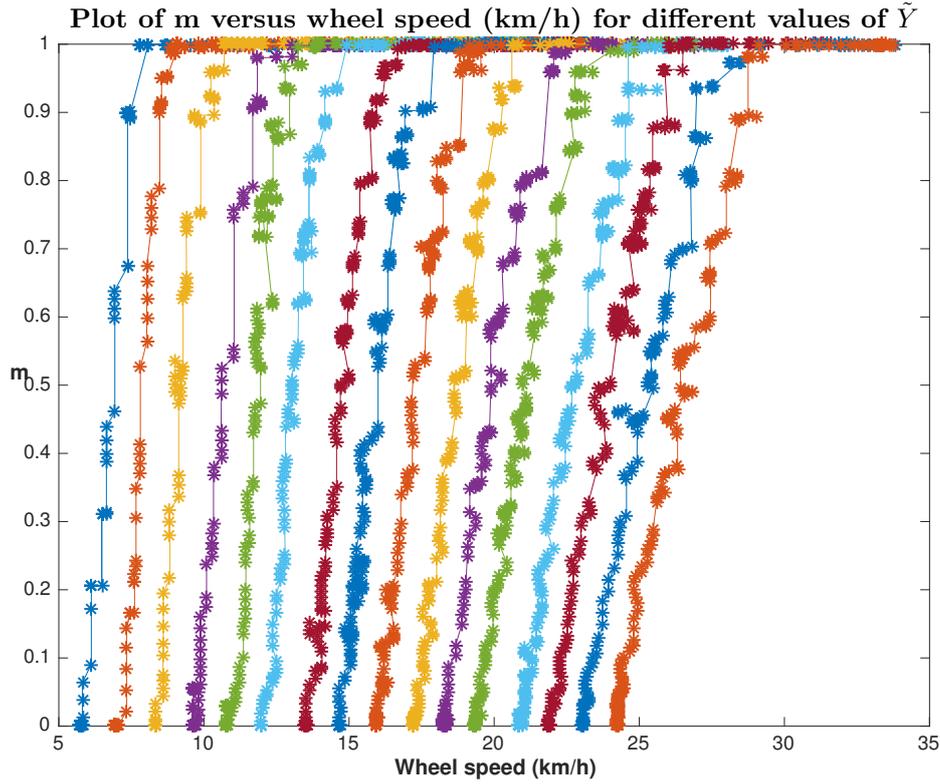}
		\caption{Plot of m versus wheel speed in km/h for different values of $\tilde{Y}$, low values of $\tilde{Y}$ are to the left, higher values of $\tilde{Y}$ are to the right.}
		\label{fig:mWheelSpeed}
	\end{figure}

	It is of interest to understand how $m$ varies with both $\tilde{Y}$ and $S_W$. The test data from Section \ref{subsec:mod_exp2} was used to calculate $m$ for each of the 16 different values of $\tilde{Y}$. Figure \ref{fig:mWheelSpeed} shows the results of this calculation.
	
	Figure \ref{fig:mWheelSpeed} shows that for any value of $\tilde{Y}$, $m$ varies from its minimum value 0, up to its maximum value 1. For each increasing value of $\tilde{Y}$, it is seen that the $\tilde{Y}$ curve shifts further to the right. The spectrum of curves of $\tilde{Y}$ shows that $m$ can take on any value between 0 and 1 over a large range of wheel speeds $S_W$. This means that by updating the value of $\tilde{Y}$, a desired value of $m$ can be set over a wide range of wheel speeds $S_W$. Interpreting, this means that the proportion of power that the motor $P_{M_w}$ and the human $P_{H_w}$ are providing to move the wheel of the bike $P_W$, can be selected over a wide range of wheel speeds $S_W$ by updating the value of $\tilde{Y}$. This fact will be made use of for many of the proposed open and closed loop use cases which are discussed in Chapter \ref{ch:openloop} and Chapter \ref{ch:closedloop}.
	
	\subsection{Extending the model} \label{subsec:mod_extensions}
	There are a number of ways in which the model presented above can be further generalised. One model improvement would be to try to include the current state of charge of the battery in the model. The state of charge of the battery was observed to have a small effect on the electrical power input provided to the motor $P_{M_e}$ which has not been included in the model. It was also assumed that the motor has a constant motor efficiency regardless of its operating point, the model could be extended to estimate the motor efficiency more accurately.
	
	Each of the experiments was conducted on a flat road in a lab environment. The performance of the model in areas with different resistance profiles and gradients has not been investigated here. The model could be extended to model these different environments. The human changing the mechanical gear would also have an effect on the model which has not been accounted for here. 
	
	The model should also be extended to take account of other interactions that the human can have with the system including twisting the hand throttle and engaging the brakes.

%% file: chapters/ch_open_loop_applications.tex
\chapter{Open Loop Applications}\label{ch:openloop}

\section{Introduction} \label{sec:ol_introduction} 
Chapter \ref{ch:modelling} derived an energy model for a smart electric bike. This model provides a basis for many different use cases. A use case refers to a specific service that the smart electric bike can provide to help people in a city. Several use cases will be discussed in Chapters \ref{ch:openloop} and \ref{ch:closedloop}. There will be a large emphasis on reducing a cyclist's inhalation of pollutants across these use cases. Use cases which rely on open loop control are discussed in this chapter, and use cases which rely on closed loop control are discussed in Chapter \ref{ch:closedloop}.

\section{Use case 1: Reduce a cyclist's inhalation of pollutants with open loop control} \label{sec:ol_pollution} 

	\subsection{Objective} \label{subsec:ol_pol_intro}
		The purpose of this use case was to propose a method by which a cyclist could reduce the amount of pollutants that they inhale while in a polluted area. As discussed in detail in Chapter \ref{ch:litreview}, air pollutants have extremely negative impacts on human health. Cyclists have been shown to inhale increased amounts of pollutants due to their elevated breathing rates. Section \ref{subsec:ol_pol_method} introduces the method that was used for the implementation. Section \ref{subsec:ol_pol_simulation} and Section \ref{subsec:ol_pol_experiment} then tested and validated the method. Finally, some conclusions on the method are drawn in Section \ref{subsec:ol_pol_conclusion}.
		
	\subsection{Method}\label{subsec:ol_pol_method}
		As discussed in Chapter \ref{ch:litreview}, a cyclist's breathing rate will increase while cycling in order to increase their uptake of oxygen to enable their muscles to provide power to move the wheel of the bike. If the cyclist is cycling through a polluted area, this means that they will breathe in more polluted air. Therefore, in order to reduce the amount of pollutants that the cyclist inhales, their breathing rate should be reduced.
		
		As established in Chapter \ref{ch:modelling}, in an electric bike, power to move the wheel $P_W$ can be provided by the motor $P_{M_w}$ or by the human $P_{H_w}$. Since a cyclist's breathing rate increases as they provide more power $P_{H_w}$, it is therefore desired to minimise the amount of power that the cyclist is providing $P_{H_w}$, in order to minimise their breathing rate, and therefore minimise the amount of pollutants that they inhale. 
		
		Chapter \ref{ch:modelling} introduced a model with a method by which the proportion of power that the human $P_{H_w}$ and the motor $P_{M_w}$ are providing to move the wheel of the bike can be calculated. This variable was expressed as
		\begin{equation*}
			m = \frac{P_{H_w}}{P_W} = \frac{P_{H_w}}{P_{H_w} + P_{M_w}} \hspace{20pt} 0 < m < 1.
		\end{equation*}	
		The value of the variable $m$ can be controlled over a wide range of wheel speeds $S_W$ by adjusting the value of the control input $\tilde{Y}$. The range of possible values of $m$ was shown in Figure \ref{fig:mWheelSpeed}. By definition, when $m = 0$, $P_{H_w} = 0$. Therefore, since the objective was to minimise $P_{H_w}$, this could be achieved by choosing a value of the control input $\tilde{Y}$ which set $m=0$. With $m=0$, this would mean that all of the power to move the wheel of the bike would be provided by the motor.
		
		However, a limitation on the target value of $m$ should now be introduced. European Union Directive 2002/24/EC states that ``Cycles with pedal assistance which are equipped with an auxiliary electric motor having a maximum continuous rated power of 0.25 kW, of which the output is progressively reduced and finally cut off as the vehicle reaches a speed of 25 km/h (16 mph) or \textbf{if the cyclist stops pedaling}''. As such, in order to  comply with the legislation, the human must provide some input power at all times. This can be implemented by ensuring that $m$ is slightly greater than 0. While this moves away from the optimal value for $m$, ensuring that $m$ is small should still ensure that the cyclist's breathing rate is reduced.
		 		
\begin{figure}[H]
		 	\centering
		 	\includegraphics[width=0.8\textwidth]{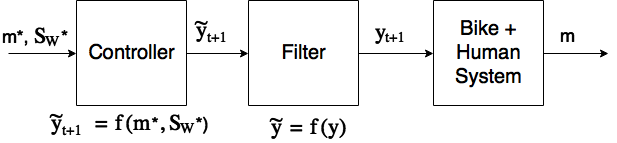}
		 	\caption{Open loop control diagram to reduce a cyclist's inhalation of pollutants.} 
		 	\label{fig:ol_min_pollution_control_circuit}
		\end{figure}

		Since there is no feedback in open loop control, a value of $\tilde{Y}$ is selected to attempt to deliver a target value of $m$, but it is not validated that the observed value of $m$ becomes equal to the target value. The target value for $m$ will be denoted as $m^*$. In order to validate that $m$ reaches its target value $m^*$, closed loop feedback control would be required. This is implemented in Section \ref{sec:cl_min_breathing}. The open loop control circuit schematic is shown in Figure \ref{fig:ol_min_pollution_control_circuit}.
		
		The target value $m^*$ should be selected to be almost equal to 0. As seen from Figure \ref{fig:mWheelSpeed}, the variable $m$ takes on a value close to 0 (let's say $m^*=0.1$) over a wide range of wheel speeds $S_W$ from approximately 5km/hr to 25km/hr. As such, a specific value of $\tilde{Y}$ can be chosen to achieve a target speed $S_W^*$. Choosing this target speed $S_W^*$ can be done based on specific policies. Note that this target speed $S_W^*$ is only applicable for environments with the same resistance profile and gradient.
		
		
	\subsection{Simulation}\label{subsec:ol_pol_simulation}
		
\begin{figure}[H]
			 	\centering
			 	\includegraphics[width=\textwidth]{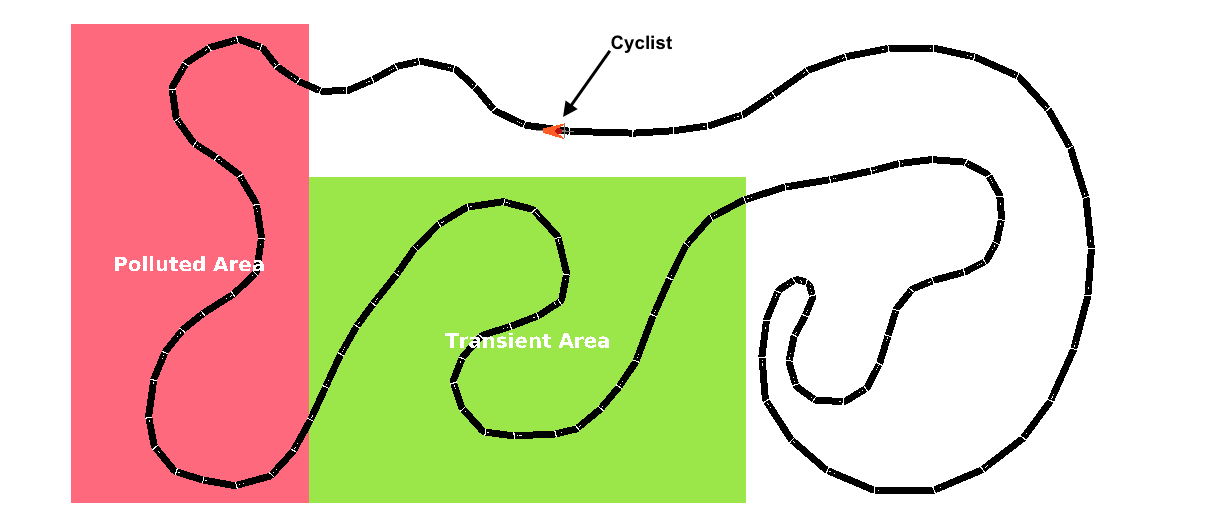}
			 	\caption{SUMO simulation of cycling route for open loop control.} 
			 	\label{fig:ol_min_pollution_circuit}
		\end{figure}

		A simulated polluted environment was created using SUMO \footnote{http://www.dlr.de/ts/en/desktopdefault.aspx/tabid-9883/16931\_read-41000/} (Simulation of Urban Mobility). In this simulation, a cycling route was created. The cycling route is shown in Figure \ref{fig:ol_min_pollution_circuit}.	
		
		Figure \ref{fig:ol_min_pollution_circuit} shows three different areas (or zones) which are 1) a non-polluted area 2) a transient area which is highlighted in green and 3) a polluted area which is highlighted in red. The polluted area represents an area with a high concentration of atmospheric pollutants. As such, it is desired that the cyclist's breathing rate is minimised in this area. However, intuitively (and proven experimentally), a person's breathing rate does not instantaneously slow down as soon as they start to reduce their power input. It is therefore necessary to include area 2) the transient area, which provides time for the cyclist's breathing rate to reduce in advance of entering the polluted area.
		
		It was required to establish communication between SUMO and the electric bike. It was necessary to transfer the speed of travel $S_W$ from the electric bike to SUMO, and to transfer a signal that indicated which of the three areas that the cyclist was cycling through from SUMO to the electric bike. Since communication between the electric bike and the smartphone was already established, it was decided to communicate between SUMO and the electric bike via the smartphone. This was achieved using the Java Socket \footnote{https://docs.oracle.com/javase/7/docs/api/java/net/Socket.html} class in the android app that was running on the smartphone.
		
		A socket was established on the same IP address and port number between SUMO and the smartphone. This required the server which was running SUMO and the smartphone to be using the same IP address for its internet connection. This was achieved using the smartphone's hotspot. When the socket was active, this meant that the two-way transfer of data between SUMO and the smartphone had been enabled.
		
	\subsection{Experiment}\label{subsec:ol_pol_experiment}
		An experiment was designed to test the method from Section \ref{subsec:ol_pol_method} in the simulated pollution environment discussed in Section \ref{subsec:ol_pol_simulation}. This experiment involved instructing a cyclist to cycle on the stationary electric bike at a constant speed $S_W$ of 20km/hr. The cyclist was not provided with any assistance from the electric motor while cycling in an unpolluted zone i.e. $m = 1$. When the cyclist entered the transient zone, the value of $\tilde{Y}$ was then increased to reduce $m$ to be very close to zero. $\tilde{Y}$ was selected to maintain the speed $S_W$ at 20km/hr. The cyclist was instructed to keep pedalling to ensure that they were still providing some power input and were in compliance with European Union Directive 2002/24/EC.
		
		It was also necessary to validate that the cyclist's breathing rate had been reduced. Measuring a cyclist's actual breathing rate using spirometry equipment while cycling a non-stationary bike is not convenient. As such, for real world cycling applications a different method should be used to validate that a cyclist's breathing rate has been reduced. Section \ref{sec:lr_validation_hr} discussed theory from previous studies showing the strong correlation between heart rate and breathing rate. For the purposes of validation in this experiment, it was assumed that a reduction in heart rate indicated a corresponding reduction in breathing rate. 
		
\begin{figure}[H]
			\centering
			\includegraphics[width=0.6\textwidth]{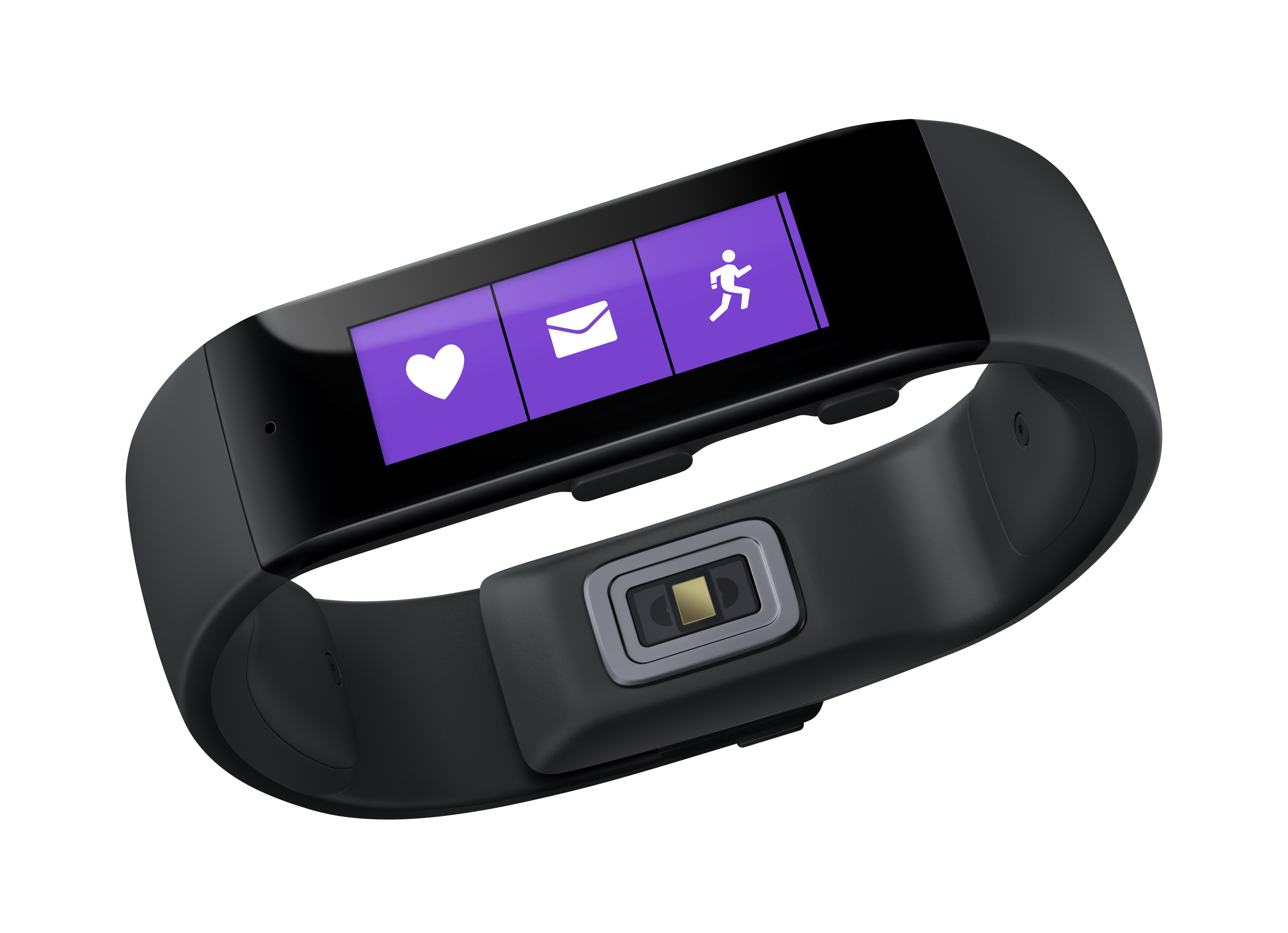}
			\caption{Microsoft Band used to measure cyclist's heart rate during experiment.} 
			\label{fig:microsoft_band}
		\end{figure}

\begin{figure}[H]
			\centering
			\includegraphics[width=\textwidth]{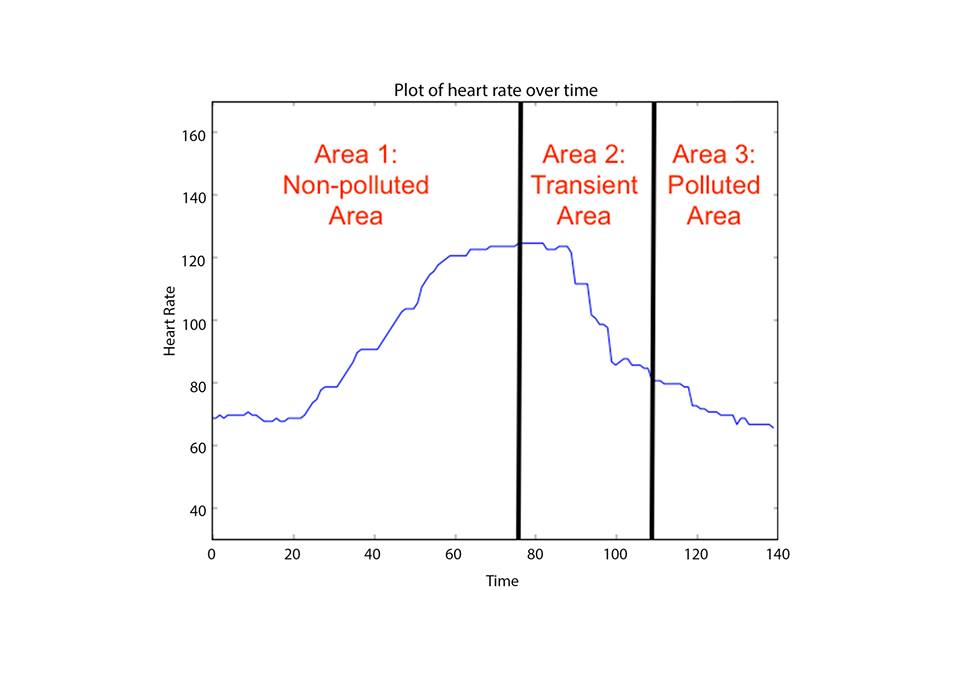}
			\caption{Graph showing heart rate during experiment.} 
			\label{fig:ol_min_pollution_graph}
		\end{figure}

		Measuring the cyclist's heart rate was a lot less intrusive than trying to measure their breathing rate. 	The cyclist's heart rate was recorded using a Microsoft Band Activity Tracker. This activity tracker is shown in Figure \ref{fig:microsoft_band}.
		The Microsoft Band transferred heart rate data by Bluetooth to the smartphone app. The results of the experiment are shown in Figure \ref{fig:ol_min_pollution_graph}. 		
		
		Figure \ref{fig:ol_min_pollution_graph} indicates the time that the cyclist spent in each specific area in the simulation. The cyclist's heart rate at the beginning of the experiment was approximately 70 beats per minute (BPM). The experiment began with the cyclist cycling in the non-polluted area. As such, $m$ = 1, which meant that the human was providing all of the power to move the wheel of the bike. As evidenced by Figure \ref{fig:ol_min_pollution_graph}, the cyclist's heart rate has increased to approximately 120 BPM when they are at the end of the non-polluted area. 
		
		When the cyclist entered the transient area, it was desired to reduce their power input in order to reduce their breathing rate. The value of $m^*$ was reduced to be close to zero and $\tilde{Y}$ was updated. The specific value of $\tilde{Y}$ was chosen to maintain a constant speed $S_W$. As can be seen from Figure \ref{fig:ol_min_pollution_graph}, the cyclist's heart rate started to fall and had reached approximately 80 BPM at the end of the transient area. This was a substantial reduction in heart rate from the peak of 120 BPM in the non-polluted area. In the polluted area the heart rate fell further to approximately 70 BPM.
		
	\subsection{Conclusion on use case}\label{subsec:ol_pol_conclusion}
		This use case proposed a new method which reduces a cyclist's inhalation of pollutants while cycling through a simulated pollution area. This was achieved by reducing the cyclist's breathing rate. The method was demonstrated and validated by simulating a pollution environment, carrying out an experiment and validating that the cyclist's breathing rate had been reduced. This use case is extended in Chapter \ref{ch:closedloop} to include closed loop control.	
		
		A simulation environment was created using SUMO because it was more appropriate for a stationary lab environment. It was also relatively easy to implement and very visually intuitive. An extension to the simulation would be to set up real life cycling test zones. A cyclist would be instructed to cycle the electric bike on a specific route. The location of the cyclist could be tracked using Google Maps location data on the smartphone. When the user arrives at a specific location, this would indicate that they had entered into either a transient or polluted zone and the method presented above to minimise their breathing rate could be used to reduce their breathing rate. These specific zone boundary locations could be defined in a number of ways, for example Google real time traffic data could be used as a data stream to indicate the density of cars in an area, which could indicate a pollution zone. A real life cycling environment simulation like this may require some modifications to be made to the model discussed in Chapter \ref{ch:modelling}.
		
		In the simulation, the size of the transient area was found by trial and error. Note that the size of the transient area depends on a number of factors, including the length of time that the cyclist has been cycling for and the individual cyclist's fitness level. Further research should be conducted into determining the size of this transient zone.
		
		As stated in Section \ref{subsec:ol_pol_intro}, the method presented above requires a specific policy to be implemented which provides a basis for selecting the desired speed of travel $S_W$. In Section \ref{subsec:ol_pol_experiment}, this desired speed of travel was selected to maintain a constant speed across the three zones. For open loop control, further work should be conducted to decide a basis for which this desired speed of travel is decided. This research could include predicting the human's desired speed of travel $S_W$. Policies could also be implemented to change the desired speed of travel based on the state of charge of the battery.
		
		A feedback mechanism should also be implemented which informs the cyclist of the zone they are travelling in and gives them advance notice of changes in the control input that are about to take place. Another extension that should be carried out to the research is to conduct significance testing which quantitatively relates the reduction in breathing rate to the reduction in the amount of pollutants that are inhaled.


%

%
%
%
%
%
%
%

%% file: chapters/ch_closed_loop_applications.tex
\chapter{Closed Loop Applications}\label{ch:closedloop}
\section{Introduction} \label{sec:cl_introduction} 
	Chapter \ref{ch:openloop} discussed use cases that were based on open loop control. This chapter will discuss use cases that implement closed loop control. Section \ref{sec:cl_min_breathing} extends the use case from Section \ref{sec:ol_pollution} which aimed to reduce a cyclist's inhalation of pollutants by reducing their breathing rate while cycling through a polluted area.
	
\section{Use case 1: Reduce a cyclist's inhalation of pollutants with closed loop control} \label{sec:cl_min_breathing} 
	\subsection{Objective} \label{subsec:min_breathing_objective}
		Section \ref{sec:ol_pollution} discussed a method to reduce the amount of pollutants that a cyclist inhaled while cycling through a polluted area using open loop control. This was achieved by reducing the cyclist's breathing rate. The objective of this use case is the same as the use case from Section \ref{sec:ol_pollution}, with the difference being that the implementation is based on closed loop control. Section \ref{subsec:cl_pol_method} introduces the method that was used for the implementation. Section \ref{subsec:cl_pol_simulation} and Section \ref{subsec:cl_pol_experiment} tests and validates the method. Finally, some conclusions on the method are drawn in Section \ref{subsec:cl_pol_conclusion}.

	\subsection{Method}\label{subsec:cl_pol_method}
\begin{figure}[H]
			\centering
			\includegraphics[width=\textwidth]{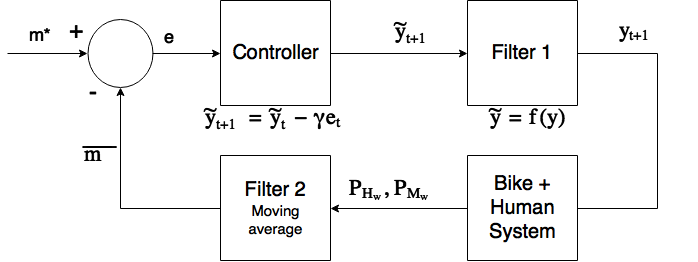}
			\caption{Closed loop control diagram to reduce a cyclist's inhalation of pollutants.} 
			\label{fig:min_breathing_cl}
		\end{figure}

		The difference in the methodology for the control between the open loop and closed loop cases was that a feedback loop was included in the closed loop case. The control schematic is shown in Figure \ref{fig:min_breathing_cl}. 
		
		The target value of $m$ is denoted as $m^*$ in Figure \ref{fig:min_breathing_cl}. This target can be set in a number of ways. For the use case being discussed, $m^*$ should be selected to be small in polluted areas to reduce the human power input $P_{H_w}$ and subsequently their breathing rate. In non-polluted areas, the value of $m^*$ can be selected to be larger. As introduced in Chapter \ref{ch:modelling}, the value of $m$ is calculated as
		\begin{equation*}
			m = \frac{P_{H_w}}{P_W} = \frac{P_{H_w}}{P_{H_w} + P_{M_w}} \hspace{20pt} 0 < m < 1.
		\end{equation*}	
		For the purposes of control, a small change can be made here which is to consider a moving average of $m$. The signals which are used to calculate $m$ (the human power and the motor power) can sometimes be noisy. Including this moving average helps to reduce the effect of this noise. As such $\bar{m}$ is introduced as
		\begin{equation*}
			\bar{m} = \frac{\bar{P}_{H_w}}{\bar{P}_W} = \frac{\bar{P}_{H_w}}{\bar{P}_{H_w} + \bar{P}_{M_w}} \hspace{20pt} 0 < m < 1.
		\end{equation*}	
		$\bar{P}_{H_w}$ was calculated as
		\begin{equation*}
			\bar{P}_{H_w} = \frac{\sum_{i=1}^{n_h}P_{H_wi}}{n_h}.
		\end{equation*}	
		And $\bar{P}_{M_w}$ was calculated as
		\begin{equation*}
			\bar{P}_{M_w} = \frac{\sum_{i=1}^{n_m}P_{M_wi}}{n_m},
		\end{equation*}	
		where $n_h$ and $n_m$ were the number of samples that were being averaged over in each case. The samples that were considered were the most recently available values. This provided a basis to calculate $\bar{m}$. As stated above, in closed loop control the actual observed value of $\bar{m}$ is compared to the target value $m^*$ in order to ensure that the actually observed value becomes equal to the target value. 
		
		As shown in Figure \ref{fig:min_breathing_cl}, an error between the target value $m^*$ and the actually observed value $\bar{m}$ was calculated. The error $e_t$ is defined as	  		  
		\begin{equation}
			e_t = m^*_t - \bar{m}_t.
			\label{eqn:cl_error_m}
		\end{equation}		
		The value of the control input $\tilde{Y}$ can then be calculated using a P controller for the next time step according to the equation
		\begin{equation*}
			\tilde{y}_{t+1} = \tilde{y}_{t} - \gamma e_{t} \hspace{20pt} \text{Where } \gamma>0,
		\end{equation*}	
		where $\gamma$ is a gain parameter. Note again that $\tilde{Y}$ is bounded between 1 and 16 which prevents it from increasing or decreasing infinitely. Note that it was also necessary to decide a sampling period which will be denoted as $T_s$. The sampling period is determined by the smartphone but due consideration should be given to how quickly the motor can change its power output, this requires a dynamic analysis.
		
		Given that $P_{H_w}$ is not controllable (it is decided by the human), it would not be expected to achieve zero error between the target value $m^*$ and the actually observed value $\bar{m}$. If $P_{H_w}$ changes unexpectedly, there will be a delay of a few time steps for $P_{M_w}$ to change to maintain $m^*$. As such, it is advised to define an acceptable tolerance $b$ between $\bar{m}$ and $m^*$.

	    




	\subsection{Simulation}\label{subsec:cl_pol_simulation}
\begin{figure}[H]
			\centering
			\includegraphics[width=\textwidth]{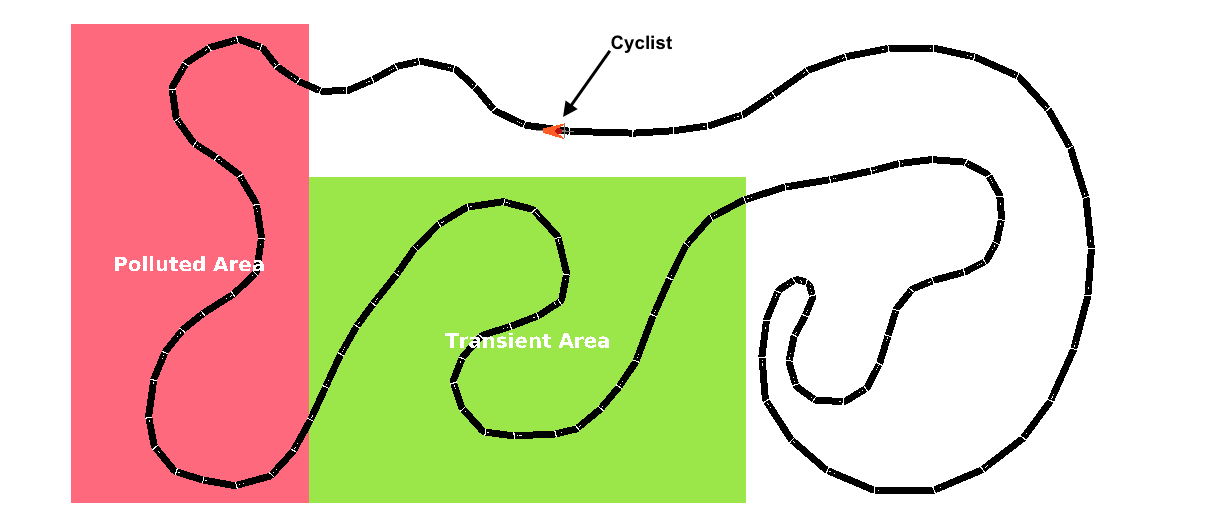}
			\caption{SUMO simulation of cycling route for closed loop control.} 
			\label{fig:cl_min_pollution_circuit}
		\end{figure}

		The simulation used to test the method above was developed in SUMO. The cycling route which was created was similar to the one used for the open loop case from Section \ref{sec:ol_pollution} and is shown in Figure \ref{fig:cl_min_pollution_circuit}.	
		The difference between the open loop simulation and the closed loop simulation was that in the closed loop simulation, a gradual change in the value of $m^*$ is introduced when the transient area was entered. This gradual transition prevented large overshoots in the control circuit. SUMO communicated a different signal to the smartphone based on the cyclist's position in the transient area which the smartphone used to gradually reduce $m^*$. This also felt much more natural for the cyclist than a sudden change in $m^*$ as had been implemented in the open loop case. 
		
		Communication between SUMO and the smartphone was the same as for the open loop case i.e. it was implemented using the Java Socket class.
		
	\subsection{Experiment}\label{subsec:cl_pol_experiment}
		A cyclist was instructed to cycle on the stationary electric bike. The speed that the cyclist was cycling at on the stationary electric bike was transferred to the SUMO simulation which updated the position of the cyclist on the screen. In the non-polluted area, the target $m$ value was set to be $m^* = 0.9$. This meant that the control circuit should aim to have the cyclist providing $90\%$ of the power to move the wheel of the bike when in the non-polluted area.
		
		The target value $m^*$ was then gradually reduced to $m^* = 0.3$ when the cyclist entered the transient region. The target value $m^*$ was maintained at $m^*=0.3$ until the cyclist left the polluted zone, at which point $m^*$ was gradually increased back to $m^*=0.9$. The closed loop control was implemented using a sampling period of $T_s =$ 1 second. This meant the request to the motor $\tilde{Y}$ was updated once per second. The gain parameter $\gamma$ was set to be 20 which was experimentally found to work well.
		
		In order to calculate $\bar{m}$, it was required to decide the values of $n_h$ and $n_m$. To calculate $\bar{P}_{H_w}$, the 20 most recent samples of $P_{H_w}$ were considered ($n_h = 20$). Since the data rate of measurements coming from the bike sensors was 5Hz, this corresponded to averaging over the last 4 seconds worth of data. In calculating $\bar{P}_{M_w}$, only the 5 most recent samples of $P_{M_w}$ were considered ($n_m = 5$). This corresponded to averaging over the last 1 second of data. More samples of $P_{H_w}$ were considered than $P_{M_w}$ because $P_{H_w}$ contained more noise than $P_{M_w}$.
		
\begin{figure}[H]
			\centering
			\includegraphics[width=0.9\textwidth]{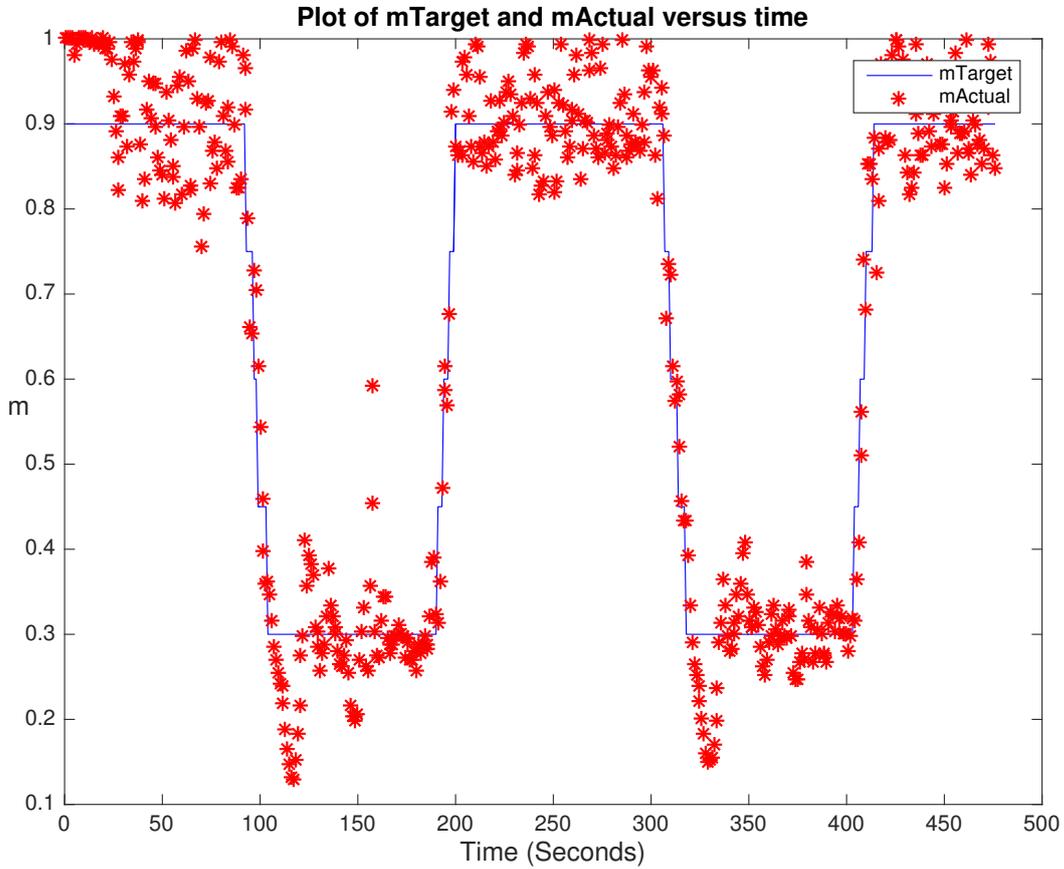}
			\caption{Target value $m^*$ (blue line) and actually observed value $\bar{m}$ (red stars) plotted versus time during closed loop control experiment.} 
			\label{fig:cl_m}
		\end{figure}

		The results of the closed loop control are shown in Figure \ref{fig:cl_m}.
		
		Figure \ref{fig:cl_m} shows $m^*$ (blue line) and $\bar{m}$ (red stars) during the experiment. The cyclist completed two laps of the simulated cycling route shown in Figure \ref{fig:cl_min_pollution_circuit}. $m^*$ updated based on the cyclist's progress through the simulation. The cycling route started with the cyclist in the non-polluted area ($m^* = 0.9$). When they entered the transient area, $m^*$ was gradually decreased to $m^* = 0.3$. This can be seen in Figure \ref{fig:cl_m}. When the cyclist left the polluted zone, $m^*$ then gradually increased to $m^* = 0.9$. The cyclist then completed one more lap of the cycling route for which the same changes in $m^*$ were implemented. 
		
		$\bar{m}$ is seen to clearly follow $m^*$. When $m^*$ is reduced from $m^* = 0.9$ to $m^* = 0.3$ there is initially seen to be a slight overshoot. This overshoot could be reduced by reducing $m^*$ more slowly or by changing the gain parameter $\gamma$. This slight overshoot was not considered to be a big problem. The cyclist reported that transitions in the amount of power that the motor was providing $P_{M_w}$ felt quite natural. 
		
\begin{figure}[H]
			\centering
			\includegraphics[width=0.9\textwidth]{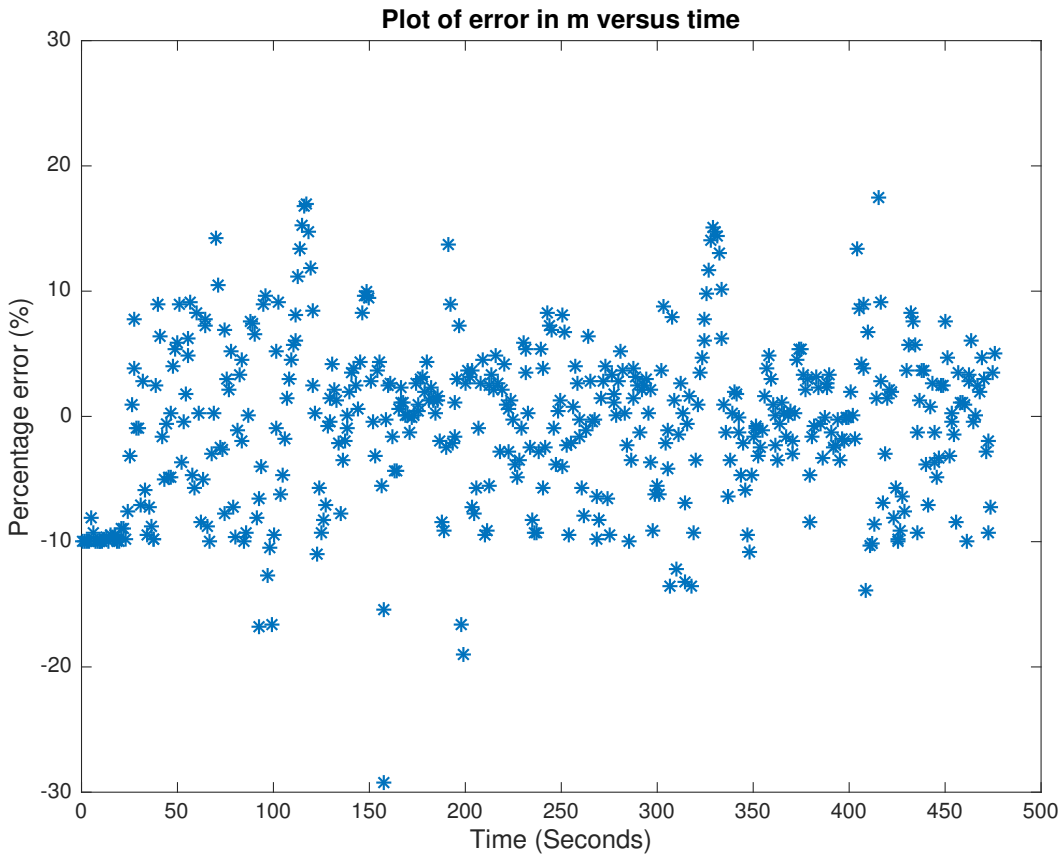}
			\caption{Distribution of error values from closed loop control experiment.} 
			\label{fig:cl_error}
		\end{figure}

		A distribution of the error $e_t$ (as defined in Equation \eqref{eqn:cl_error_m}) is shown in Figure \ref{fig:cl_error}. This error is calculated from Figure \ref{fig:cl_m} as the difference between $m^*$ and $\bar{m}$ for every value of time $t$.
		
		The error distribution is summarised based on percentiles in Table \ref{table:cl_mError_distribution}.
		\begin{table}[H]
			\centering 
			\begin{tabular}{||c|c||} 
				\hline
				\bfseries Percentile  & \bfseries $\%$ Error \\ \hline
				5 &  -10.0\% \\\hline 
				10 &  -9.46\% \\\hline 
				25 &  -4.79\% \\     \hline           
				Median &  0\% \\        \hline 
				75 &  3.33\% \\        \hline  
				90 &  7.59\% \\        \hline 
				95 & 9.53\% \\\hline    
			\end{tabular}
			\caption{Summary of error distribution from closed loop control experiment.}
			\label{table:cl_mError_distribution}
		\end{table}
		From Table \ref{table:cl_mError_distribution} it is seen that the middle 50\% of data was contained on an error interval [-4.79\%, 3.33\%] and the middle 90\% of data was contained on an error interval [-10.0\%, 9.53\%]. It would not be expected to achieve zero error in this system given that the system is attempting to track a target value of $m^*$ which changes with time, and $\bar{m}$ depends on the variables $P_{H_w}$ and $P_{M_w}$. The power that the human is providing $P_{H_w}$, can change based on the human's desire to speed up or slow down which is not known ahead of time. $P_{M_w}$ then has to track this change which leads to errors. As such the nature of the error is understandable and should be expected from a cyber-physical system such as the one being considered.
		
\begin{figure}[H]
			\centering
			\includegraphics[width=\textwidth]{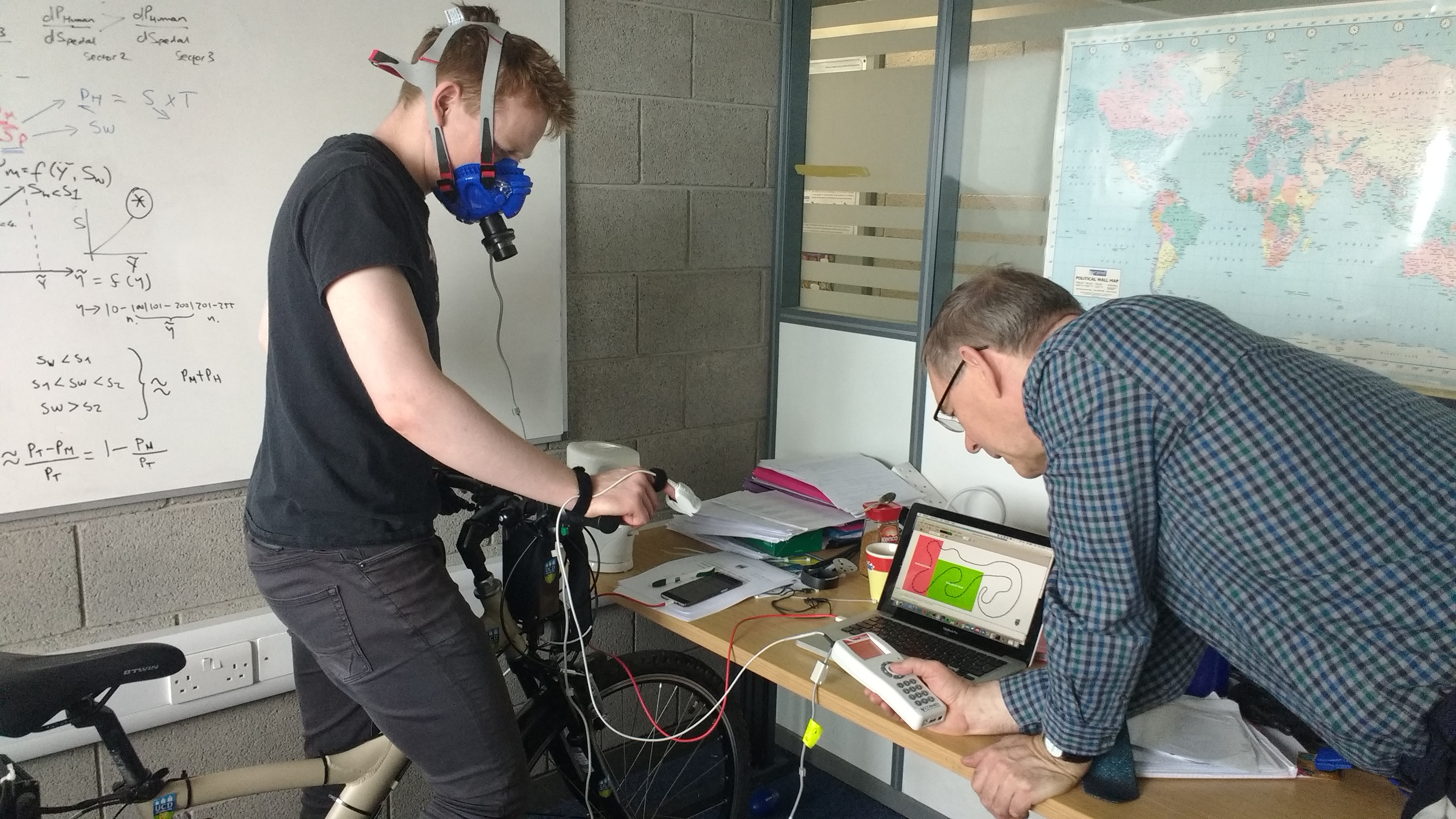}
			\caption{Photograph showing cyclist equipped with spirometry equipment during experiment.} 
			\label{fig:cl_pol_test_photo}
		\end{figure}

		During the experiment, the cyclist's breathing rate was monitored. This was done using spirometry equipment. A Spiropalm 6MWT hand-held spirometer  \footnote{http://carestreammedical.com/wp-content/uploads/Carestream-Cosmed-Spiropalm-6MWT.pdf} was purchased from COSMED. The cyclist was equipped with a face mask  \footnote{http://www.bodpod.com/images/pdf/productliterature/PG\_Facemask\_C04211-02-93\_A4\_web\_EN.pdf} during the experiment which enabled their breathing rate to be measured by the spirometer. The spirometer used a turbine flow meter to record the cyclist's inhaled minute volume of air in litres per minute. A finger pulse oximeter was also used to record the cyclist's pulse rate. A photograph taken during the experiment showing the cyclist equipped with the spirometry equipment is shown in Figure \ref{fig:cl_pol_test_photo}. 
		
\begin{figure}[H]
			\centering
			\includegraphics[width=0.9\textwidth]{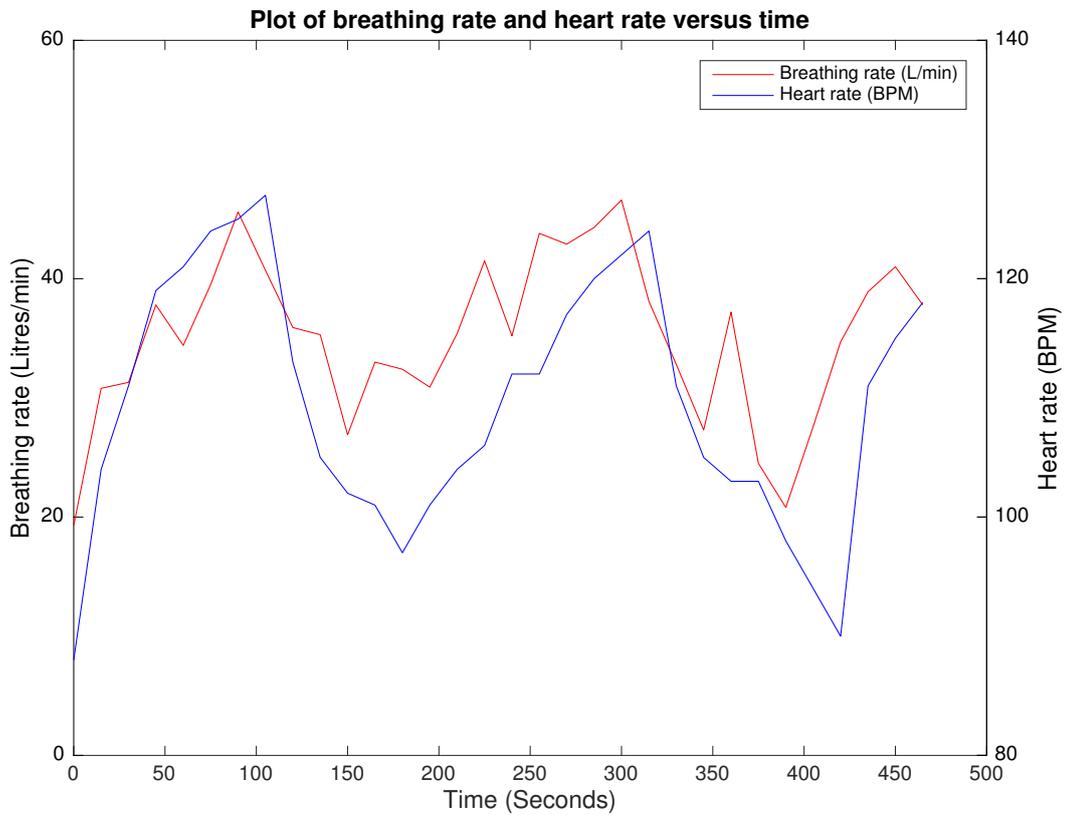}
			\caption{Heart rate (blue line) and breathing rate (red line) from spirometer plotted versus time.} 
			\label{fig:cl_pol_test_hr_br}
		\end{figure}

		The results recorded by the spirometry equipment are shown in Figure \ref{fig:cl_pol_test_hr_br}. 
		This breathing rate is shown as the red data series in Figure \ref{fig:cl_pol_test_hr_br}. The cyclist's heart rate is shown as the blue data series. As can clearly be seen, there were changes in both the cyclist's breathing rate and heart rate during the experiment. The two signals also seem to approximately track each other and the breathing rate signal seems to be slightly noisier than the heart rate signal. Each data point in Figure \ref{fig:cl_pol_test_hr_br} is spaced 15 seconds apart. These values were calculated by the spirometer.	
		
\begin{figure}[H]
			\centering
			\includegraphics[width=0.9\textwidth]{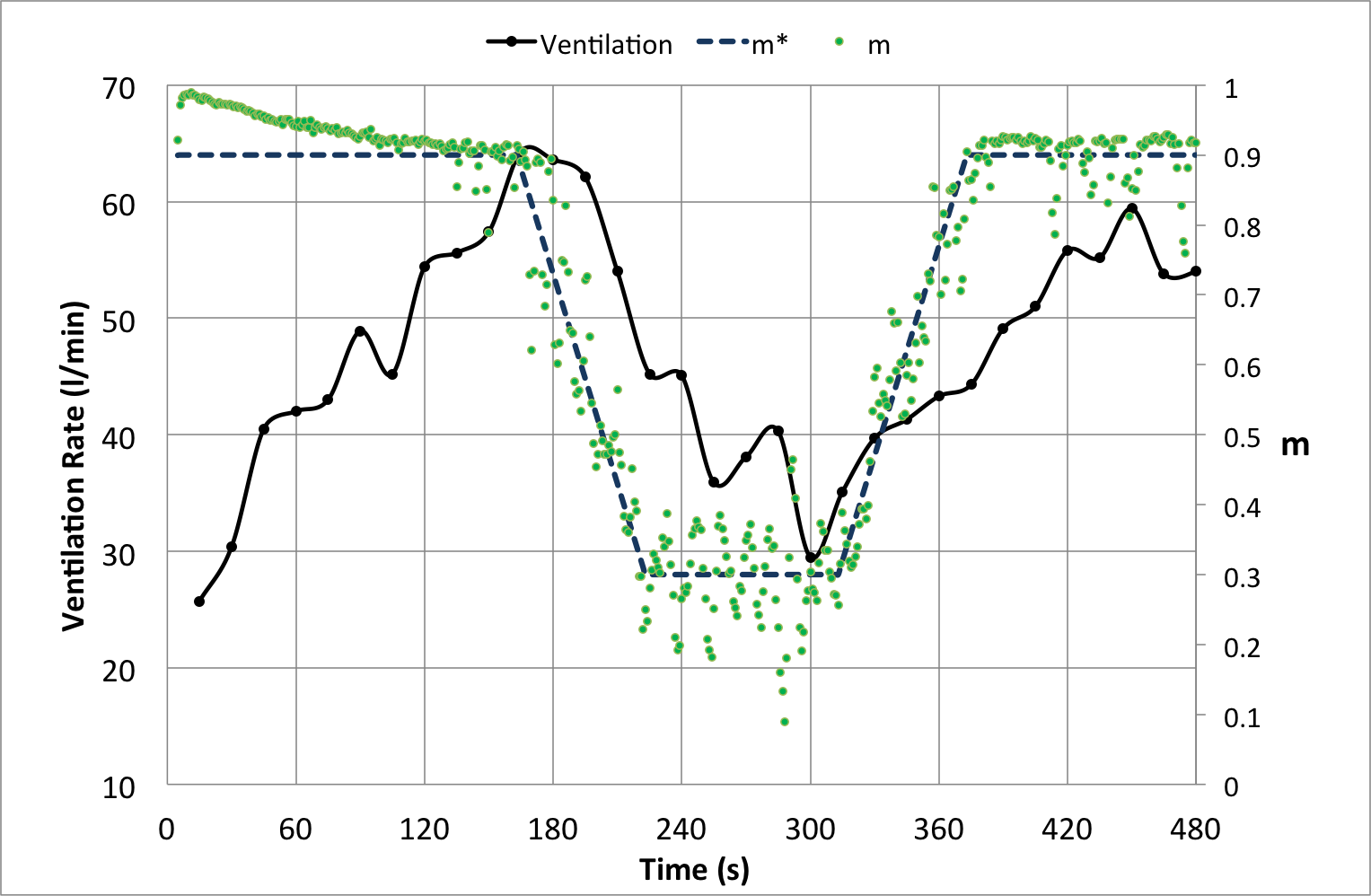}
			\caption{Ventilation rate (black line), target value of $m^*$ (blue dashed line) and actual value of $\bar{m}$ (green dots) plotted versus time.} 
			\label{fig:cl_pol_test_m_br}
		\end{figure}

		It is of particular interest to understand how the cyclist's breathing rate varied as the cyclist travelled through the different zones of the simulation. Figure \ref{fig:cl_pol_test_m_br} plots the cyclist's breathing rate during the experiment versus the target value $m^*$.
		As can be seen from Figure \ref{fig:cl_pol_test_m_br}, changes in the breathing rate (black) approximately track changes in the target value $m^*$. The experiment started with $m^*=0.9$ meaning that the human was providing 90\% of the power required to move the wheel and the motor was only providing 10\% of the power. At the start of the experiment, the cyclist's breathing rate was approximately 25 litres/minute. When they started to cycle in the simulated non-polluted area ($m^*=0.9$), their breathing rate was seen to rise to a peak of 65 litres/minute.
		
		The gradual reduction in the blue dashed curve from $m^*=0.9$ to $m^*=0.3$ corresponded to the cyclist entering the transient zone. In the transient zone (and the polluted zone), the human was only providing 30\% of the power required to move the wheel of the bike with the other 70\% being provided by the motor. As clearly seen from Figure \ref{fig:cl_pol_test_m_br}, when $m^*$ drops to 0.3, the breathing rate decreased to approximately 30 litres/minute.

	\subsection{Conclusion on use case}\label{subsec:cl_pol_conclusion}
		This use case built on the use case from Section \ref{sec:ol_pollution}. The objective of both use cases was to propose a method by which the amount of pollutants that a cyclist inhaled could be reduced while cycling through a polluted area. The method implementation was different from the open loop case, in that it involved implementing closed loop control. The closed loop control relied heavily on the model derived in Chapter \ref{ch:modelling}. The objective of the control circuit was to ensure that the actually observed value $\bar{m}$ tracked the target value of $m^*$. It was shown (see Figure \ref{fig:cl_m}) that the control circuit was able to track the target value $m^*$ with an error within $\pm$ 10\%, 90\% of the time. It was not expected to be able to achieve perfect tracking between the actual value $\bar{m}$ and the target value $m^*$ due to the nature of the system.
		
		The effect these changes in $m^*$ had on the cyclist's breathing rate was monitored using a spirometer. It was found that the cyclist's breathing rate approximately halved when the target value $m^*$ was reduced from 0.9 to 0.3. This proves the hypothesis that was being tested which was to show that the cyclist's breathing rate could be reduced by decreasing the value of $m^*$. In a polluted area, if the cyclist's breathing rate is reduced, the amount of air pollutants that they inhale are also reduced. As such, a new method has been presented that reduces the amount of pollutants that a cyclist inhales while cycling through a polluted area.
		
		Many of the extensions to the use case that were listen in Section \ref{subsec:ol_pol_conclusion} are also applicable here, particularly the next steps to expand the simulation by carrying out a real world cycling experiment. Significance testing should also be carried out based on the reduction in breathing rate and the associated reduction in inhaled air pollutants. Further refinement of the specific parameters in the control circuit could also be carried out including 1) the sampling period 2) the gain parameter and 3) the way in which $m^*$ is reduced and increased when entering the transient zone and leaving the polluted zone.
		
		Using the spirometer to record the breathing rate and the finger pulse oximeter to record the heart rate also helped to validate that heart rate was a good indicator of breathing rate. As seen from Figure \ref{fig:cl_pol_test_hr_br}, both signals approximately track each other.  This meant that for real life non-stationary cycling simulations, measuring the cyclist's heart rate using an activity tracker would be acceptable to validate that the cyclist's breathing rate had been reduced. This was done for the open loop case in Section \ref{subsec:ol_pol_experiment}. The transient between the breathing rate and heart rate should be studied further and incorporated into the methodology to determine the size of the transient area in the simulation.
		
		This use case will be presented at the Eighteenth Yale Workshop on Adaptive and Learning Systems \footnote{https://www.eng.yale.edu/css/} at Yale University, USA in June 2017.

%% file: chapters/ch_conclusions.tex
\chapter{Conclusions} \label{ch:conclusions}
\setstretch{1.4}
	This project had one main objective which was to demonstrate different services that electric bikes can provide in smart cities to help and protect people. In order to demonstrate these services, it was first required to define and implement the concept of a smart electric bike. A smart electric bike was defined as an electric bike that is situation aware and can use this awareness to respond to its surroundings to deliver a service to help and protect people.
	
	A number of design criteria were stated in Section \ref{sec:design_considerations}. A smart electric bike which met those design criteria was designed and fully instrumented. The hardware components of the smart electric bike were discussed in detail in Section \ref{sec:hardware} and the software components were discussed in Section \ref{sec:software}. A key design decision was to connect the electric bike and the smartphone by Bluetooth in order to widen the range of services that the smart electric bike could offer. This meant that the electric bike could provide services using any data stream that was available on the internet. A number of proposed improvements to the system are discussed in Section \ref{sec:final_design_considerations}.
	
	When the system was fully instrumented, it was then possible to show how the electric bike could provide different services in cities to help and protect people. Chapter \ref{ch:modelling}, Chapter \ref{ch:openloop} and Chapter \ref{ch:closedloop} focused on these services. Section \ref{sec:ol_pollution} proposed a method which aimed to reduce the amount of pollutants that a cyclist inhaled when they were cycling through a polluted area. This method was based on the concept that in order to reduce the amount of pollutants that a cyclist inhaled that their breathing rate should be reduced. This method was tested and validated in Section \ref{subsec:ol_pol_experiment}. 
	
	The open loop use case presented above was demonstrated as a proof of concept for Dublin City Council and the Fraunhofer-Morgenstadt Initiative as part of a smart transport solutions demonstration in Trinity College Dublin in mid-March 2017. When this proof of concept was established, it was decided to expand on this use case by implementing closed loop control. In order to implement closed loop control it was necessary to have a reliable model of the system behaviour, this was developed in Chapter \ref{ch:modelling}. Section \ref{sec:cl_min_breathing} then relied heavily on this model to implement closed loop control.
	
	For the closed loop use case in Section \ref{sec:cl_min_breathing}, the variable $m$ introduced in Chapter \ref{ch:modelling} was made use of to manage the proportion of power that the cyclist and the motor were providing to move the wheel of the bike. It was demonstrated that it was possible to regulate the observed value $\bar{m}$ so that it tracked a target value $m^*$. The error between the target and the actual value of $m$ was maintained within $\pm$ 10\%, 90\% of the time. It is believed that this could be improved by further refining the control parameters including the gain $\gamma$, the sampling period $T_s$ and the way in which the target value $m^*$ changes when entering and leaving a polluted area.
	
	It was validated that the cyclist's breathing rate was reduced in different ways for both the open loop and the closed loop cases. In the closed loop case, the cyclist was equipped with spirometry equipment to measure the volumetric rate at which they were breathing in air. They were also equipped with a finger pulse oximeter which monitored their heart rate. It was found that the user's breathing rate more than doubled from 25 litres/minute to 65 litres/minute when they were cycling from rest in a non-polluted area ($m^*=0.9$). By reducing the value of $m^*$ to 0.3, it was found that this reduced the cyclist's breathing rate to approximately 30 litres/minute. 
	
	It was also found that for the steady state case, there was a strong correlation between the breathing rate and heart rate signals which agreed with the review of literature discussed in Section \ref{sec:lr_validation_hr}. This fact meant that (to a first approximation) it was possible to use a measurement of the heart rate to validate that the cyclist's breathing rate had been reduced. This was made use of for validation in the open loop case where the cyclist wore an activity tracker which measured their heart rate. This meant that for non-stationary cycling tests, the cyclist would not need to be equipped with spirometry equipment. The transient length in the breathing rate signal and the heart rate signal seem to be different and should be studied further. If they are different, this would impact the size of the required transient zone.
	
	Many extensions of this use case are possible. This includes changes to the model from Chapter \ref{ch:modelling} to generalise it further. The model should be expanded to include different cycling environments particularly including roads with different positive and negative gradients. The use case could be demonstrated further by carrying out a real life cycling experiment where a user would cycle on a specified path through an area with a specific part of the area being designated as a polluted zone. The system could then use location data available on the smartphone to update the value of $m^*$ while in that area. This  use case will be presented at the Eighteenth Yale Workshop on Adaptive and Learning Systems at Yale University in the USA in June 2017.
	
	It is possible to demonstrate many other use cases using the smart electric bike. An example would be a service to manage the way in which energy is delivered to the cyclist over a cycling trip. This would involve the system becoming aware of the final destination of the cyclist, and then using that knowledge to optimally deliver the energy in the battery to the cyclist over the course of the journey. This could involve saving energy in the battery for times when the cyclist encounters large hills on their route and therefore needs extra assistance. This use case could be particularly useful for vulnerable groups of road users like older people or people who have physical disabilities.
	
	Another interesting use case would be to examine the potential of the smart electric bike for smart routing applications. In this use case, the system would become aware of the cyclist's final destination on their journey and would then use smart routing to recommend the optimal route for the cyclist to travel on to achieve some objective. This objective might be to go on a route that minimises the cyclist's inhalation of pollutants, or that has the best cycling paths or the least number of cars. 
	
	Another use case would be to make use of some of the data from the activity tracker (discussed in Section \ref{sec:ol_pollution}). Activity trackers like the Microsoft Band estimate the amount of metabolic calories that the user is burning off. It would be possible to implement a closed loop control circuit whereby a cyclist could set a target rate that they wish to burn calories at over a journey and the control circuit could regulate the amount of power that the motor is delivering to the human in order to ensure that they stay on track with their target rate of calories burned. It would also be valuable to investigate the correlation between this estimation of metabolic calories burned and the human power input at the pedals. If the activity tracker calories burned measurement and the human power input measurement tracked each other well, it might be possible to substitute the THUN X-CELL RT sensor for an activity tracker for some use cases which would cut down on costs.
	
	The use case regarding reducing a cyclist's inhalation of pollutants generated particular interest which was why it was focused on so much in Chapters \ref{ch:openloop} and \ref{ch:closedloop}. In general, there are a large number of potential services that electric bikes can provide in our cities to help and protect people. They can also help to solve or lessen a number of problems in cities like congestion due to cars, parking problems and to reduce the concentration of air pollutants in cities which would have a direct impact on improving human health.